\documentclass[aps,pra,twocolumn,preprintnumbers]{revtex4}

\usepackage{amsmath,amssymb,bm,graphicx}
\usepackage{subfigure}
\usepackage{appendix}
\usepackage{natbib}

\begin{document}

\title{Compressibility, zero sound, and effective mass of a fermionic dipolar gas at finite temperature}
\author{J.~P.~Kestner and S.~Das Sarma}
\affiliation{Condensed Matter Theory Center, Department of Physics, University of Maryland, College Park, MD 20742}

\begin{abstract}
The compressibility, zero sound dispersion, and effective mass of a gas of fermionic dipolar molecules is calculated at finite temperature for one-, two-, and three-dimensional uniform systems, and in a multilayer quasi-two-dimensional system.  The compressibility is nonmonotonic in the reduced temperature, $T/T_F$, exhibiting a maximum at finite temperature.  This effect might be visible in a quasi-low-dimensional experiment, providing a clear signature of the onset of many-body quantum degeneracy effects.  The collective mode dispersion and effective mass show similar nontrivial temperature and density dependence.  In a quasi-low-dimensional system, the zero sound mode may propagate at experimentally attainable temperatures.
\end{abstract}

\maketitle
\section{Introduction}
The study of ultracold atomic systems has received much attention in recent years, motivated largely by the prospect of realizing novel strongly correlated many-body physics.  A series of remarkable experimental breakthroughs have produced an extremely well-controlled physics playground~\citep{Lewenstein07, Bloch08, Ketterle08}.  The latest breakthrough is the ability to associate atoms via a Feshbach resonance to form ultracold molecules in the ro-vibrational ground state~\citep{Deiglmayr08, Lang08}, and the JILA group has achieved a nearly degenerate gas of ground state polar molecules~\citep{Ospelkaus08, Ni08, Ni09, Ospelkaus09, Ni10}.  This is a true milestone in the program, since strong dipolar interactions between degenerate molecules in the presence of an external electric field allow for the design of exotic Hamiltonians~\citep{Micheli06, Ortner09} and are expected to give rise to exciting phenomena including roton softening~\citep{Santos03, Ronen07, Wang08}, supersolidity~\citep{Sengupta05, Boninsegni05, Capogrosso09, Pollet09, Burnell09}, artificial photons~\citep{Tewari06}, bilayer quantum phase transitions~\citep{Wang07}, and multi-layer self-assembled chains~\citep{Wang06} for bosonic molecules, and spontaneous interlayer superfluidity \citep{Lutchyn09}, itinerant ferroelectricity~\citep{Lin09}, Fermi liquid anisotropy~\citep{Sogo09, Miyakawa08, Chan10}, fractional quantum Hall effect~\citep{Baranov05}, Wigner crystallization~\citep{Baranov08a}, biaxial nematic phase~\citep{Fregoso09}, topological superfluidity~\citep{Cooper09}, and a $Z_2$ topological phase~\citep{Sun09} for fermionic molecules.

Most, if not all, of these novel quantum phases will require temperatures on the order of $0.1 T_F$ or less (with $T_F$ the Fermi temperature), which will require further experimental advance.  Efforts to overcome the current obstacles of collisional instability and insufficient cooling are already underway, though, and given the rate of progress in reaching the current state-of-the-art, the future is bright~\citep{Baranov08b, Pupillo08, Carr09}. However, it is worthwhile to ask what interesting effects one might be able to observe in the immediate future with the temperature on the order of $T_F$.

From a condensed matter perspective, the very idea of having a $1/r^3$ interaction rather than the usual Coulomb interaction is intriguing.  Such a system has no parallel in solid state materials, and study of the previously unmotivated problem of many-body physics in a system interacting via a $1/r^3$ potential is still in its infancy.  Of course, the actual form of the inter-molecular interaction potential is quite complicated, particularly at short range, and the scattering and stability of the molecules are sensitive to these details \citep{Micheli10}.  Further investigation is required to allow full quantitative comparison between theory and experiment.  For our purposes, though, it is sufficient to take a $1/r^3$ interaction with some undetermined short-range cutoff; the low-energy many-body physics of a stable, dilute gas should not depend qualitatively on the short-range details.  It is then appropriate to consider what quantum many-body effects might be calculated for a gas with $1/r^3$ interactions that could be observed in cold polar molecule experiments.

In this paper, we calculate the compressibility (or equivalently, the ordinary sound dispersion), zero sound dispersion, and effective mass at finite temperatures comparable to those achieved in current experiments.  These quantities should be readily accessible to experiment, and we present the first calculations to include thermal effects, as well as some trap effects.  As these calculations are carried out within a leading order perturbation approximation in the dipolar coupling constant, we are working within the standard weak-coupling theory in the sense of the Landau Fermi liquid theory.  This is the first theory one must do before one does anything else for the dipolar systems.

For dipolar interactions, in contrast to the familiar Coulomb case, the weak-coupling regime corresponds to the \emph{low}-density limit.  This is fortuitous, since current experiments cannot achieve high densities without significant loss rates.  For all practical purposes, the leading-order perturbative results are exact at typical experimental densities.  However, lower densities correspond to higher $T/T_F$, necessitating the calculations be done for finite temperature.  As the density is increased, the leading-order perturbative results should remain qualitatively correct as long there is no phase transition to break adiabaticity.  For very high densities one will enter the interesting strong-coupling regime and the weak-coupling theory will fail in a systematic way, but this regime is completely inaccessible at present in dipolar molecular systems.

In two-dimensional semiconductor-based electron systems and in graphene, studying compressibility experimentally \citep{Eisenstein94, Rahimi03, Martin08} and theoretically \citep{Hwang07, DasSarma09} has been an important tool for studying quantum many-body effects in Coulomb systems.  In these two-dimensional condensed matter systems with the $1/r$ Coulomb interaction, the Hartree-Fock approximation works remarkably well for understanding compressibility, even in the strongly interacting regime.  This is a general result of the frequency independence of the compressibility, and not specific to the Coulomb interaction.

We allow for different trap geometries by considering uniform three-dimensional (3D), two-dimensional (2D), and one-dimensional (1D) gases, as well as a nonuniform 3D gas in a periodic potential along $z$.  We find that under certain conditions the compressibility varies nonmonotonically with temperature.  A closely related quantity with the same temperature dependence, $dE_F/d\mu$ (with $E_F$ the Fermi energy and $\mu$ the chemical potential), also varies nonmonotonically with density, even for temperatures on the order of $T_F$.  The zero sound speed and effective mass also exhibit nontrivial dependencies.  However, in 3D, propagation of zero sound requires lower temperatures than currently feasible.  In 2D and 1D, though, we find that zero sound propagates undamped even at experimentally realistic temperatures, assuming that the intermolecular potential behaves like $1/r^3$ at short enough distances.  In the low-dimensional situations, the geometry with dipoles aligned perpendicular to their separations is particularly relevant as this is the most stable configuration conducive to our approximation of a $1/r^3$ interaction, so we will focus on this case.

The layout of the paper is as follows:  In section~\ref{sec:Vq} we review the form of the dipolar interaction in momentum space.  In section~\ref{sec:results} we present analytic low-temperature expansions for the self-energy, compressibility, and effective mass of a uniform system, as well as numerical results for arbitrary temperature.  We also present numerical results for the zero sound mode.  In section~\ref{sec:multilayer} we numerically obtain the compressibility and low-lying collective modes for the spatially inhomogeneous case of a 1D periodic external potential dividing the 3D cloud into multiple quasi-2D layers.  In section~\ref{sec:exp}, we discuss experimental observation of quantum many-body effects in the compressibility, and we conclude in section~\ref{sec:conclusions}.  In Appendix A, we briefly compare with the case of Coulomb interactions.

\section{Dipole potential in momentum space}\label{sec:Vq}
We consider a gas of one-component fermionic molecules with number density $n$, each possessing dipole moment $\mathbf{d}$, aligned by an external electric field $\mathbf{E}$ so that the inter-molecular interaction potential at large separation $\mathbf{r}$ is given by
\begin{equation}\label{eq:V3Dr}
V_{3D}\left(\mathbf{r}\right) = \frac{d^2}{r^3}\left(1 - 3 \cos^2 \theta_r \right)
\end{equation}
where $\theta_r$ is the angle between $\mathbf{r}$ and $\mathbf{E}$.  We will parameterize the interaction using the dimensionless ratio $\lambda = d^2/r_0^3 E_F$, where $r_0$ is the average interparticle distance.  For the different dimensionalities we have
\begin{equation}
\lambda_{3D} = \frac{m d^2 k_{F0}}{3\pi^2 \hbar^2}, \quad \lambda_{2D} = \frac{m d^2 k_{F0}}{4\pi^{3/2} \hbar^2}, \quad \lambda_{1D} = \frac{2 m d^2 k_{F0}}{\pi^3 \hbar^2} ,
\end{equation}
where $k_{F0}$ is the noninteracting Fermi wavevector.  For fermionic KRb ($d = 0.57$ Debye, $m = 127$ amu) at densities around $10^{12} cm^{-3}$ \citep{Ospelkaus08,Ni08,Ni09} (or $10^{8} cm^{-2}$ or $10^{4} cm^{-1}$ in lower dimensions), $\lambda \sim 0.1$.

Strictly speaking, the Fourier transform of $V_{3D}\left(\mathbf{r}\right)$ has ultraviolet and infrared divergences.  However, recall that anyways this potential is not valid for arbitrarily short range and the system has some finite size.  Thus, as in Ref.~\citep{Chan10}, we introduce a short-range cutoff, $\epsilon$, and a long-range cutoff, $R$.  The Fourier transform can then be performed to obtain
\begin{multline}\label{eq:V3Dq}
V_{3D}\left(\mathbf{q}\right) = 8\pi d^2 P_2 \!\left(\cos \theta_{q}\right) \left[\frac{j_1 \left(q \epsilon \right) }{q\epsilon} - \frac{j_1 \left(q R \right) }{qR} \right]
\\
\overset{q\epsilon \rightarrow 0, qR \rightarrow \infty}{\longrightarrow} \frac{8\pi}{3} d^2 P_2 \!\left( \cos \theta_{q} \right) = 16 \pi^3 \lambda_{3D} P_2 \!\left( \cos \theta_{q} \right) \frac{E_F}{k_{F0}^3}
\end{multline}
where $P_2 \!\left(x\right) = \left(3 x^2-1 \right)/2$ is the second Legendre polynomial, $\theta_q$ is the angle between $\mathbf{q}$ and $\mathbf{E}$, and $j_1 \left(x\right)$ is the spherical bessel function of the first kind.

If there is strong confinement along $z$, then assuming a fixed gaussian profile in that direction $n\left( k_z \right) = e^{-k_z^2 w^2/4}$, and integrating Eq.~\ref{eq:V3Dq} over this momentum profile, the effective 2D interaction is
\begin{multline}\label{eq:V2Dq}
V_{2D}\left(\mathbf{q}\right) = 16\pi^{5/2} \lambda_{2D} \biggl[ \frac{4}{3 k_{F0}\sqrt{\pi} w} P_2 \!\left(\cos \theta_{E}\right)
\\
- \frac{q}{k_{F0}} \left( P_2 \!\left(\cos \theta_{E}\right) - \frac{1}{2} \sin^2 \theta_E \cos 2\phi_q \right) \biggr] \frac{E_F}{k_{F0}^2}
\end{multline}
where $\theta_E$ is the angle between $\mathbf{E}$ and the $z$-axis, $\phi_q$ is the azimuthal angle between $\mathbf{q}$ and the projection of $\mathbf{E}$ onto the $x-y$ plane, and we have taken $q w \rightarrow 0$.  Alternatively, fixing $\mathbf{r}$ to lie in the $x-y$ plane and taking the 2D Fourier transform of Eq.~\ref{eq:V3Dr} with a short-range cutoff, $w$, as in Ref.~\citep{Chan10} yields the same result with a numerical factor of order unity in front of $w$.

Similarly, when there is strong confinement and a fixed gaussian profile $n\left( k_{\perp} \right) = e^{-k_{\perp}^2 w^2}$ in the radial direction, integrating out the radial degree of freedom in Eq.~\ref{eq:V3Dq} yields an effective 1D interaction
\begin{multline}\label{eq:V1Dq}
V_{1D}\left(q\right) = \pi^3 \lambda_{1D} P_2 \!\left(\cos \theta_E \right) \biggl[-\frac{1}{3 w^2}
\\
 + q^2 e^{q^2 w^2} \Gamma\! \left(0, q^2 w^2\right) \biggr] \frac{E_F}{k_{F0}^3}
\\
\overset{qw \rightarrow 0}{\longrightarrow} -\pi^3 \lambda_{1D} P_2 \!\left(\cos \theta_{E}\right) \left[ \frac{1}{3 w^2} + q^2\left(\gamma + 2\ln |q w| \right) \right] \frac{E_F}{k_{F0}^3}
\end{multline}
where $\theta_E$ is again the angle between $\mathbf{E}$ and the $z$-axis, $\Gamma\!\left(a, x\right)$ is the incomplete gamma function, and $\gamma \approx 0.577$ is Euler's constant. (Here again, performing a 1D Fourier transform of a $1/z^3$ potential with a short-distance cutoff $w$ gives a similar dependence on $q$ and $w$, although the numerical factors are different.)  Unique to the quasi-1D case, the momentum dependent part of the effective interaction is not independent of the transverse width, even for very strong confinement.

\section{Uniform self-energy, compressibility, and collective modes at finite temperature}\label{sec:results}
To calculate the compressibility, $\kappa = \frac{1}{n^2} \frac{dn}{d\mu}$~\citep{Mahan}, we first need to obtain the chemical potential in the presence of the dipolar interactions.  In the uniform case, to first order, $\mu = \mu_0 + \Sigma^{iso} \left( k_{F0} \right)$, where $\mu_0$ is the noninteracting chemical potential and $\Sigma^{\text{iso}} \left( k_{F0} \right)$ is the isotropic part of the Hartree-Fock self-energy at the unperturbed Fermi surface,
\begin{equation}\label{eq:sigma}
\Sigma \left( \mathbf{k_{F0}} \right) = 1/\left(2\pi\right)^3 \int d^3 \mathbf{k} n_0 \!\left(k\right) \left( V \left(0\right) - V \left( |\mathbf{k_{F0}-k}| \right) \right)
\end{equation}
with $n_0 \!\left(\mathbf{k}\right) = 1/\left(e^{\left(\hbar^2 k^2/2m - \mu_0\right)/k_B T} + 1\right)$ the noninteracting Fermi distribution.  The zero-temperature self-energy has recently been calculated in Ref.~\citep{Chan10}.  Below we extend this to finite temperature and obtain the inverse compressibility, effective mass, and zero sound dispersion.

\subsection{3D}\label{subsec:3D}
\subsubsection{Compressibility and effective mass}
\begin{figure*}[]
  \includegraphics[width=.66\columnwidth]{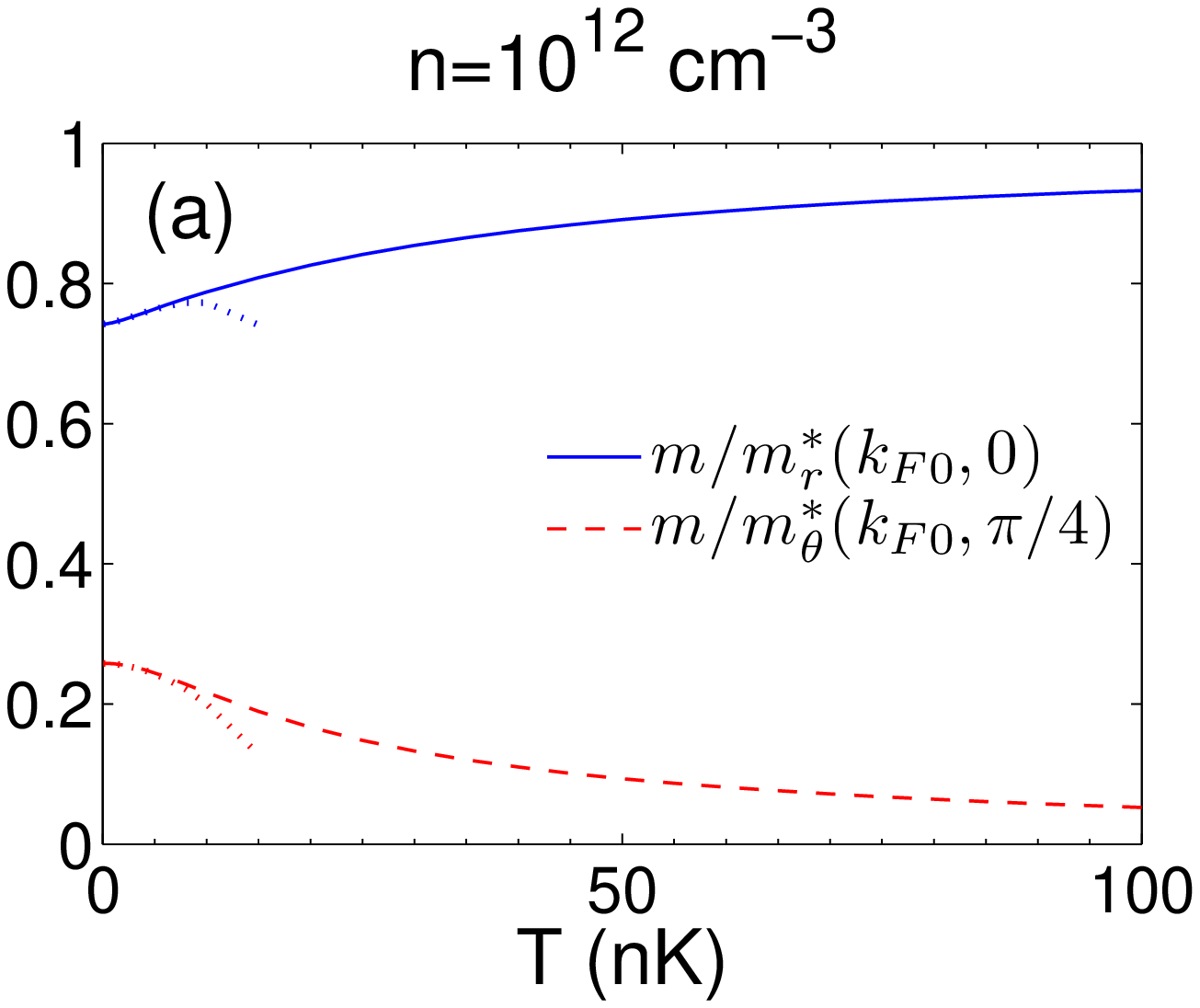}\label{fig:3DmeffvsT}
  \includegraphics[width=.66\columnwidth]{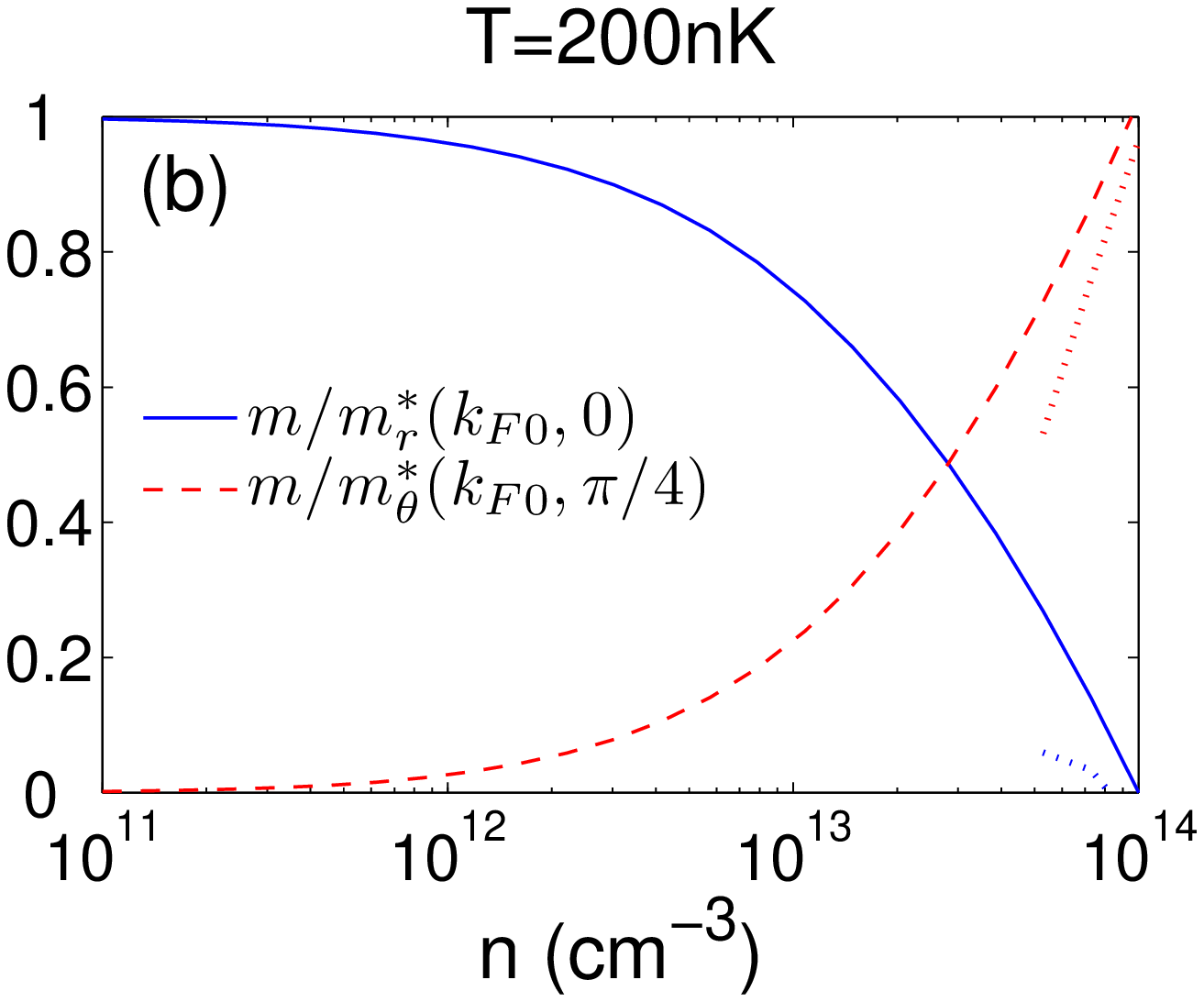}\label{fig:3Dmeffvsn}
  \includegraphics[width=.66\columnwidth]{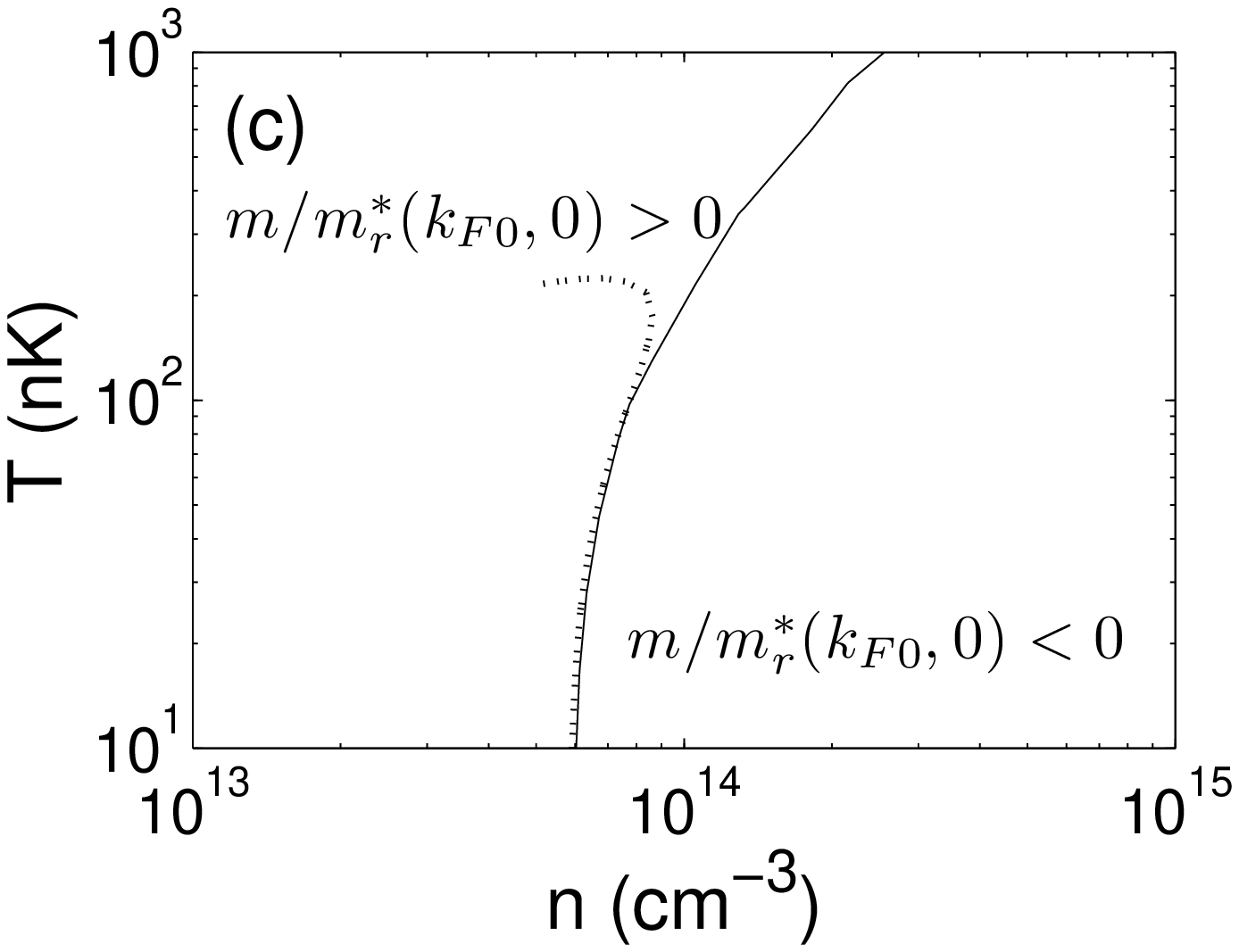}\label{fig:3Dstability}
\caption{Effective mass at the Fermi surface for KRb in 3D (a) vs temperature at fixed density and (b) vs density at fixed temperature. (c) Regions of positive and negative radial effective mass at the poles of the Fermi surface. In all panels, the dotted lines show the low $T/T_F$ approximations.}\label{fig:3Dmeff}
\end{figure*}
Let us begin by considering the 3D case at finite temperature $t \equiv k_B T/E_F$, with $\tilde{\mu_0} \equiv \mu_0/E_F  \approx 1 - \pi^2 t^2/12$,
\begin{widetext}
\begin{align}
\Sigma_{3D} \left( \mathbf{k} \right) &= \lambda_{3D} \frac{E_F}{k_{F0}^3} \int_0^{\infty} k'^2 dk' n_0 \!\left(k'\right) \int_{4\pi} d\Omega_{\mathbf{k'}} \left[ 1 - \frac{3 \left(k \cos \theta_{\mathbf{k}} - k' \cos \theta_{\mathbf{k'}} \right)^2}{k^2 + k'^2 -2 k k' \left( \cos \theta_{\mathbf{k}} \cos \theta_{\mathbf{k'}} + \sin \theta_{\mathbf{k}} \sin \theta_{\mathbf{k'}} \cos \phi_{\mathbf{k'}} \right) } \right] \label{eq:3Dsigma1}
\\
&= 2\pi\lambda_{3D} E_F \frac{k^3}{k_{F0}^3}  P_2 \!\left(\cos \theta_{\mathbf{k}} \right) \int_0^{\infty} \frac{dx}{e^{\left(x^2 k^2/k_{F0}^2 - \tilde{\mu_0} \right)/t} + 1} \left[ -\frac{5}{2}x^2 + \frac{3}{2} x^4 + \frac{3}{4} x \left(x^2-1\right)^2 \ln \left| \frac{x-1}{x+1} \right| \right] \label{eq:3Dsigma2}
\\
\Sigma_{3D} \left( \mathbf{k_{F0}} \right) &= 2\pi\lambda_{3D} E_F P_2 \!\left(\cos \theta_{\mathbf{k_{F0}}} \right) \Biggl[ \frac{ 5 \sqrt{\pi}}{8} t^{3/2} \text{Li}_{\frac{3}{2}} \left(-e^{\tilde{\mu_0}/t}\right) - \frac{ 9 \sqrt{\pi} }{16} t^{5/2} \text{Li}_{\frac{5}{2}} \left(-e^{\tilde{\mu_0}/t}\right) \notag
\\
& \qquad + \frac{3}{4} \int_0^{\sqrt{\tilde{\mu_0}} } x dx \left(x^2-1\right)^2 \ln \left| \frac{x-1}{x+1} \right| + \frac{3}{4} \int_0^{\infty} \frac{dx \text{sgn} \left( x^2 - \tilde{\mu_0} \right) }{e^{\left|x^2 - \tilde{\mu_0} \right|/t} + 1} x \left(x^2-1\right)^2 \ln \left| \frac{x-1}{x+1} \right| \Biggr] \label{eq:3Dsigma3}
\\
&=  2\pi\lambda_{3D} E_F P_2 \!\left(\cos \theta_{\mathbf{k_{F0}}} \right) \Biggl\{ \frac{3 \tilde{\mu_0}^{5/2}}{10} - \frac{5 \tilde{\mu_0}^{3/2}}{6} + \frac{9 \pi^2 \tilde{\mu_0}^{1/2}}{48} t^2 - \frac{5 \pi^2}{48 \tilde{\mu_0}^{1/2}} t^2 - \frac{2}{15} \notag
\\
& \qquad + \frac{\pi^2}{16} t^2 \left[- \left(1 - \tilde{\mu_0} \right) \left(1 + 2\zeta' \left(-1\right) + 2 \ln \frac{4\pi}{\left( \sqrt{\tilde{\mu_0}} + 1 \right)^2} \right) + \frac{\left(1 - \tilde{\mu_0} \right)^2}{\tilde{\mu_0} + \sqrt{\tilde{\mu_0}}} \right] + O\left(t^4\right) + O\left(\left(1 - \tilde{\mu_0} \right)^3 \right) \Biggr\} \label{eq:3Dsigma4}
\\
&= 4\pi\lambda_{3D} E_F P_2 \!\left(\cos \theta_{\mathbf{k_{F0}}} \right) \left[ -\frac{1}{3} + \frac{\pi^2}{16} t^2 + O\left(t^4 \ln t\right) \right] \label{eq:3Dsigma5}
\end{align}
\begin{align}
\frac{\partial \Sigma_{3D} \left( \mathbf{k} \right) }{\partial k} |_{k=k_{F0}} &= \frac{1}{k_{F0}} \left(3 - 2 t \frac{\partial}{\partial t} - 2 \tilde{\mu_0} \frac{\partial}{\partial \tilde{\mu_0}} \right) \Sigma_{3D} \left( \mathbf{k_{F0}} \right) \label{eq:3Ddsigma1}
\\
&= - 2\pi\lambda_{3D} \frac{E_F}{k_{F0}} P_2 \!\left(\cos \theta_{\mathbf{k}} \right) \left[1 +\frac{\pi^2}{4} t^2 \ln t + \left(\frac{3}{8} + 3 \zeta' \left(-1\right) + \frac{1}{4} \ln \pi \right)\pi^2 t^2 + O\left(t^4 \ln t\right) \right] \label{eq:3Ddsigma2}
\end{align}
\end{widetext}
where $\theta_{\mathbf{k}}$ is the angle between $\mathbf{k}$ and $\mathbf{E}$, $\text{Li}_a \left(x\right)$ is the polylogarithm function, and $\zeta' \left(x\right)$ is the derivative of the Riemann zeta function.

In this case, the self-energy has a purely $d$-wave angular dependence and does not affect the compressibility, which is given in units of the zero-temperature, noninteracting compressibility $\kappa_0 \equiv \frac{1}{n^2} \frac{d n}{d E_F} $ by
\begin{equation}
\frac{\kappa_0}{\kappa} \approx 1 + \frac{\pi^2}{12} t^2.
\end{equation}

The temperature-dependent effective masses along the radial and angular directions can be written to first-order as in the zero-temperature case \citep{Chan10}:
\begin{equation}\label{eq:3Dmeffr}
\frac{m}{m_{r}^{\ast} \left(\mathbf{k}\right)} = 1 + \frac{m}{\hbar^2 k } \frac{\partial \Sigma_{3D} \left( \mathbf{k} \right) }{\partial k}
\end{equation}
\begin{equation}\label{eq:3Dmefft}
\frac{m}{m_{\theta}^{\ast} \left(\mathbf{k}\right) } = \frac{m}{\hbar^2 k^2 } \frac{\partial \Sigma_{3D} \left( \mathbf{k} \right) }{\partial \theta_{\mathbf{k}}}.
\end{equation}
For general temperatures we must evaluate Eq.~\eqref{eq:3Dsigma2} and its derivative numerically. At low temperatures, though, Eqs.~\eqref{eq:3Dsigma5} and \eqref{eq:3Ddsigma2} give analytic approximations on the Fermi surface.  We have plotted the numerical inverse effective mass results at the Fermi surface as functions of temperature and density in Fig.~\ref{fig:3Dmeff}, as well as the low-temperature expansions.

Note that the deviation of the inverse effective radial mass from the inverse bare mass is proportional to $-P_2 \!\left(\cos \theta_{\mathbf{k}} \right)$ (the integrand in Eq.~\eqref{eq:3Dsigma2} is always negative), so the effective radial mass is smaller than the bare mass at the equator and larger (or negative) at the poles.  Also, the effective angular mass is proportional to $\csc 2\theta_{\mathbf{k}}$.  The average effective mass over the Fermi sphere is thus unchanged by the interaction, but when the effective radial mass at the poles of the Fermi surface is negative, the system is unstable.  In the low-temperature limit we have the stability condition $\lambda_{3D} < 1/\pi$, in excellent agreement with previous numerical results using a variational approach which gave $\lambda_{3D} < 0.32$ \citep{Sogo09}.  Extending this result to finite temperature, we obtain the critical line for KRb in the $n-T$ plane plotted in Fig.~\ref{fig:3Dmeff}(c).  However, the system actually becomes unstable at densities an order of magnitude lower due to formation of density waves perpendicular to the electric field \citep{Ronen10,Chan10}.

\subsubsection{Zero sound mode}\label{subsubsec:3Dzerosound}
\begin{figure*}
  \includegraphics[width=\columnwidth]{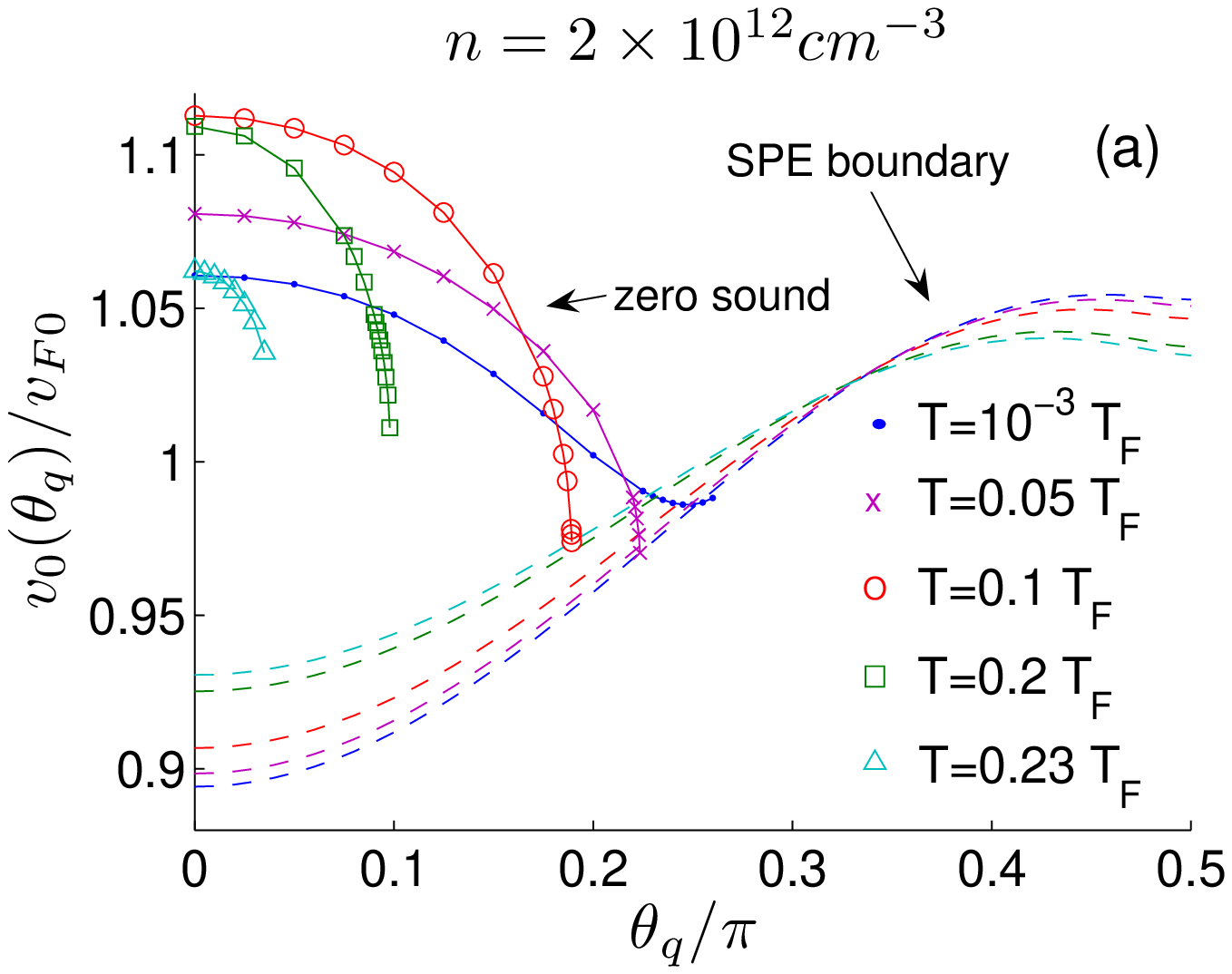}
  \includegraphics[width=\columnwidth]{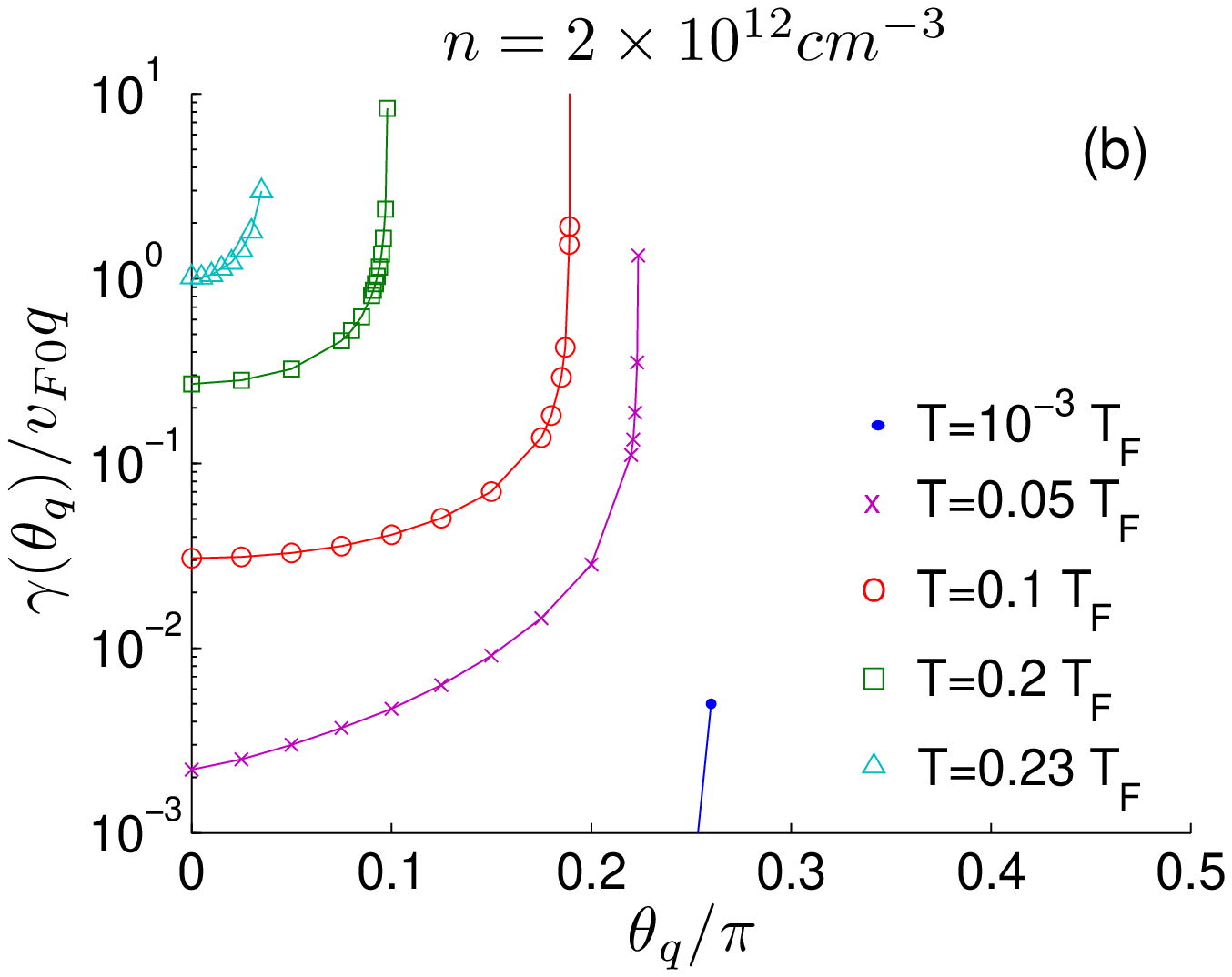}
  \caption{(Color online.) Zero sound speed (a) and damping (b) vs angle of propagation for finite temperatures.  Dashed lines show the boundary of the single-particle excitation continuum.}\label{fig:3Dzerosound}
\end{figure*}
We now turn our attention to the zero sound mode, which has recently been calculated in the zero-temperature limit \citep{Ronen10,Chan10}.  We anticipate that zero sound will only propagate at low temperatures, so we assume that the Landau Fermi liquid description
\begin{equation}
\frac{\delta E}{V} = \!\int\! \frac{d^3 k}{\left(2\pi\right)^3} \epsilon\! \left( \mathbf{k} \right) \delta n\! \left( \mathbf{k} \right) +\! \int\! \frac{d^3 k d^3 k'}{\left(2\pi\right)^6} f\!\left( \mathbf{k, k'} \right) \delta n\! \left( \mathbf{k} \right) \delta n\! \left( \mathbf{k'} \right)
\end{equation}
remains valid, with the temperature dependence entering through the single-particle energy, $\epsilon \!\left(\mathbf{k}\right) = \hbar^2 k^2/2m + \Sigma_{3D} \left(\mathbf{k}\right)$, and the Fermi distribution function, $n \!\left(\mathbf{k}\right) = 1/\left(e^{\left(\epsilon \!\left(\mathbf{k}\right) - \mu_0\right)/k_B T} + 1\right)$. The temperature dependence of the Hartree-Fock quasi-particle interaction, $f\!\left( \mathbf{k, k'} \right) = V_{3D} \left(\mathbf{q} \rightarrow 0\right) - V_{3D} \left(\mathbf{k-k'}\right)$, is negligible since the interaction near the Fermi surface is independent of the magnitude of the momenta in the limit $\left| \mathbf{k-k'} \right| \ll k_F $ \citep{Pethick73}.  Note that, although we have not made the notation explicit, the interaction depends on the direction of momentum transfer.

Neglecting collisions, the linearized Boltzmann equation \citep{Negele88} is
\begin{equation}
\delta \nu \left( \mathbf{k} \right) = \frac{\mathbf{q \cdot \nabla_k} \epsilon \left( \mathbf{k} \right)}{\omega - \mathbf{q \cdot \nabla_k} \epsilon \left( \mathbf{k} \right)} \frac{\partial n \left( \mathbf{k} \right)}{\partial \epsilon \left( \mathbf{k} \right)} \int \frac{d^3 k'}{\left(2\pi\right)^3} f \left( \mathbf{k,k'} \right) \nu \left( \mathbf{k'} \right),
\end{equation}
where $ \delta \nu \left( \mathbf{k} \right)$ is the deviation of the quasi-particle distribution from the equilibrium anisotropic distribution, $n \left( \mathbf{k} \right)$.  Integrating out the radial degree of freedom and decomposing into spherical harmonics, $Y_{lm} \left( \Omega_{\mathbf{k}}^{\mathbf{q}} \right)$, with the angle in the argument measured relative to the direction of momentum transfer, $\mathbf{\hat{q}}$,  we can rewrite the above as
\begin{equation}\label{eq:3Dmateq}
\delta \nu_{lm} = \sum_{l'' l'm'} \chi_{ll'',mm'}\left(\mathbf{q},\omega \right) f_{l''l',m'} \delta \nu_{l'm'},
\end{equation}
where
\begin{equation}
\delta \nu_{lm} = \int \frac{d^3 k}{\left(2\pi\right)^3} Y_{lm}^{\ast} \left( \Omega_{\mathbf{k}}^{\mathbf{q}} \right) \delta \nu \left( \mathbf{k} \right),
\end{equation}
\begin{equation}
f \!\left( \mathbf{k,k'} \right)\! \simeq\! f\!\left( \Omega_{\mathbf{k}}, \Omega_{\mathbf{k'}} \right) \!=\! \frac{E_F}{k_{F0}^3} \sum_{ll'm} f_{ll',m}  Y_{lm}^{\ast} \left( \Omega_{\mathbf{k}}^{\mathbf{q}} \right) Y_{l'm} \left( \Omega_{\mathbf{k'}}^{\mathbf{q}} \right),
\end{equation}
and
\begin{widetext}
\begin{multline}\label{eq:3Dchilm}
\chi_{ll',mm'}\left(\mathbf{q},\omega \right) = \frac{E_F}{k_{F0}^3} \int \frac{d^3 k}{\left(2\pi\right)^3} Y_{lm}^{\ast} \left( \Omega_{\mathbf{k}}^{\mathbf{q}} \right) \frac{\mathbf{q \cdot \nabla_k} \epsilon \left( \mathbf{k} \right) }{\omega - \mathbf{q \cdot \nabla_k} \epsilon \left( \mathbf{k} \right) +i0} \frac{\partial n \left( \mathbf{k} \right)}{\partial \epsilon \left( \mathbf{k} \right)} Y_{l'm'} \left( \Omega_{\mathbf{k}}^{\mathbf{q}} \right)
\\
= \frac{1}{k_{F0}^3} \int \frac{d^3 k}{\left(2\pi\right)^3} Y_{lm}^{\ast} \left( \Omega_{\mathbf{k}}^{\mathbf{q}} \right) \frac{e^{\left(k^2/k_{F0}^2 + \Sigma_{3D}^0 \left(\mathbf{k}\right)/E_F - \tilde{\mu_0} \right)/t}}{t \left(e^{\left(k^2/k_{F0}^2 + \Sigma_{3D}^0 \left(\mathbf{k}\right)/E_F - \tilde{\mu_0} \right)/t}+1\right)^2} \left(1 - \frac{\frac{\omega}{q v_{F0}} }{\frac{\omega}{q v_{F0}} - \frac{k}{k_{F0}} g\left(\mathbf{k},\theta_q\right) + i0} \right) Y_{l'm'} \left( \Omega_{\mathbf{k}}^{\mathbf{q}} \right),
\end{multline}
with $v_{F0}=\hbar^2 k_{F0}/m$ the noninteracting Fermi velocity and
\begin{equation}
g\left(\mathbf{k},\theta_q\right) = \frac{\mathbf{\hat{q}\cdot \nabla_k} \epsilon\! \left(\mathbf{k}\right)}{\hbar^2 k/m} = \left( \cos{\theta_k}\cos{\theta_q} + \sin{\theta_k}\cos{\phi_k}\sin{\theta_q} \right) \frac{m}{m_r^{\ast 0} \left(\mathbf{k}\right)} - \frac{3}{2} \left(\cos{\theta_k}\cos{\phi_k}\sin{\theta_q} - \sin{\theta_k}\cos{\theta_q}\right) \frac{m}{m_{\theta}^{\ast 0} \left(\mathbf{k}\right)} .
\end{equation}
\end{widetext}
In the preceding we have used the Hartree-Fock single-particle energy in order to account for the anisotropy and obtain results consistent with Refs.~\citep{Ronen10,Chan10}.  Since the self-energy also appears in the Fermi distribution, the integrand is evaluated in some thermally broadened range of momenta around the \emph{distorted} Fermi surface.  However, note that in order to consistently include only first-order corrections to the noninteracting single-particle energy, the self-energy terms themselves must be evaluated around the \emph{unperturbed} Fermi surface.  Hence the added superscript denoting the subtle distinction, and for $\mathbf{k}$ not too far below the Fermi surface one can simply take $\Sigma_{3D}^0 \left(\mathbf{k}\right) \equiv \Sigma_{3D} \left(\mathbf{k-k_F+k_{F0}}\right)$, where the latter is given by Eq.~\eqref{eq:3Dsigma2} and the three vectors in the argument are collinear.

We will keep only the coupling between the $s$-wave and longitudinal $p$-wave modes as in Ref.~\citep{Chan10}, which yields good agreement with the results of Ref.~\citep{Ronen10}.  The Landau parameters for the dipolar interaction have previously been calculated \citep{Fregoso09,Chan10}, and in our notation and choice of basis the relevant nonzero parameters are
\begin{equation}
f_{00,0} = 2^6\pi^4 \lambda_{3D} P_2 \!\left(\cos \theta_{\mathbf{k}} \right), \; f_{11,0} = \frac{3}{5} 2^5 \pi^4 \lambda_{3D} P_2 \!\left(\cos \theta_{\mathbf{k}} \right) .
\end{equation}
The energy, $\Omega$, and damping, $\gamma$, of the collective modes for a given momentum, $\mathbf{q}$, are then given by the solutions of $\det \left|I - M\left(\mathbf{q},\Omega - i \gamma \right) \right| = 0$, where
\begin{equation}
M\left(\mathbf{q},\omega \right) = \begin{pmatrix}
                 \chi_{00,00}\left(\mathbf{q},\omega \right) f_{00,0} & \chi_{10,00}\left(\mathbf{q},\omega \right) f_{11,0} \\
                 \chi_{10,00}\left(\mathbf{q},\omega \right) f_{00,0} & \chi_{11,00}\left(\mathbf{q},\omega \right) f_{11,0} \\
               \end{pmatrix}.
\end{equation}

We numerically find solutions corresponding to underdamped zero sound propagation.  In that case, the dispersion is given by
\begin{equation}\label{eq:3Dfreq}
\text{Re} \det \left|I - M\left(\mathbf{q} \rightarrow 0,\Omega \rightarrow v_0 \left(\theta_q\right) q \right) \right| = 0
\end{equation}
and the damping, assumed to be small compared to $\Omega$, is given by
\begin{equation}\label{eq:3Ddamping}
\gamma = \frac{\text{Im} \det \left|I - M\left(\mathbf{q} \rightarrow 0,\Omega \rightarrow v_0 \left(\theta_q\right) q \right) \right|}{\frac{\partial}{\partial \omega}|_{\omega = v_0 \left(\theta_q\right) q} \text{Re} \det \left|I - M\left(\mathbf{q} \rightarrow 0, \omega \right) \right|}
\end{equation}
The dispersion and damping for various temperatures are shown in Figs.~\ref{fig:3Dzerosound}(a) and \ref{fig:3Dzerosound}(b), respectively, with the density chosen such that $\lambda_{3D} = 1/\pi^2$ for comparison with Refs.~\citep{Ronen10,Chan10}.  For temperatures up to about $0.1 T_F$, the thermal effect is to increase the axial propagation speed while decreasing the range of angles for which zero sound propagates.  As the temperature increases further, the axial propagation speed decreases.  Note that at finite temperature, the mode can become overdamped long before it enters the single-particle excitation (SPE) continuum, and Eq.~\eqref{eq:3Dfreq} ceases to have a solution.  For temperatures above $0.2 T_F$ the damping is large even at $\theta_q =0$, and above $\sim 0.23 T_F$ there is no solution to Eq.~\eqref{eq:3Dfreq} for any $\theta_q$.  Clearly, the assumption of $\gamma \ll \Omega$ quickly breaks down with increasing temperature and propagation off the $z$-axis.

\subsection{2D}\label{subsec:2D}
\subsubsection{Compressibility and effective mass}
In the remainder of this manuscript our primary focus will be on dipoles aligned perpendicular to their motional degrees of freedom, so it will be unnecessary to distinguish between distorted and undistorted Fermi surfaces, $\mathbf{k_F}$ and $\mathbf{k_{F0}}$, and we will drop the extra subscript.

For the 2D case, with $\tilde{\mu_0} = t \ln \left( e^{1/t} -1 \right)$, $\theta_E$ the angle between $\mathbf{E}$ and the normal to the plane, and $\phi_{\mathbf{k_{F}}}$ the angle between the Fermi wavevector and the projection of $\mathbf{E}$ onto the plane, from the isotropic part of the interaction we obtain:
\begin{widetext}
\begin{align}
\Sigma_{2D}^{\text{iso}} \left(\mathbf{k}\right) &= 4\sqrt{\pi} \lambda_{2D} P_2 \!\left(\cos \theta_E\right) \frac{E_F}{k_{F}^3} \int_0^{\infty} k' dk' n_0 \!\left(k'\right) \int_0^{2\pi} d\phi_{k'} \sqrt{k'^2+k^2-2k' k \cos \left(\phi_{k'}-\phi_{\mathbf{k}}\right) } \label{eq:2Dsigmaiso1}
\\
&= 16\sqrt{\pi} \lambda_{2D} P_2 \!\left(\cos \theta_E\right) E_F \int_0^{\infty} \frac{x dx}{e^{ \left(x^2 - \tilde{\mu_0} \right)/t} + 1} \left(\frac{k}{k_{F}} +x\right) E\left(\frac{2\sqrt{x k_{F}/k}}{1+x k_{F}/k}\right) \label{eq:2Dsigmaiso2}
\\
\Sigma_{2D}^{\text{iso}} \left(\mathbf{k_{F}}\right) &= 16\sqrt{\pi} \lambda_{2D} P_2 \!\left(\cos \theta_E\right) E_F  \Biggl\{ \int_0^{\tilde{\mu_0}} x dx \left(1+x\right) E\left( \frac{2\sqrt{x}}{1+x} \right) \notag
\\
& \qquad \qquad \qquad \qquad \qquad + \int_0^{\infty} \frac{x dx \, \text{sgn} \left(x^2-\tilde{\mu_0}\right) }{e^{ \left|x^2 - \tilde{\mu_0} \right|/t } + 1} \left(2 - \left(1-x\right) + O\left[\left(1-x\right)^2 \ln |1-x| \right]\right) \Biggr\} \label{eq:2Dsigmaiso3}
\\
&= 16\sqrt{\pi} \lambda_{2D} P_2 \!\left(\cos \theta_E\right) E_F  \left(\frac{8}{9}+\frac{\pi^2}{24} t^2 + O\left[t^4 \ln t\right] \right), \label{eq:2Dsigmaiso4}
\end{align}
\begin{align}
\frac{d\Sigma_{2D}^{\text{iso}}\left( \mathbf{k} \right) }{dk} |_{k=k_{F}} &= 8\sqrt{\pi} \lambda_{2D} P_2 \!\left(\cos \theta_E\right) \frac{E_F}{k_{F}}  \int_0^{\infty} \frac{x dx}{e^{ \left(x^2 - \tilde{\mu_0} \right)/t} + 1} \left[\left(1 +x\right) E\left(\frac{2\sqrt{x}}{1+x}\right) + \left(1-x\right) K\left(\frac{2\sqrt{x}}{1+x}\right) \right] \label{eq:2Ddsigmaisodk1}
\\
&= 16\sqrt{\pi} \lambda_{2D} P_2 \!\left(\cos \theta_E\right) \frac{E_F}{k_{F}}  \left(\frac{2}{3} + \frac{\pi^2}{48}t^2 \left(1 + \ln \frac{\pi}{4} + 12 \zeta'\left(-1\right)\right) + \frac{\pi^2}{48}t^2 \ln t + O\left[t^4 \ln t\right] \right) \label{eq:2Ddsigmaisodk2}
\end{align}

\begin{equation}\label{eq:2Ddsigmaiso}
\frac{d\Sigma_{2D}^{\text{iso}}\left( \mathbf{k_{F}} \right) }{dn} = \frac{k_{F}}{2n} \frac{d\Sigma_{2D}^{\text{iso}}\left( \mathbf{k} \right) }{dk} |_{k=k_{F}} + 16\sqrt{\pi} \lambda_{2D} P_2 \!\left(\cos \theta_E\right) \frac{E_F}{n} \left(\tilde{\mu_0} - t \frac{d \tilde{\mu_0}}{d t} \right) \int_0^{\infty} dx \frac{e^{\left(x^2 - \tilde{\mu_0} \right)/t}/t}{\left(e^{\left(x^2 - \tilde{\mu_0} \right)/t} + 1\right)^2} x \left(1+x\right) E\left(\frac{2\sqrt{x}}{1+x}\right)
\end{equation}
\end{widetext}
with $K\left(x\right)$ and $E\left(x\right)$ the complete elliptic integrals of the first and second kind~\citep{Gradshteyn}.  Although we will focus on the most stable scenario of dipoles aligned perpendicular to the plane, for completeness we also give the form of the self-energy arising from the anisotropic part of the interaction:
\begin{widetext}
\begin{align}
\Sigma_{2D}^{\text{ani}} \left(\mathbf{k}\right) &= -2\sqrt{\pi} \lambda_{2D} \sin^2 \theta_E \frac{E_F}{k_{F}^3} \int_0^{\infty} \frac{k' dk'}{\left(2\pi\right)^2} n_0 \!\left(k'\right) \int_0^{2\pi} d\phi_{k'} \frac{k^2 \cos 2\phi_{k} +k'^2 \cos 2\phi_{k'} -2k k' \cos \left(\phi_{k} + \phi_{k'} \right)}{\sqrt{k^2+k'^2-2k k' \cos \left(\phi_{k} - \phi_{k'} \right) }} \label{eq:2Dsigmaani1}
\\
&= -\frac{8\sqrt{\pi}}{3} \lambda_{2D} \frac{E_F k^3}{k_{F}^3} \sin^2 \theta_E  \cos 2\phi_{\mathbf{k}} \int_0^{\infty} \frac{x dx \left(1+x\right)}{e^{ \left(x^2 k^2/k_{F}^2 - \tilde{\mu_0} \right)/t } + 1} \left[\left(2-x^2\right) E\left(\frac{2\sqrt{x}}{1+x}\right) + \left(1-x\right)^2 K\left(\frac{2\sqrt{x}}{1+x}\right) \right] \label{eq:2Dsigmaani2}
\\
\Sigma_{2D}^{\text{ani}} \left(\mathbf{k_{F}}\right) &= -\frac{8\sqrt{\pi}}{3} \lambda_{2D} E_F \sin^2 \theta_E \cos 2\phi_{\mathbf{k_{F}}} \Biggl\{ \int_0^{\tilde{\mu_0}} x dx \left(1+x\right) \left[\left(2-x^2\right) E\left(\frac{2\sqrt{x}}{1+x}\right) + \left(1-x\right)^2 K\left(\frac{2\sqrt{x}}{1+x}\right) \right] \notag
\\
& \qquad \qquad \qquad \qquad \qquad \qquad \qquad + \int_0^{\infty} x dx \frac{\text{sgn} \left(x^2 - \tilde{\mu_0} \right)}{e^{\left|x^2 - \tilde{\mu_0} \right|/t} + 1} \left(1+2 \left(1-x\right) + O\left[\left(1-x\right)^2 \ln |1-x| \right]\right) \Biggr\} \label{eq:2Dsigmaani3}
\\
&= -\frac{8\sqrt{\pi}}{3} \lambda_{2D} E_F \sin^2 \theta_E \cos 2\phi_{\mathbf{k_{F}}} \left(\frac{8}{5}- \frac{\pi^2}{8} t^2 + O\left[t^4 \ln t \right] \right), \label{eq:2Dsigmaani4}
\end{align}
\end{widetext}

In Eqs.~\eqref{eq:2Dsigmaiso3} and \eqref{eq:2Dsigmaani3} we have noted that at low temperature the integrand in the temperature-dependent term is only appreciable near $x=1$ and expanded the elliptic integrals about that point, permitting analytic integration.  We have shown the derivative of the self-energy with respect to density since the inverse compressibility is $\kappa^{-1} = n^2 \frac{\partial \mu}{\partial n}$.  For arbitrary temperatures, the compressibility is obtained via numerical integration from Eq.~\eqref{eq:2Ddsigmaiso}. For low temperatures, recognizing that the Fermi energy, $E_F$, and reduced temperature, $t$, and coupling strength, $\lambda$, all depend on density,  we can obtain an analytic result by differentiating the self-energy given in Eq.~\eqref{eq:2Dsigmaiso4} according to
\begin{align}
\frac{\kappa_0}{\kappa} &= \frac{\partial \mu_0}{\partial E_F} + \left( \frac{\partial}{\partial E_F} - \frac{t}{E_F} \frac{ \partial }{\partial t} + \frac{\lambda}{2 E_F} \frac{\partial}{\partial \lambda} \right) \Sigma^{\text{iso}}\left(\mathbf{k_{F}}\right) \label{eq:Kratio}
\\
&= 1 + 16\sqrt{\pi} \lambda_{2D} P_2 \!\left(\cos \theta_E\right) \left(\frac{4}{3}-\frac{\pi^2}{48} t^2 \right) + O\left(t^4 \ln t \right) . \label{eq:2DKratio}
\end{align}
\begin{figure*}[tbp]
  \includegraphics[width=.9\columnwidth]{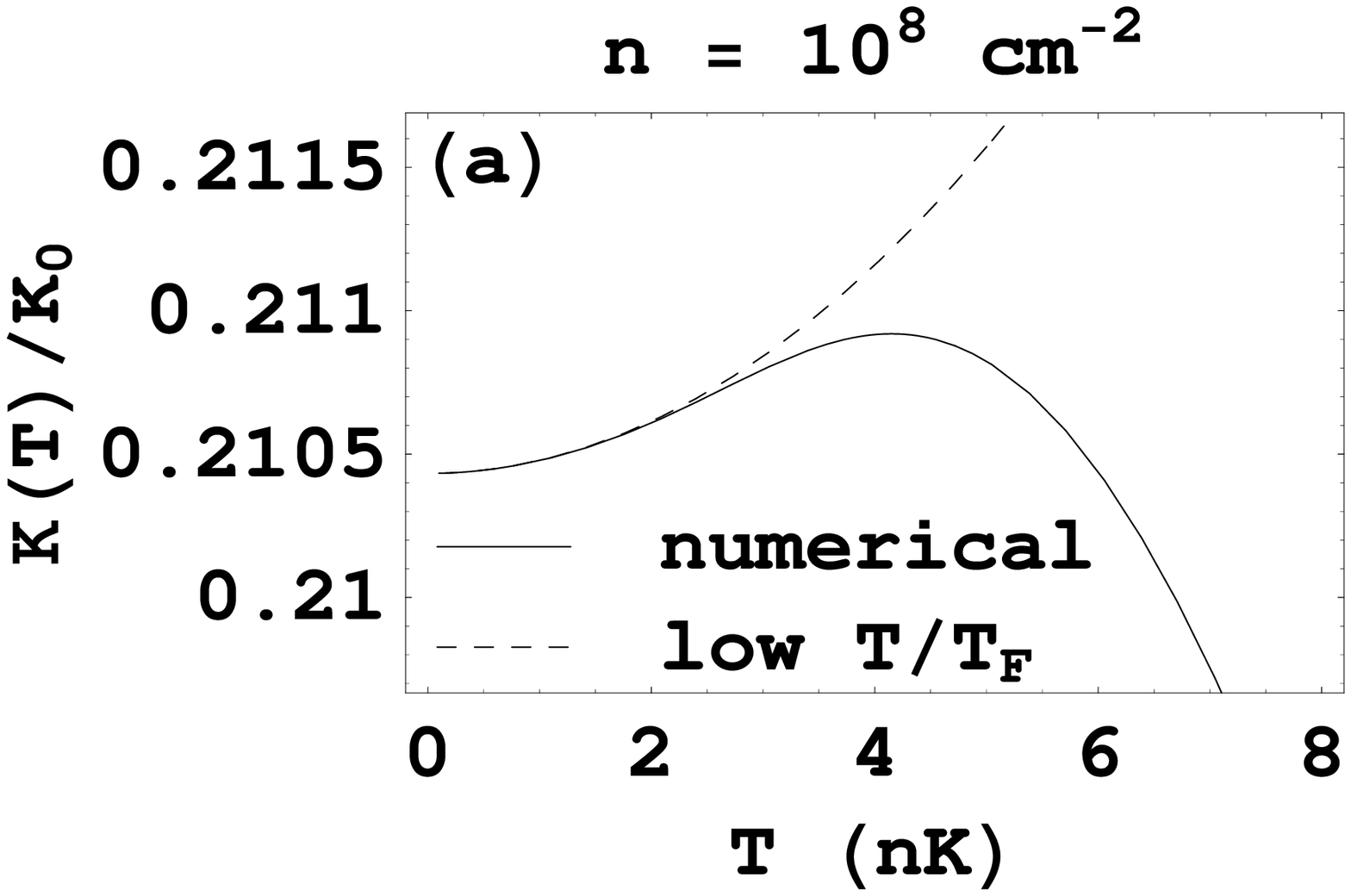}
  \includegraphics[width=.9\columnwidth]{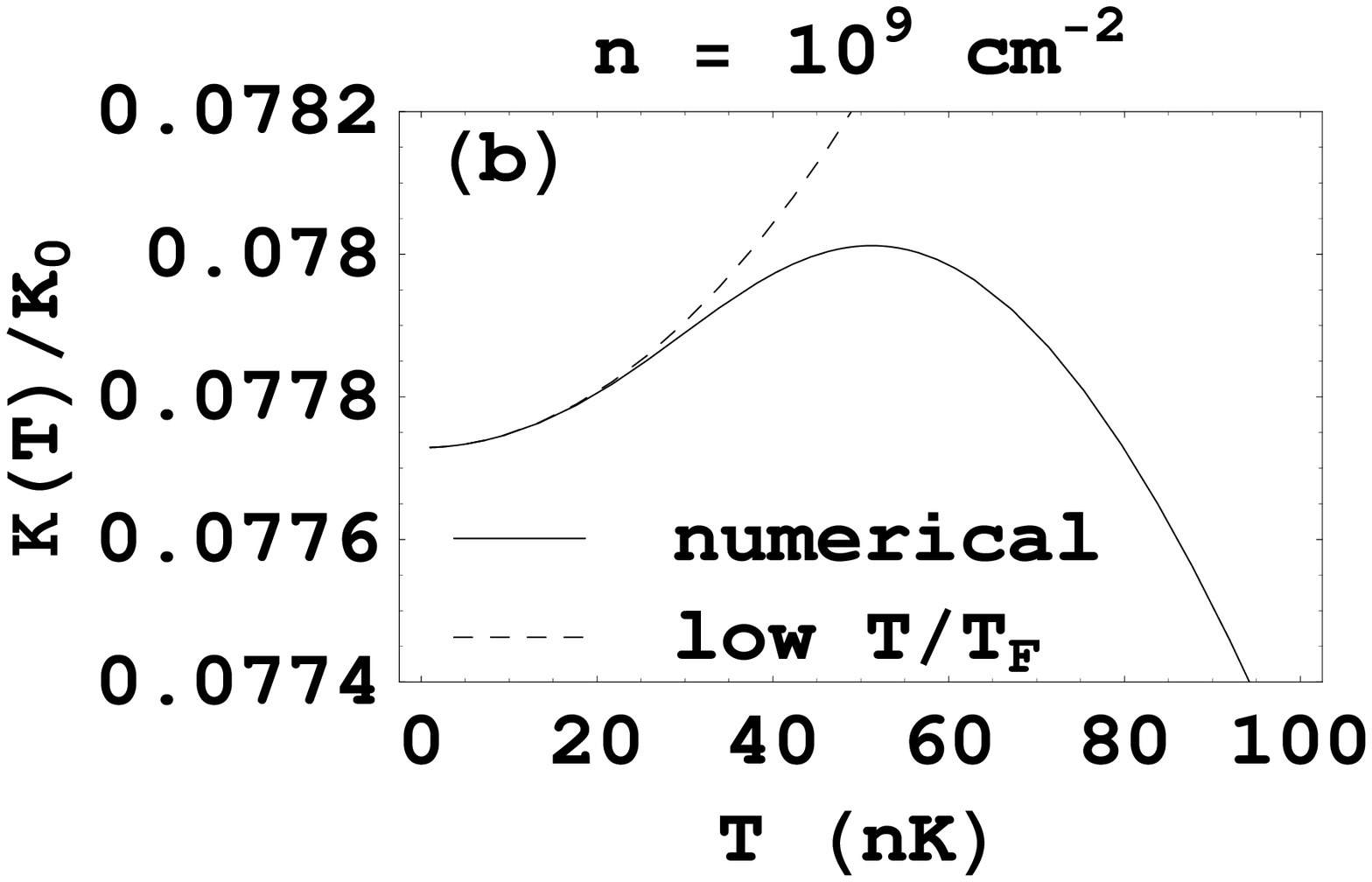}
  \\
  \includegraphics[width=.9\columnwidth]{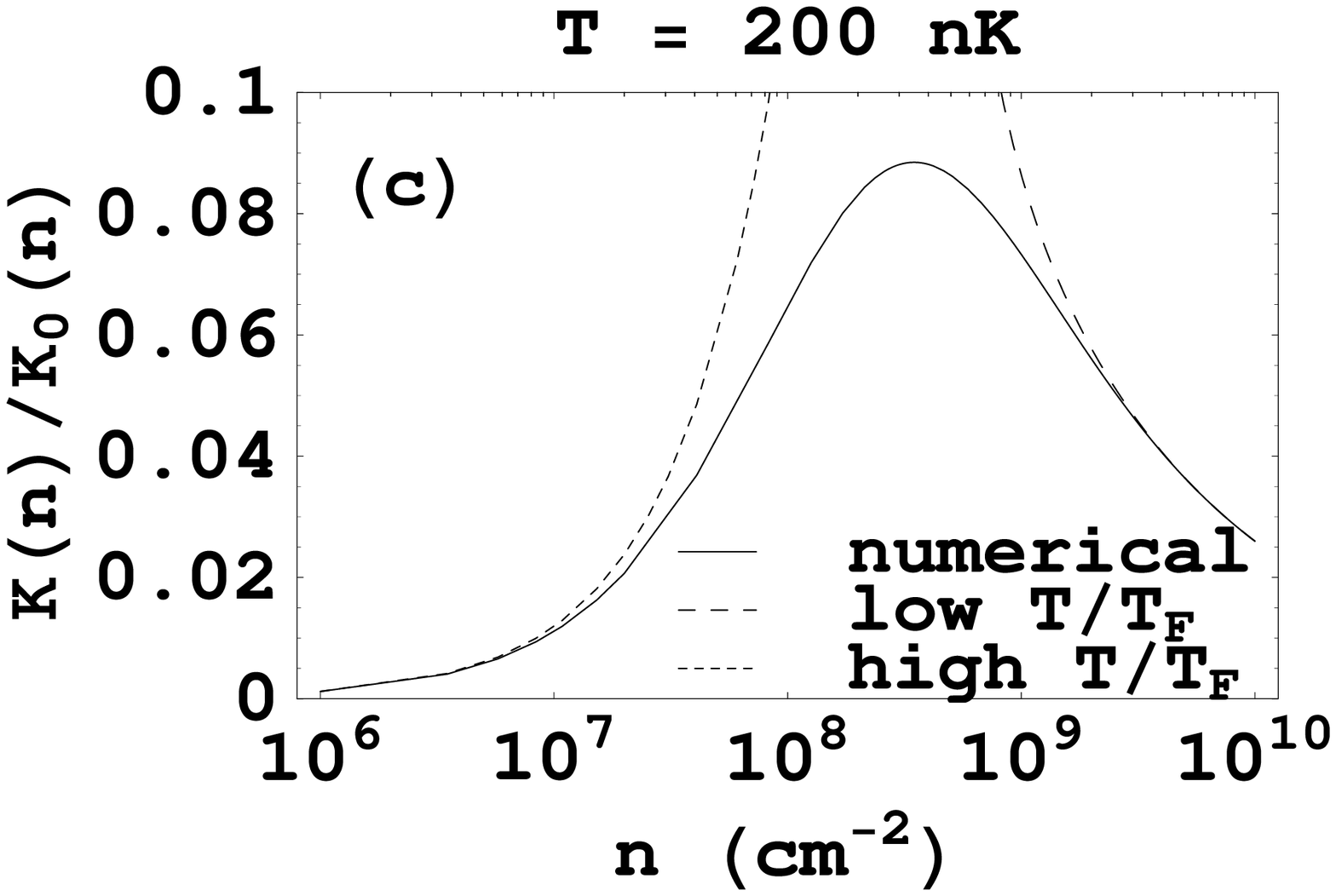}
  \includegraphics[width=.9\columnwidth]{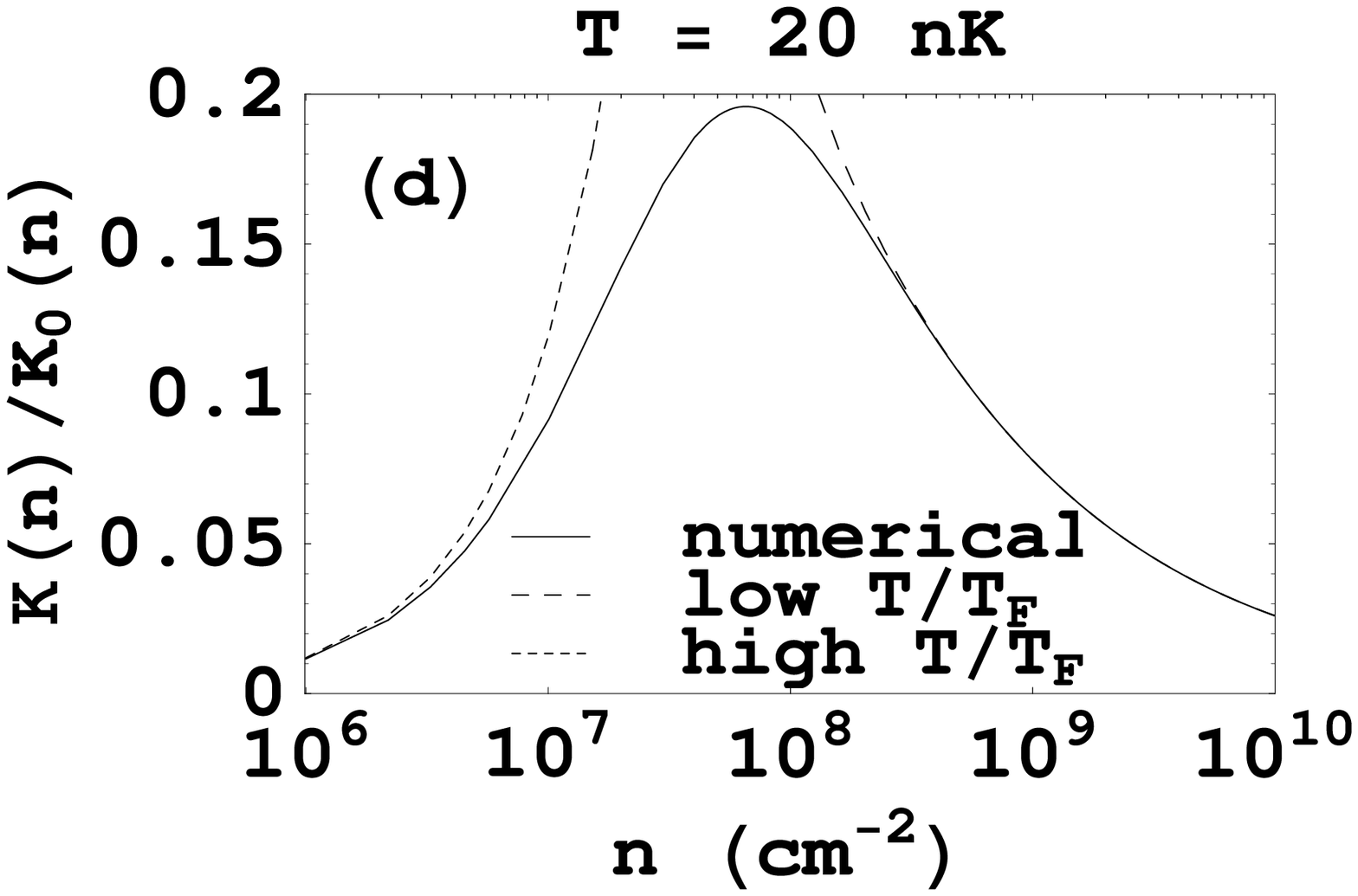}
  \\
  \includegraphics[width=.9\columnwidth]{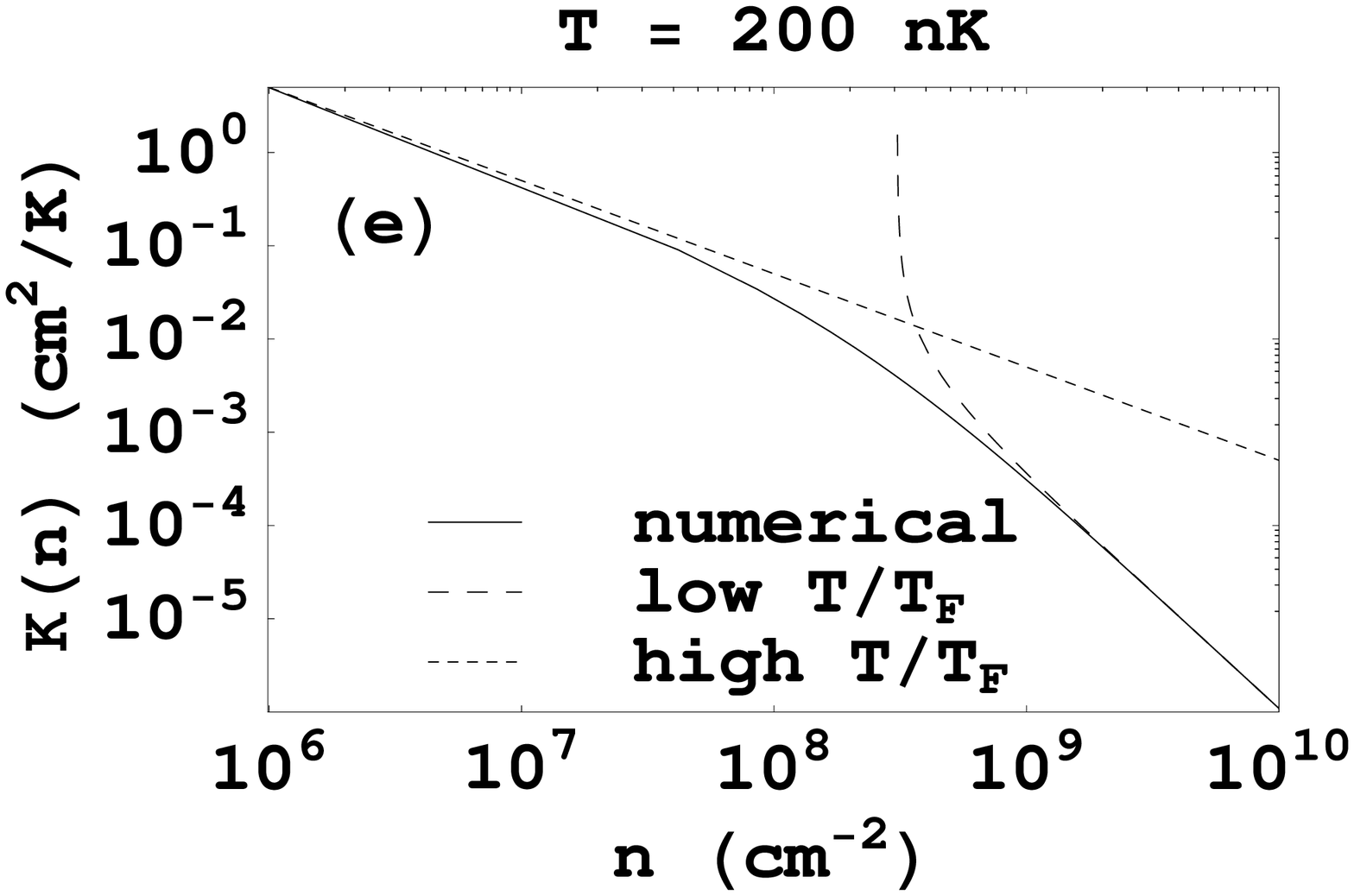}
  \includegraphics[width=.9\columnwidth]{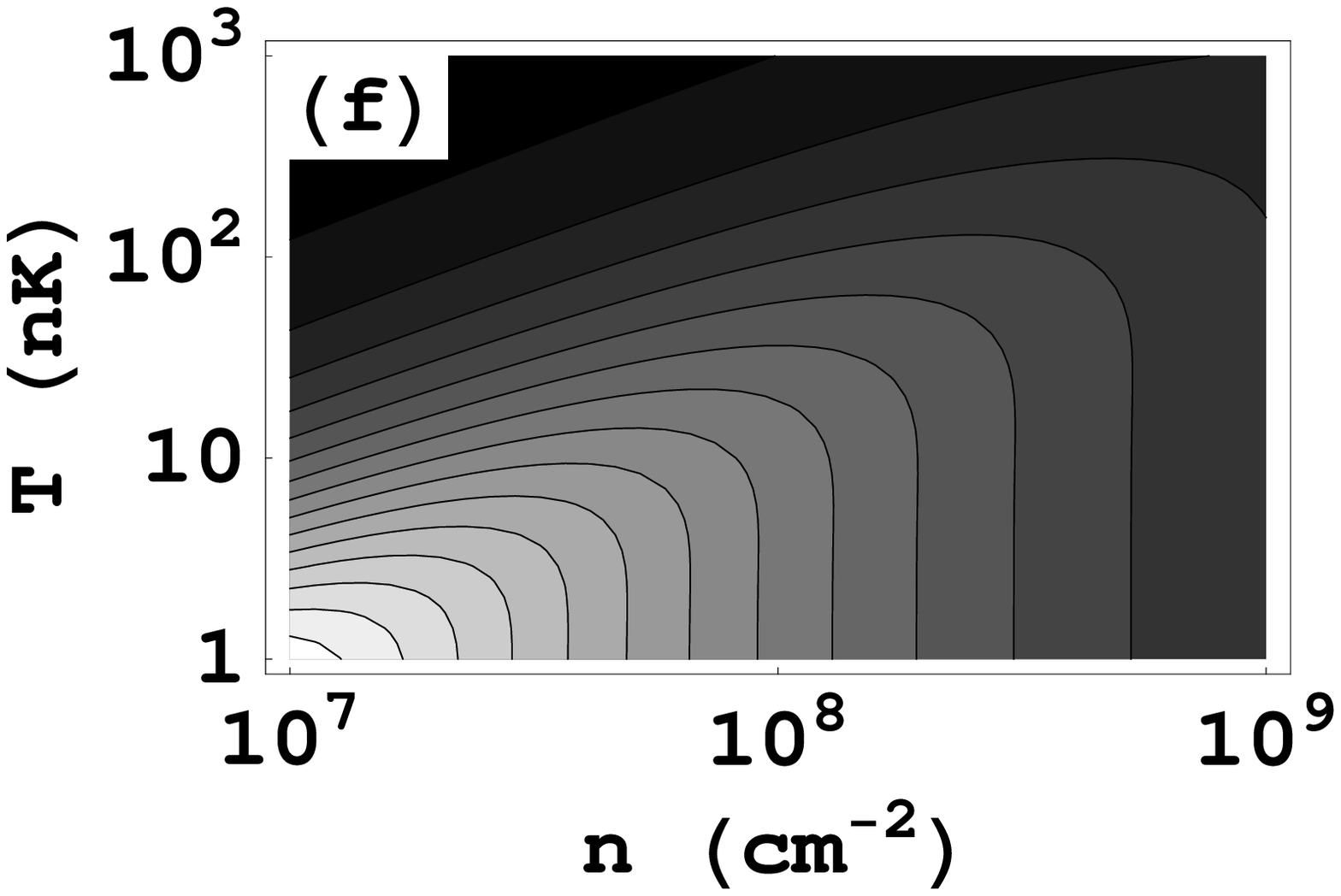}
\caption{Compressibility for KRb in 2D for $\theta_{E}=0$, corresponding to dipoles aligned perpendicular to the plane. (a-b) $\kappa/\kappa_0$ vs temperature at fixed density; (c-d) $\kappa/\kappa_0$ (\textit{i.e.}, $dn/d\mu$) vs density at fixed temperature; (e) $\kappa$ vs density at fixed temperature; (f) contour plot of $\kappa/\kappa_0$ vs temperature and density (lighter is higher).}\label{fig:2D}
\end{figure*}

When the orienting electric field is perpendicular to the plane ($\theta_E = 0$), the compressibility is always positive (assuming the gas is not so dense that the $1/r^3$ approximation to the potential breaks down).  As shown in Figs.~\ref{fig:2D}(a) and \ref{fig:2D}(b), the compressibility is nonmonotonic in temperature for experimentally relevant densities.  The nonmonotonicity is due to the interaction-dependent term, $d\Sigma^{\text{iso}}\left(k_{F}\right)/dn$, which decreases quadratically with $t$ before turning over and increasing with $t$ for fixed density.  Since $t \sim m T/n$ and $\lambda_{2D} \sim m d^2 \sqrt{n}$, both the temperature at which the peak occurs and the height of the peak are larger for higher density.  In Fig.~\ref{fig:2D}(c), we show how the compressibility varies as the density is changed.  Note that we do not use the natural unit of $\kappa_0 = \frac{m}{2\pi \hbar^2 n^2}$ in this plot, since it is density-dependent.  In the low-density (high-$t$) limit, the compressibility behaves classically, as expected, and is drastically reduced in the high-density (low-$t$) limit.

The nonmonotonicity in the compressibility shown in Figs.~\ref{fig:2D}(a) and \ref{fig:2D}(b) occurs well below typical experimental temperatures of $\sim 200$ nK.  However, we may also consider the ratio of the finite-temperature interacting and zero-temperature noninteracting compressibilities, $\kappa/\kappa_0 \sim n^2 \kappa \sim dn/d\mu$, which has the same temperature dependence as the compressibility.  Figures \ref{fig:2D}(c) and \ref{fig:2D}(d) show that this quantity also displays a nonmonotonic density dependence.  Moreover, this nonmonotonic behavior is evident even at relatively high temperature, with $T\sim T_F$.  [For the homogeneous system considered here, $T_F =$ 24nK (240nK) for $n=10^8$ cm$^{-2}$ ($10^9$ cm$^{-2}$).]  In this scenario, the peak is not due to nonmonotonicity in the interaction-dependent term, $d\Sigma^{\text{iso}}\left(k_{F}\right)/dn$. Both $d\Sigma^{\text{iso}}\left(k_{F}\right)/dn$ and $d\mu_0/dn$ are monotonic in the density, but while $d\Sigma^{\text{iso}}\left(k_{F}\right)/dn$ increases with density, $d\mu_0/dn$ decreases.  The competition between the interacting and noninteracting density dependencies is the source of the nonmonotonicity.

Figure \ref{fig:2D}(e) gives a sense of how $\kappa/\kappa_0$ changes as temperature and density are simultaneously changed in some manner.  Although, it is hard to see on the logarithmic scale, the vertical portions of the lines actually bulge outwards slightly, consistent with the fixed density results plotted in Figs.~\ref{fig:2D}(a) and \ref{fig:2D}(b).  Clearly, nonmonotonic behavior is more pronounced if the measurements are performed at fixed temperature and variable density than the other way around.  We will discuss measurement in Sec.~\ref{sec:exp}.

\begin{figure}[tbp]
  \includegraphics[width=\columnwidth]{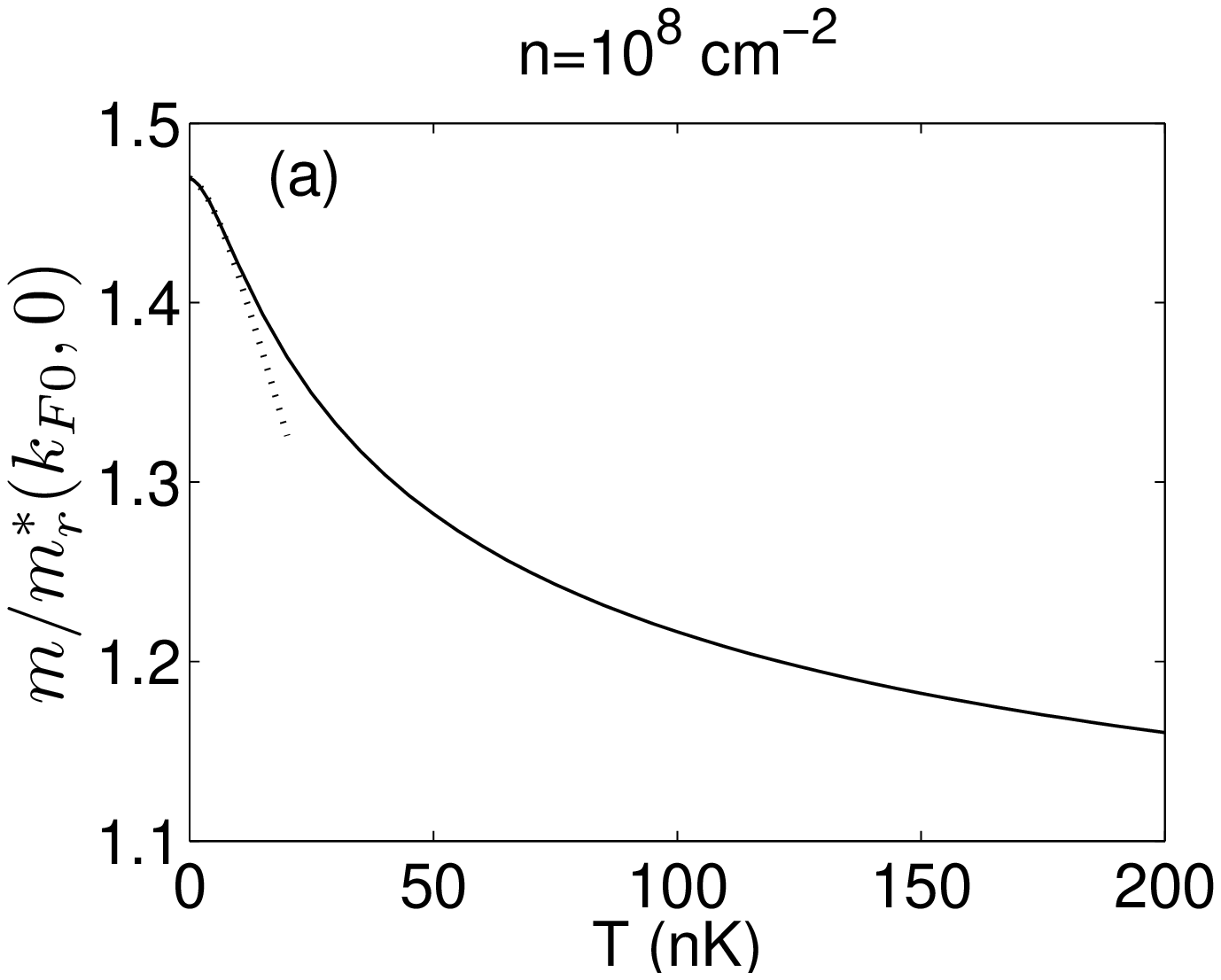}
  \\
  \includegraphics[width=\columnwidth]{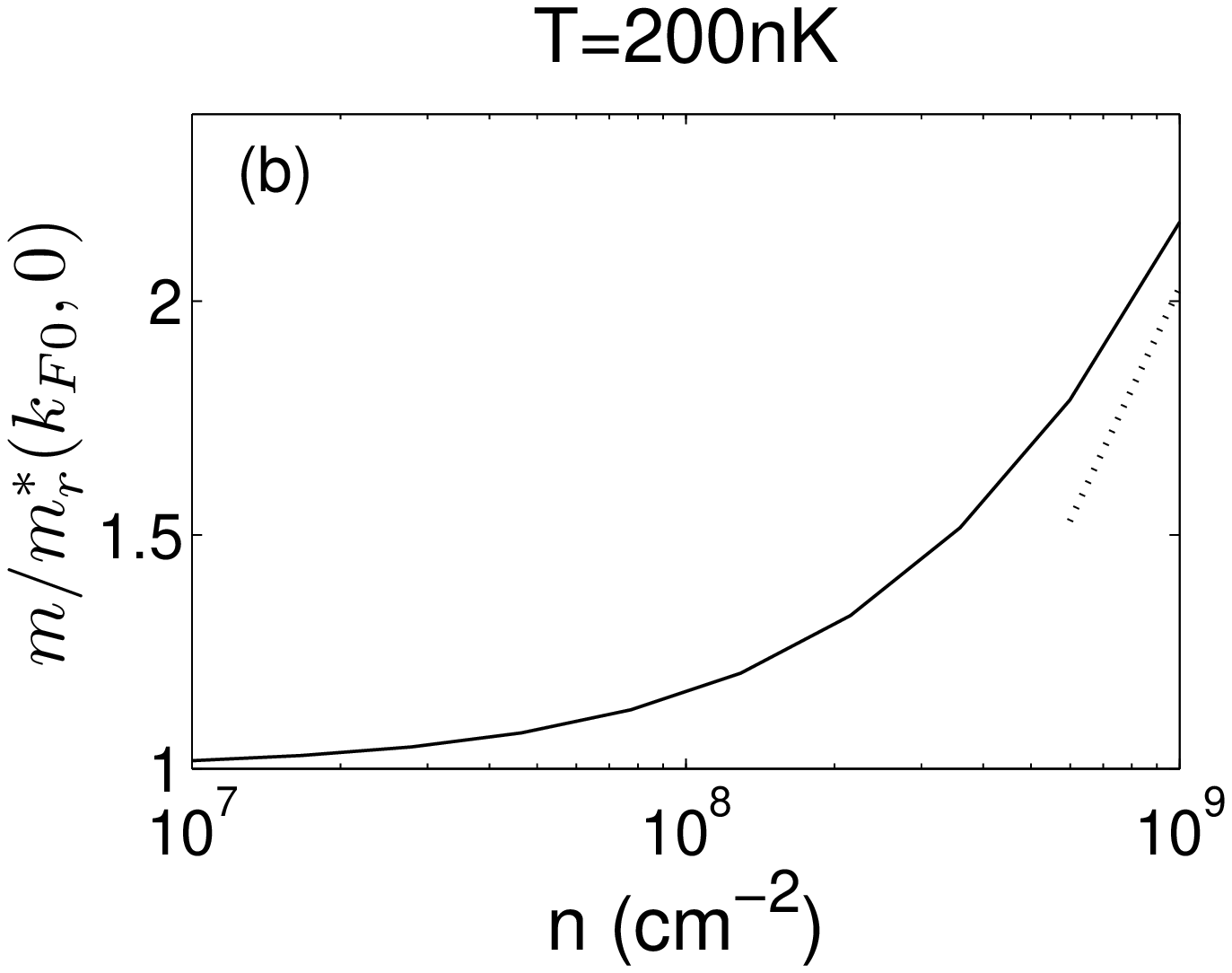}
\caption{Effective mass at the Fermi surface for KRb in 2D (a) vs temperature at fixed density and (b) vs density at fixed temperature. Dotted lines show the low $T/T_F$ approximations.}\label{fig:2Dmeff}
\end{figure}

With the radial and angular effective masses, $m_r^{\ast}$ and $m_{\phi}^{\ast}$, defined as in Eqs.~\eqref{eq:3Dmeffr} and \eqref{eq:3Dmefft}, we have plotted the radial mass at the Fermi surface for the isotropic case in Fig.\ref{fig:2Dmeff}.  We see that at low temperatures or high densities the effective mass is diminished by the interaction while at high temperatures or low densities the effect of the interaction is small and the effective mass approaches its bare value.  The analytic low temperature results from Eq.~\eqref{eq:2Ddsigmaisodk2} are shown as dotted lines.

\subsubsection{Zero sound mode}\label{subsubsec:2Dzerosound}
\begin{figure}[tbp]
  \includegraphics[width=\columnwidth]{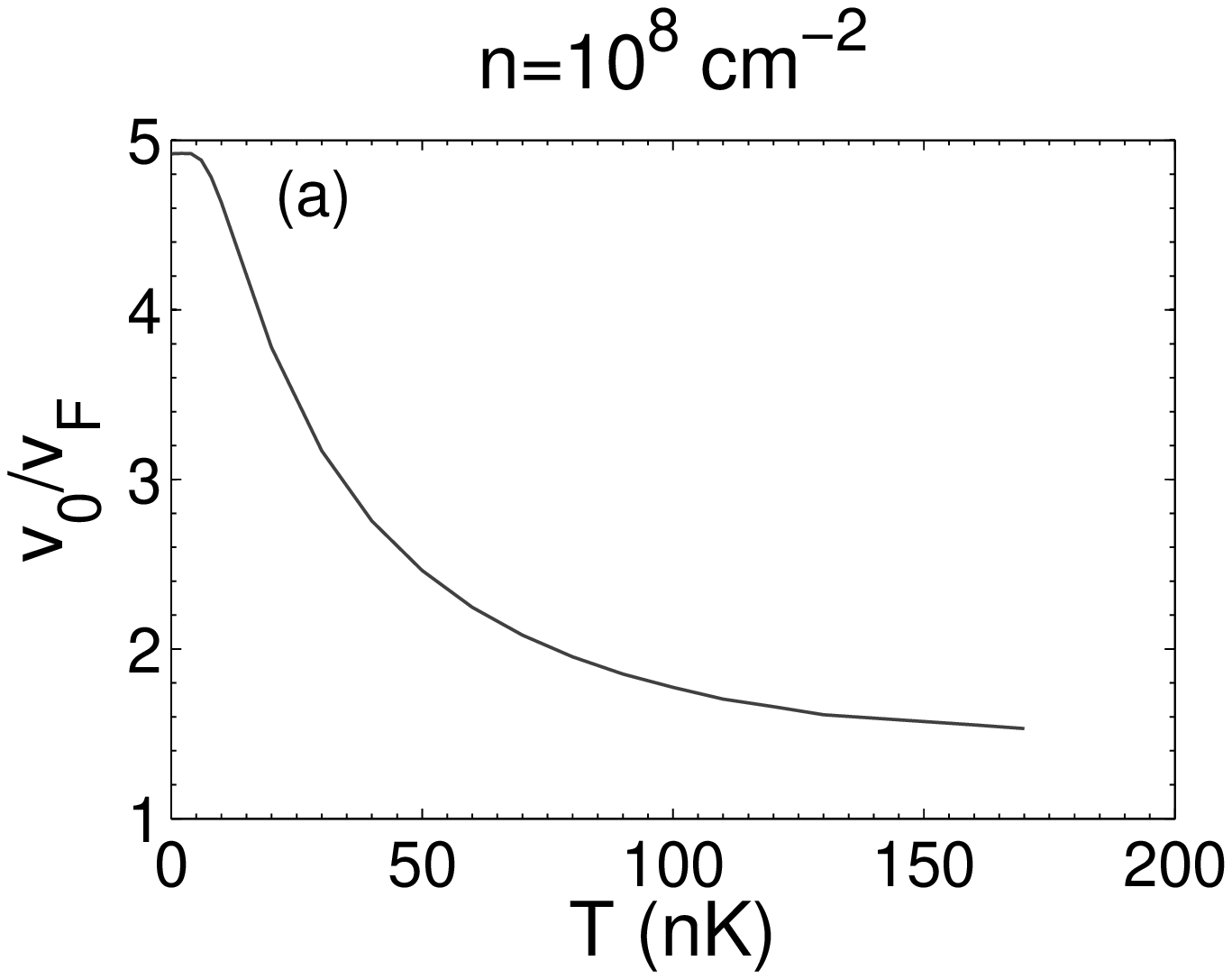}
  \\
  \includegraphics[width=\columnwidth]{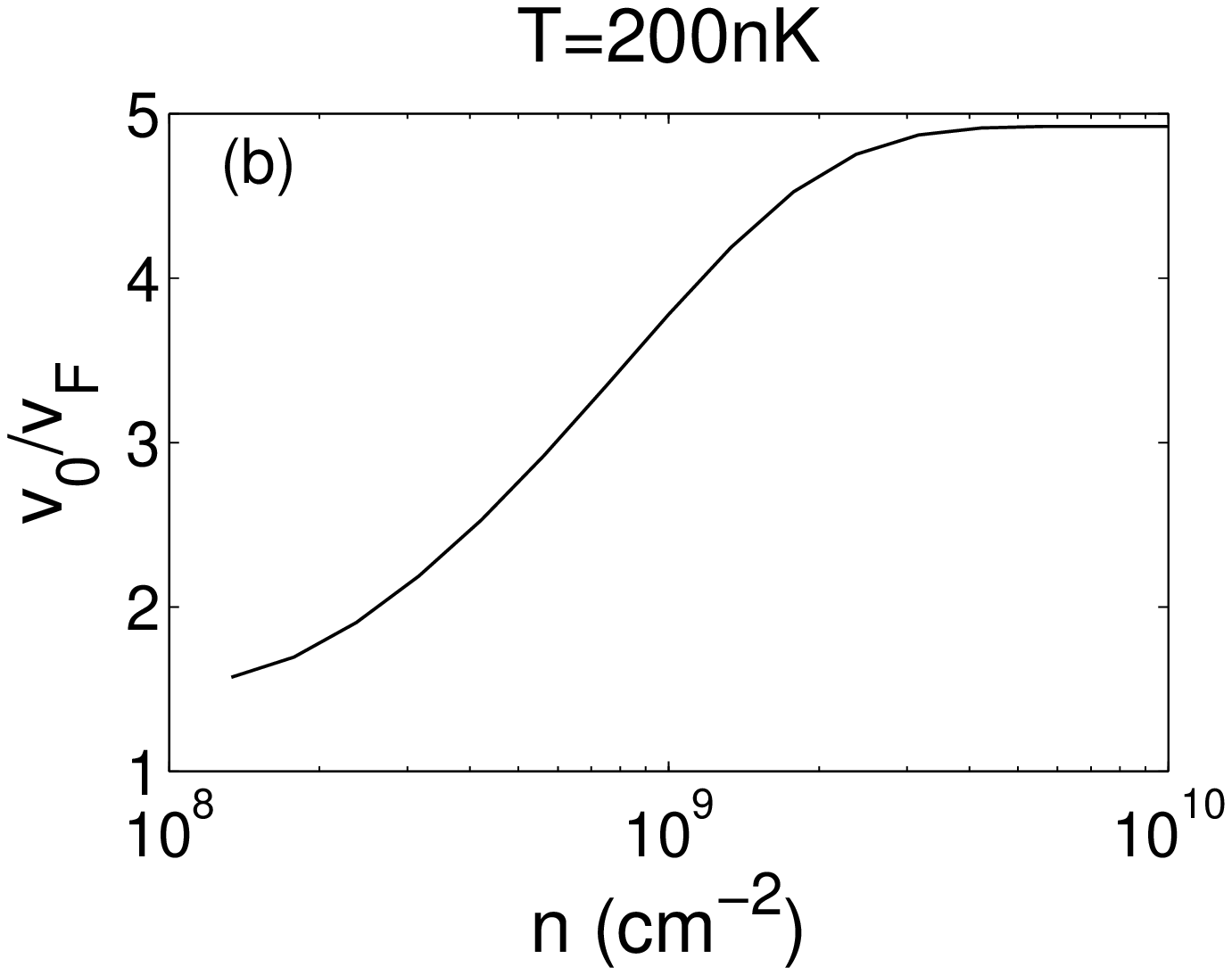}
\caption{Zero sound speed for a 10 nm thick monolayer of KRb (a) vs temperature at fixed density and (b) vs density at fixed temperature.}\label{fig:2Dzerosound}
\end{figure}

For dipoles aligned perpendicular to the plane, the 2D interaction is isotropic in contrast to the 3D case, and we need not include coupling to higher partial waves when calculating the zero sound.  Likewise, since the single-particle energy is also isotropic, to leading order we can simply use the bare single-particle energy.  The zero sound dispersion is then simply given by the solution of $1-V_{2D}\left(0\right) \chi \left(q\rightarrow 0, \omega\rightarrow v_0 q\right) = 0$, where
\begin{multline}
\chi \left(q\rightarrow 0, v_0 q \right) = \int \frac{d^2 k}{\left(2\pi\right)^2} \frac{\partial n \left( \mathbf{k} \right)}{\partial \epsilon \left( \mathbf{k} \right)} \frac{\mathbf{\hat{q}} \cdot \frac{\mathbf{k}}{k_{F}} }{\frac{v_0}{v_{F}} - \mathbf{\hat{q}} \cdot \frac{\mathbf{k}}{k_{F}} +i0}
\\
= \frac{k_F^2}{E_F} \int \frac{x d x}{2\pi} \frac{- e^{\left(x^2 - \tilde{\mu_0} \right)/t}}{t \left[e^{\left(x^2 - \tilde{\mu_0} \right)/t}+1\right]^2} \!\left[1 - \frac{\frac{v_0}{v_{F}}}{\sqrt{\frac{v_0^2}{v_{F}^2} - x^2 - i0}} \right]
\end{multline}
Note from Eq.~\eqref{eq:V2Dq} that the thickness of the monolayer plays a critical role in determining the zero sound speed.  The dispersion is plotted in Fig.~\ref{fig:2Dzerosound} as a function of temperature and density for a thickness of $w = 10$nm.  In the limit $T/T_F \rightarrow 0$, one can show that $v_0/v_{F} = \sqrt{2 d^2 m/3\sqrt{\pi}w\hbar^2}$ \citep{Li10}, and we numerically recover this limit to better than 2\%.  The damping is negligible, and we have not shown it.  The mode propagates even at quite high temperatures due to the strong repulsive delta-function core of the effective 2D interaction \eqref{eq:V2Dq}.  In fact, it persists to arbitrarily high temperature for tight enough confinement.  This is an artifact of the ultraviolet divergence of the 2D Fourier transform of the $1/r^3$ interaction, and in experiment this robustness will be limited by the range, $r_*$, at which the actual intermolecular potential deviates from a $1/r^3$ behavior.  However, provided $k_F r_* \ll 1$ and $r_* \ll w$, these results will remain valid and the collective mode will be robust at experimentally feasible temperature and density.

\subsection{1D}\label{subsec:1D}
\subsubsection{Compressibility and effective mass}
\begin{figure*}[tbp]
  \includegraphics[width=.66\columnwidth]{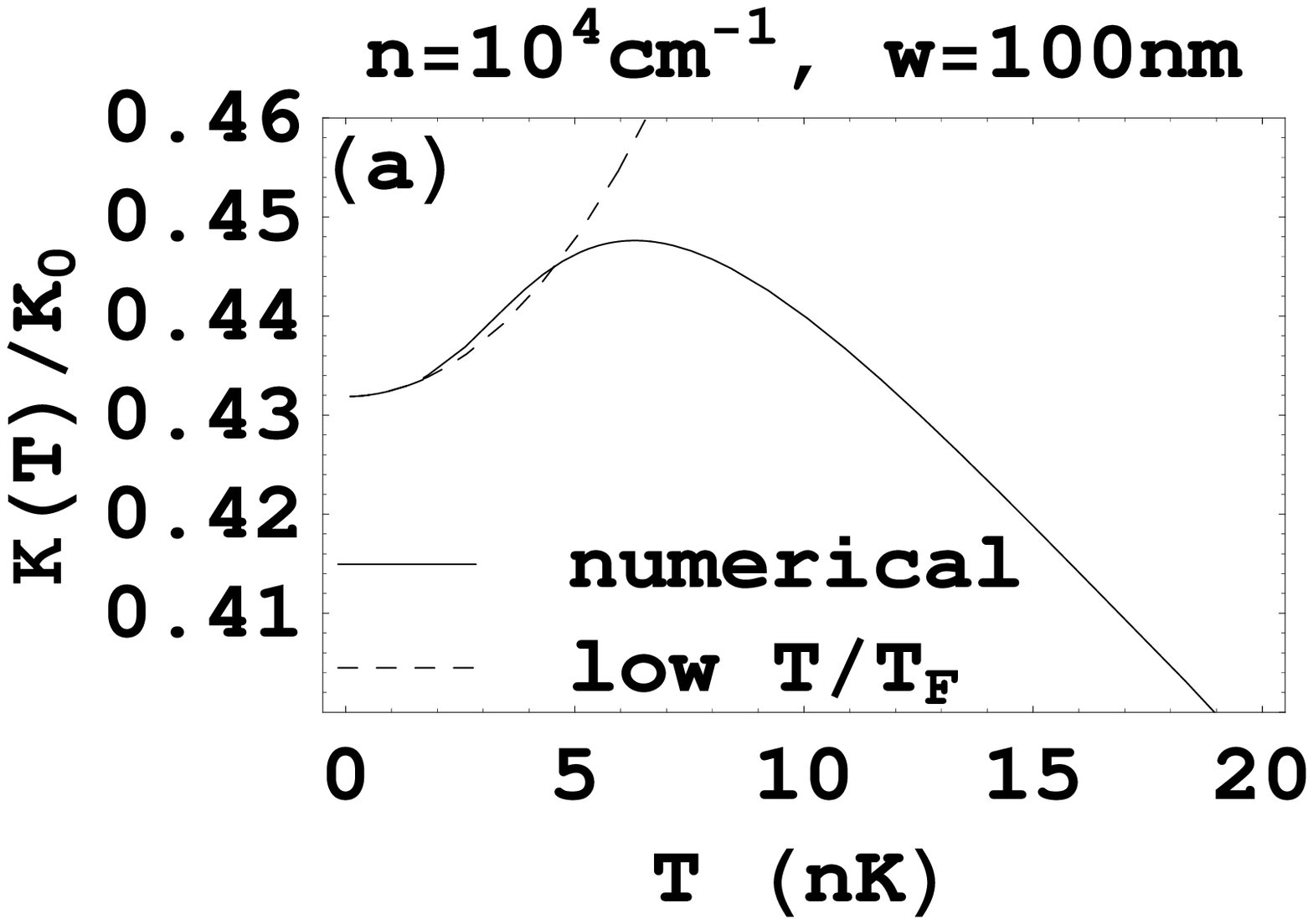}
  \includegraphics[width=.66\columnwidth]{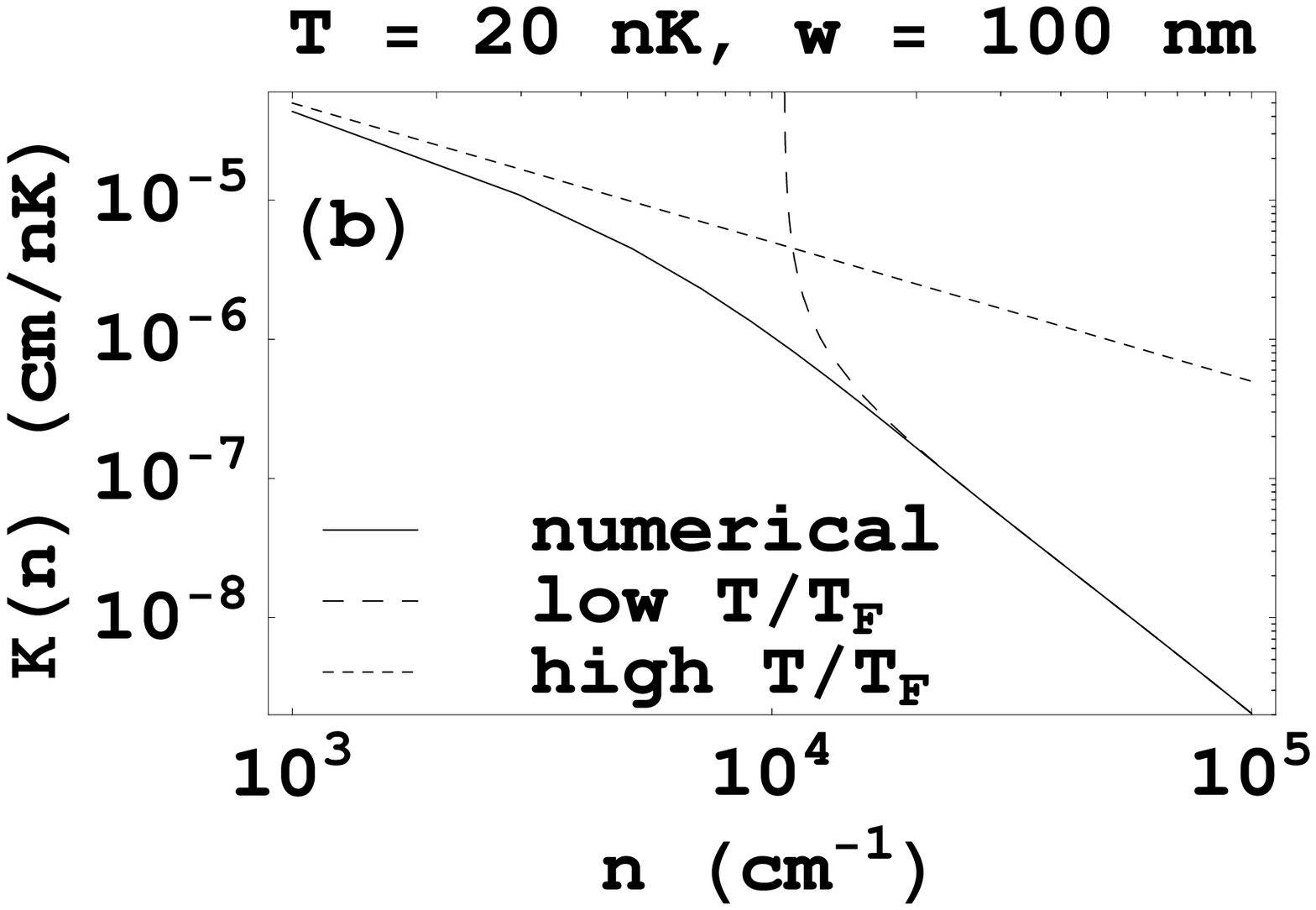}
  \includegraphics[width=.66\columnwidth]{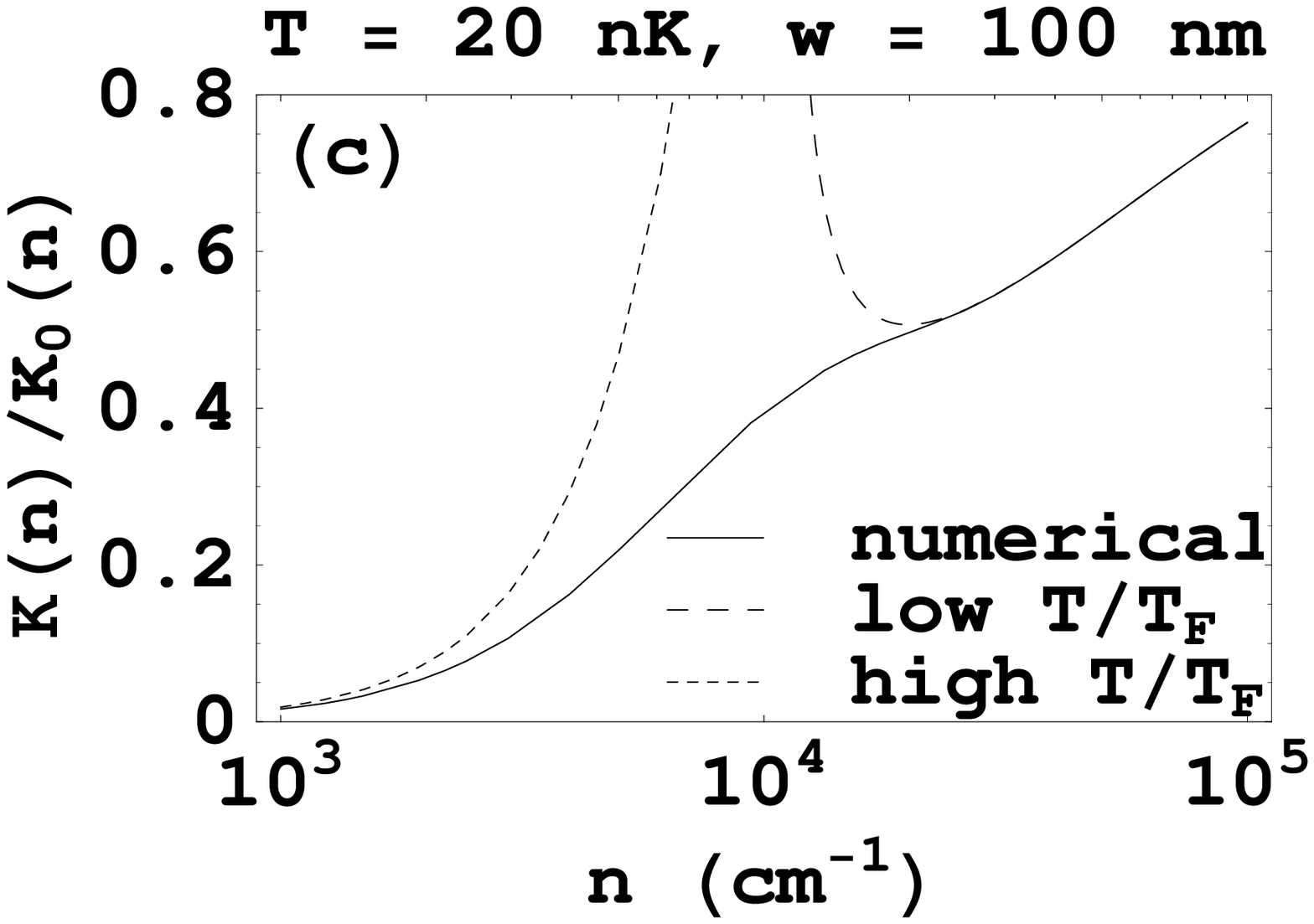}
  \\
  \includegraphics[width=.66\columnwidth]{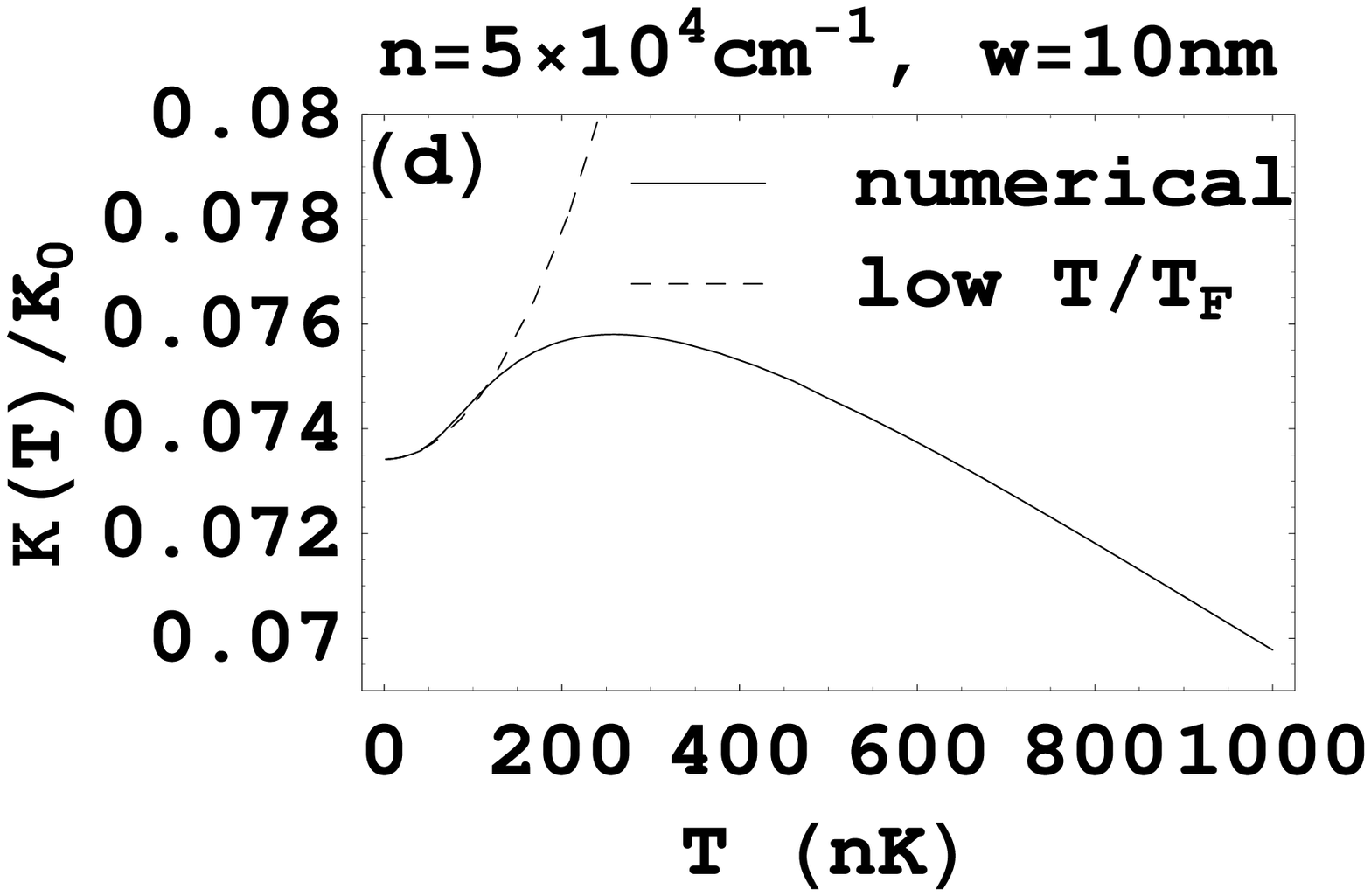}
  \includegraphics[width=.66\columnwidth]{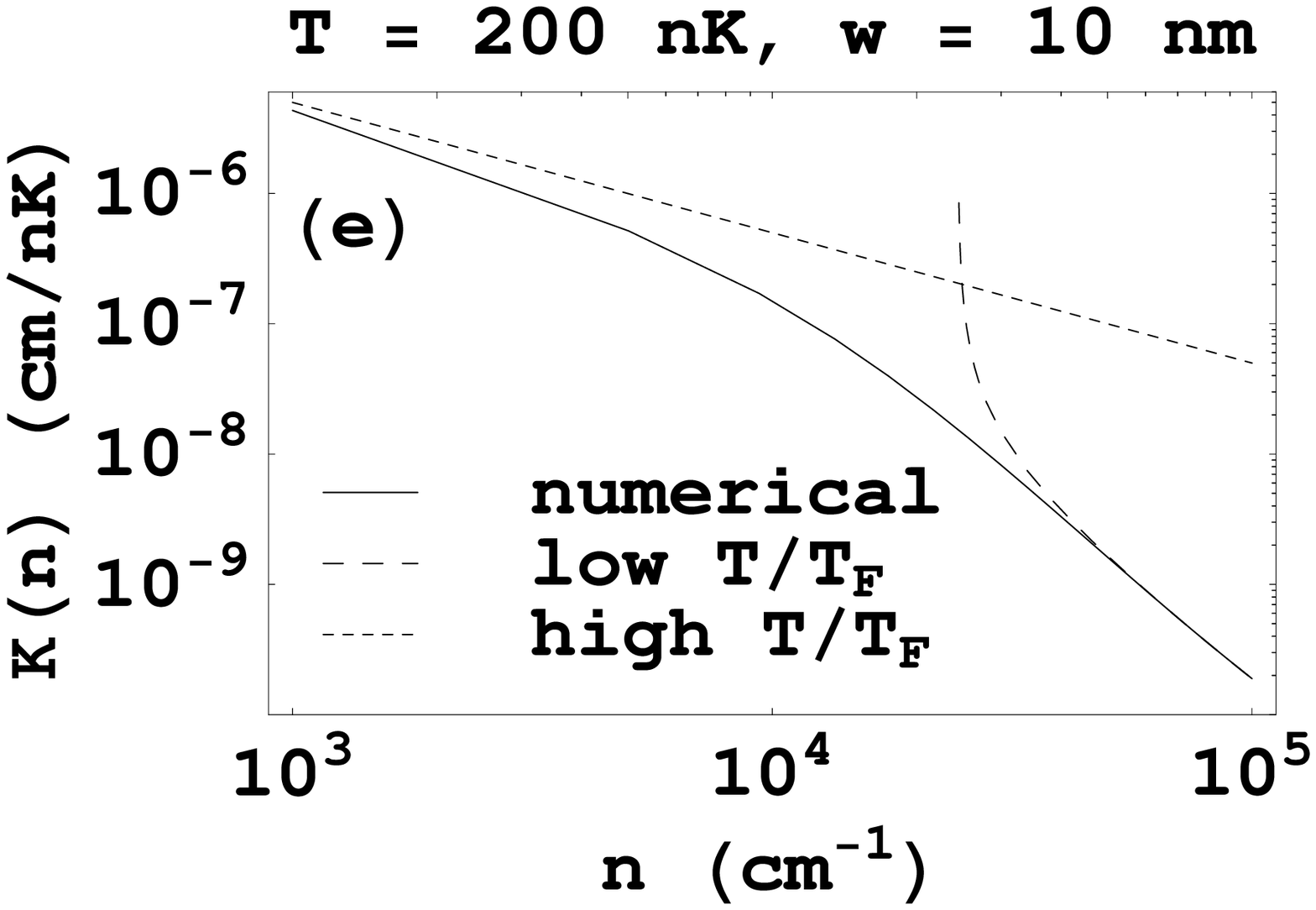}
  \includegraphics[width=.66\columnwidth]{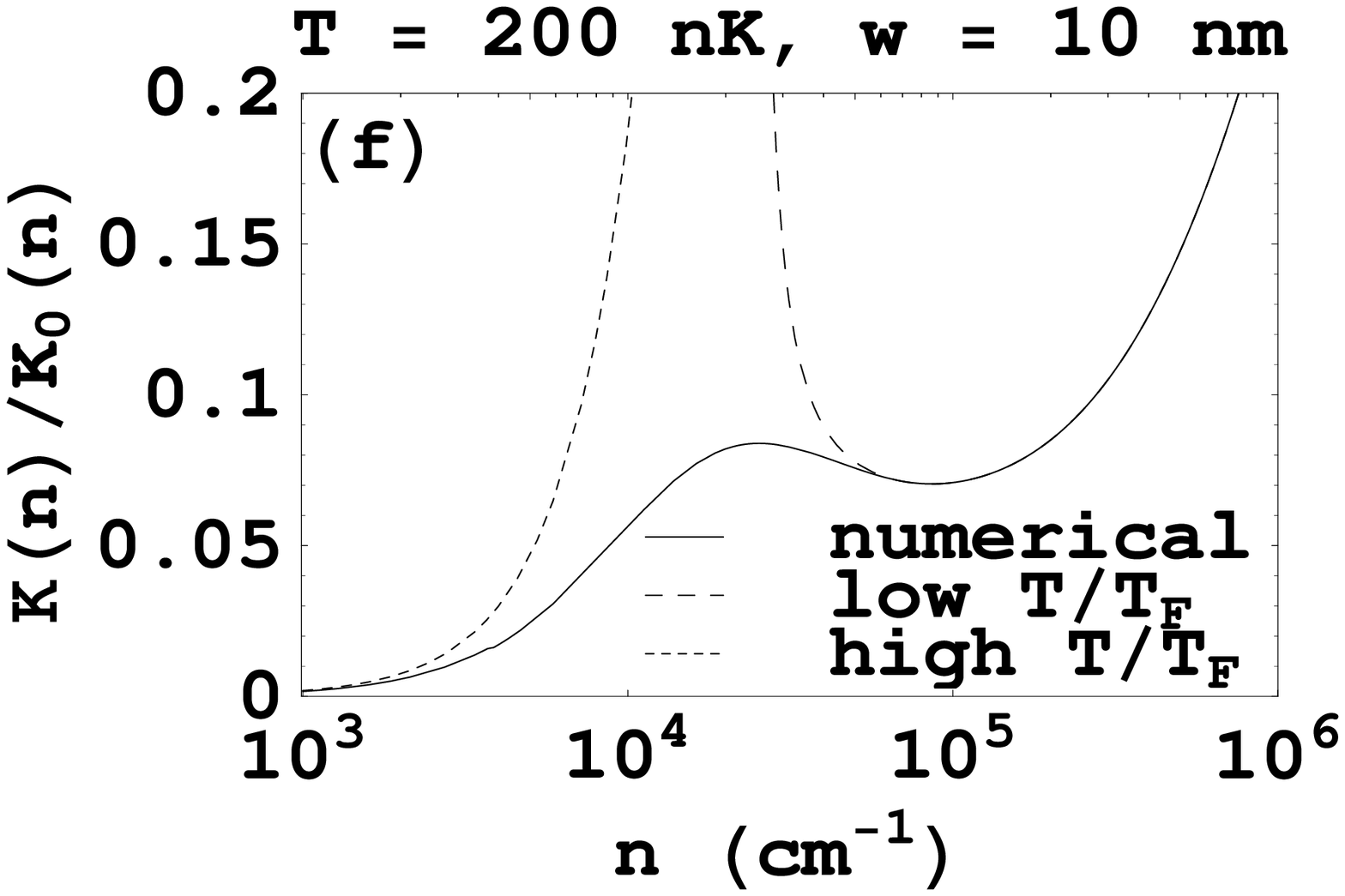}
  \\
  \includegraphics[width=.66\columnwidth]{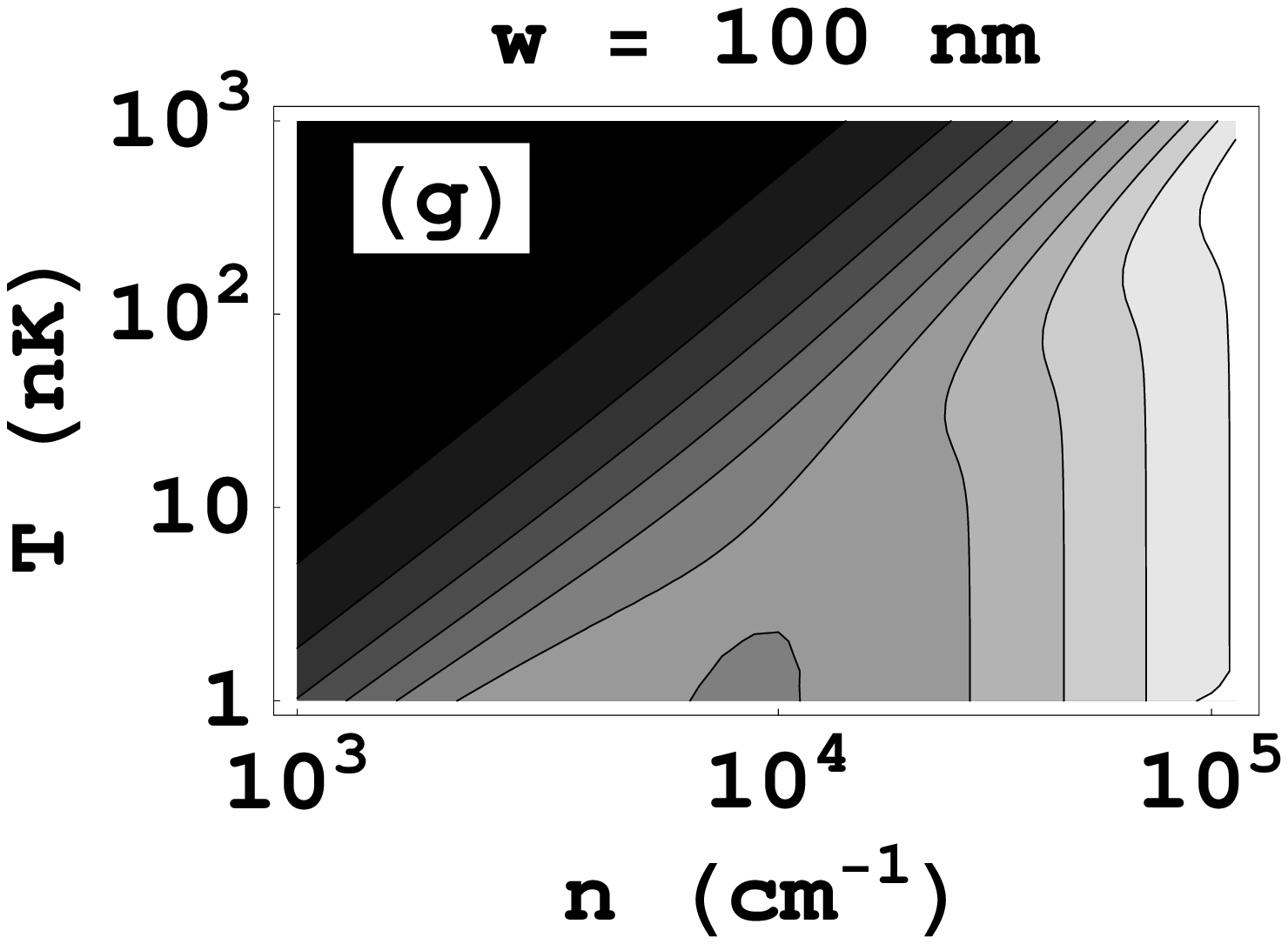}
  \includegraphics[width=.66\columnwidth]{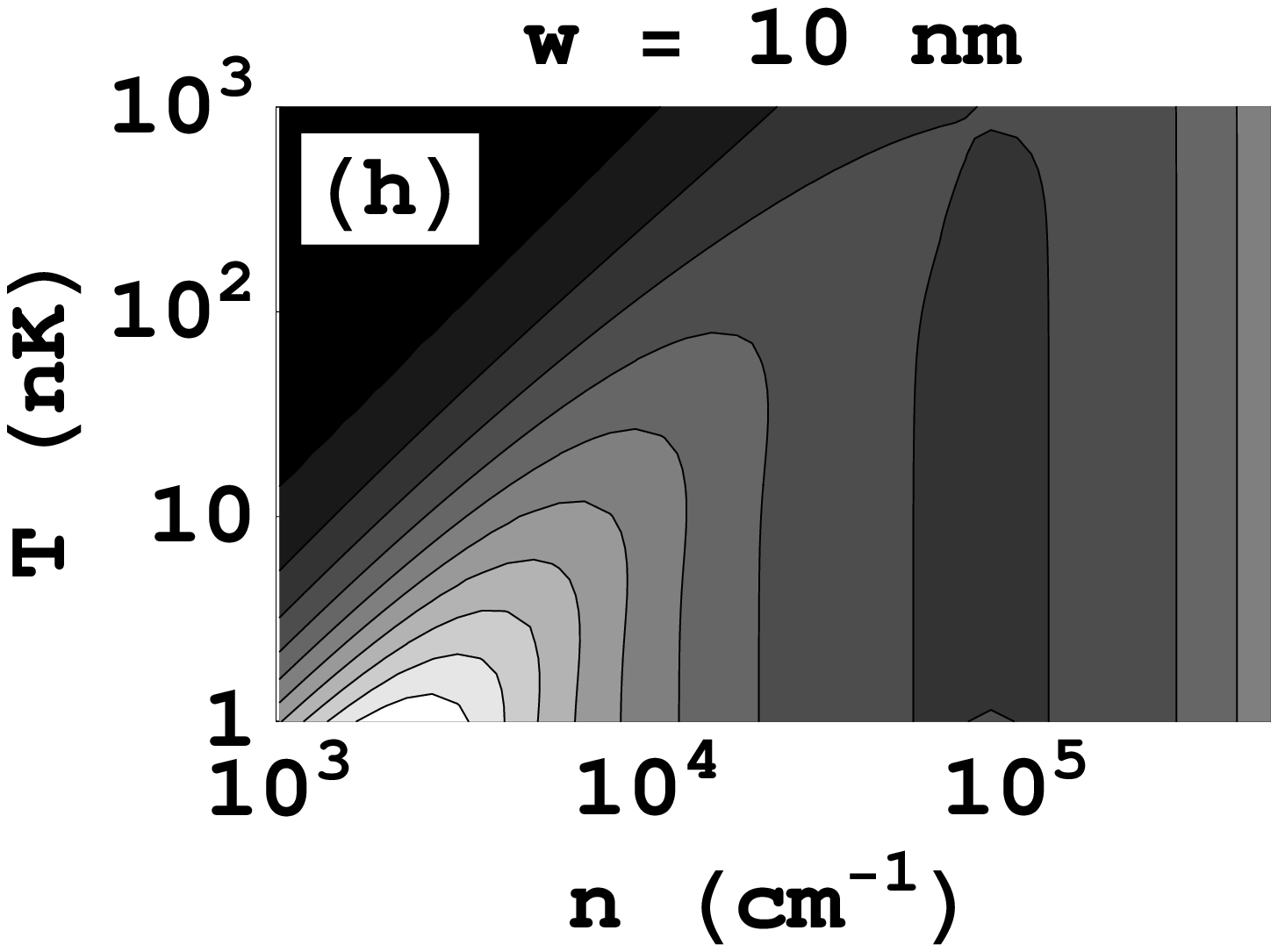}
\caption{Compressibility for KRb at $\theta_E=\pi/2$, corresponding to dipoles aligned perpendicular to the axis in quasi-1D. (a,d) $\kappa/\kappa_0$ (\textit{i.e.}, $dE_F/d\mu$) vs temperature at fixed density; (b,e) $\kappa$ vs density at fixed temperature; (c,f) $\kappa/\kappa_0$ vs density at fixed temperature; (g,h) contour plot of $\kappa/\kappa_0$ vs temperature and density (lighter regions are higher).}\label{fig:1D}
\end{figure*}
For the 1D case, $\tilde{\mu_0} \approx 1 + \pi^2 t^2/12$, and $\theta_E$ is the angle between $\mathbf{E}$ and the trap axis.  In this case, recall from Eq.~\eqref{eq:V1Dq} that the results will depend explicitly on the transverse width (or short range interaction cutoff), $w$.  In the limit $k_F^2 w^2 \ll 1$, the Hartree-Fock self-energy is
\begin{widetext}
\begin{align}
\Sigma_{1D} \left(\mathbf{k}\right) &= \frac{\pi^2}{2} \lambda_{1D} \frac{E_F}{k_F^3} P_2 \!\left(\cos \theta_E \right) \int_{-\infty}^{\infty} dk' n_0 \!\left(k'\right) \left(k-k'\right)^2 \left(\gamma + 2 \ln \left|\left(k-k'\right)w \right| \right) \label{eq:1Dsigma1}
\\
\Sigma_{1D} \left(\mathbf{k_F}\right) &= \frac{\pi^2}{2} \lambda_{1D} E_F P_2 \!\left(\cos \theta_E \right) \biggl[ \left( \gamma + 2 \ln \left(k_F w\right)\right) \left(2 - \frac{\sqrt{\pi} t^{3/2}}{2} \text{Li}_{\frac{1}{2}}\!\left( -e^{- \tilde{\mu_0}/t} \right) \right) + 2 \int_{-\infty}^{\infty} \frac{dx \left(1-x\right)^2}{e^{\left(x^2 - \tilde{\mu_0} \right)/t} + 1}  \ln \left|1-x \right| \biggr] \label{eq:1Dsigma2}
\\
&= \frac{\pi^2}{2} \lambda_{1D} E_F P_2 \!\left(\cos \theta_E \right) \left[ \frac{8}{3}\left( \gamma - \frac{2}{3} + 2 \ln \left(2 k_F w\right) \right) + \left( \gamma  + 1 + 2 \ln \left(2 k_F w\right) \right) \frac{\pi^2}{6} t^2 \right] + O\left[t^4 \ln t\right] \label{eq:1Dsigma3}.
\end{align}
\end{widetext}
Differentiating with respect to density yields
\begin{multline}\label{eq:1DKratio1}
\frac{\kappa_0}{\kappa} = 1 - \frac{\pi^2}{12} t^2 + \frac{\pi^2}{4} \lambda_{1D} P_2 \!\left(\cos \theta_E\right) \biggl[8 \left(\gamma + 2\ln \left(2 k_F w\right) \right)
\\
- \left(\gamma -1 + 2\ln \left(2 k_F w\right) \right) \frac{\pi^2}{6} t^2 \biggr] + O\left[t^4 \ln t\right] .
\end{multline}
However, even if we take transverse size $w \sim 100$nm, corresponding to a very strong optical dipole trap, then at $n \sim 10^{4}$ cm$^{-1}$ although the system is quasi-1D ($\hbar^2/m w^2 \gg k_B T, E_F$) for $T \lesssim 40$ nK, $k_F w$ is actually of order unity.  This tends to diminish interaction effects since less repulsive off-axis interactions must be included in the effective 1D interaction.  (We did not hesitate to take $k_{F} w \ll 1$ in the 2D case \eqref{eq:V2Dq} because there the working assumption is that several independent quasi-2D layers will be formed by an optical lattice, which can be made considerably tighter than a dipole trap.  In any event, the general treatment of the multilayer quasi-2D in Sec.~\ref{sec:multilayer} includes finite thickness effects.)  For arbitrary values of $k_F w$, then, a similar calculation without assuming $q w \ll 1$ in Eq.~\eqref{eq:V1Dq} yields
\begin{widetext}
\begin{multline}
\Sigma_{1D} \left(k_F\right) = \frac{\pi^2}{2} \lambda_{1D} E_F P_2 \!\left(\cos \theta_E \right) \Biggl\{ 4 G^{2,2}_{2,3}\!\left( 4 k_F^2 w^2 \Bigg| {-1/2,0 \atop 0,0,-3/2} \right) + \frac{\pi^2}{6} t^2 \Biggl[ -1 + 4 k_F^2 w^2 e^{4 k_F^2 w^2} \Gamma\!\left(0, 4 k_F^2 w^2\right)
\\
+\frac{3}{2} G^{2,2}_{2,3}\!\left( 4 k_F^2 w^2 \Bigg| {-1/2,0 \atop 0,0,-3/2} \right) - 4 k_F^2 w^2 G^{2,2}_{2,3}\!\left( 4 k_F^2 w^2 \Bigg| {-3/2,-1 \atop -1,0,-5/2} \right) \Biggr] \Biggr\} + O\left(t^4 \ln t\right)
\end{multline}
\begin{multline}\label{eq:1Ddsigmadk}
\frac{d\Sigma_{1D}\left( k \right) }{dk} \bigg|_{k=k_{F}} = \frac{\pi^2}{2} \lambda_{1D} \frac{E_F}{k_F}  P_2 \!\left(\cos \theta_E \right) \biggl\{ -4 e^{4 k_{F}^2 w^2} \Gamma\! \left(0, k_F^2 w^2\right)
+ \frac{\pi^2}{12} t^2 \biggl[ \gamma -\frac{12}{\pi^2} \zeta'\left(2\right) + 8 k_F^2 w^2 - \ln k_F^2 w^2
\\
- \left(1 + 20 k_F^2 w^2 +32 k_F^4 w^4 \right) e^{4 k_F^2 w^2} \Gamma\! \left(0, 4 k_F^2 w^2\right) \biggr]  -\frac{\pi^2}{6} t^2 \ln t + O\left(t^4 \ln t\right) \biggr\}
\end{multline}
\begin{multline}\label{eq:1DKratio2}
\frac{\kappa_0}{\kappa} = 1 - \frac{\pi^2}{12} t^2 - \frac{\pi^2}{4} \lambda_{1D} P_2 \!\left(\cos \theta_E \right) \Biggl\{ 8 e^{4 k_F^2 w^2} \Gamma\!\left(0, 4 k_F^2 w^2\right) - \frac{\pi^2}{6} t^2 \Biggl[ -1 - e^{4 k_F^2 w^2} \Gamma\!\left(0, 4 k_F^2 w^2\right) \left(3 + 4 k_F^2 w^2  + 32 k_F^4 w^4\right)
\\
+ 8 k_F^2 w^2 \left(1 + e^{4 k_F^2 w^2} \Gamma\!\left(-1, 4 k_F^2 w^2\right) \right) + 6 G^{2,2}_{2,3}\!\left( 4 k_F^2 w^2 \Bigg| {-1/2,0 \atop 0,0,-3/2} \right) -16 k_F^2 w^2 G^{2,2}_{2,3}\!\left( 4 k_F^2 w^2 \Bigg| {-3/2,-1 \atop -1,0,-5/2} \right) \Biggr] \Biggr\} + O\left(t^4 \ln t\right)
\end{multline}
\end{widetext}
where $G^{2,2}_{2,3}\!\left( x \Big| {a_1, a_2 \atop b_1, b_2, b_3} \right)$ is a Meijer G-function~\citep{Gradshteyn}.  The more general Eq.~\eqref{eq:1DKratio2} was used for the low temperature approximation in Fig.~\ref{fig:1D}.  For the second row of plots, with $w = 10$ nm corresponding to a tight optical lattice forming an array of independent quasi-1D tubes, the simpler initial approach gives the same results except for at high densities.

Figures \ref{fig:1D}(a) and \ref{fig:1D}(d) show a maximum in the compressibility at finite temperature as in the 2D case.  Now, though, $t \sim mT/n^2$, so the temperature at which the peak occurs increases with density more rapidly than in 2D.  For $w = 10$nm, we have plotted results for higher densities and temperatures since these values can be reached while maintaining low-dimensionality.  In contrast to the 2D case, in a tightly confined system at reasonable density, the peak occurs at higher temperature as shown in Fig.~\ref{fig:1D}(d).  However, it is important to note that in 1D this is no longer a signature of interaction effects since the compressibility exhibits a maximum even in a noninteracting system due to the nonmonotonic temperature dependence of $\partial \mu_0/ \partial n$.  In fact, the main effect of interactions is to add a constant shift to the inverse compressibility, making the peak relatively \emph{less} dramatic.

As in the 2D case, we can also look for nonmonotonicity in the density dependence of $\kappa/\kappa_0 = dE_F/d\mu \sim n^3 \kappa$.  In the 2D case, we found a maximum in $\kappa/\kappa_0$ in the presence of interactions.  Here there is a weak maximum in the absence of interactions, due to the nonmonotonicity of $\mu_0$.  In the presence of interactions, this feature washes out, but the interactions may give rise to a more pronounced local maximum.  Unlike the 2D case, here the nonmonotonicity in density is predominantly due to nonmonotonic behavior of $d\Sigma\left(k_{F}\right)/dn$ rather than competition between the interacting and noninteracting terms.  For dipoles perpendicular to the trap axis and $w\sim 100$ nm, the density dependence at fixed temperature above a few nK is monotonic as in Fig.~\ref{fig:1D}(c).  However, for tighter radial confinement, the interaction effects become more pronounced and nonmonotonicity emerges at higher temperatures, as shown in Fig.~\ref{fig:1D}(f).  At high densities, the effective 1D interaction \eqref{eq:V1Dq} at the Fermi surface saturates due to the dependence on the finite transverse size.  As a result $\kappa/\kappa_0$ begins rising again.  Actually, for $w=100$nm ($w=10$nm), one-dimensionality is already beginning to break down for $n>2\times 10^4$ cm$^{-1}$ ($n>2\times 10^5$ cm$^{-1}$), since there $E_F \sim \hbar^2/mw^2$.

\begin{figure}[tbp]
  \includegraphics[width=\columnwidth]{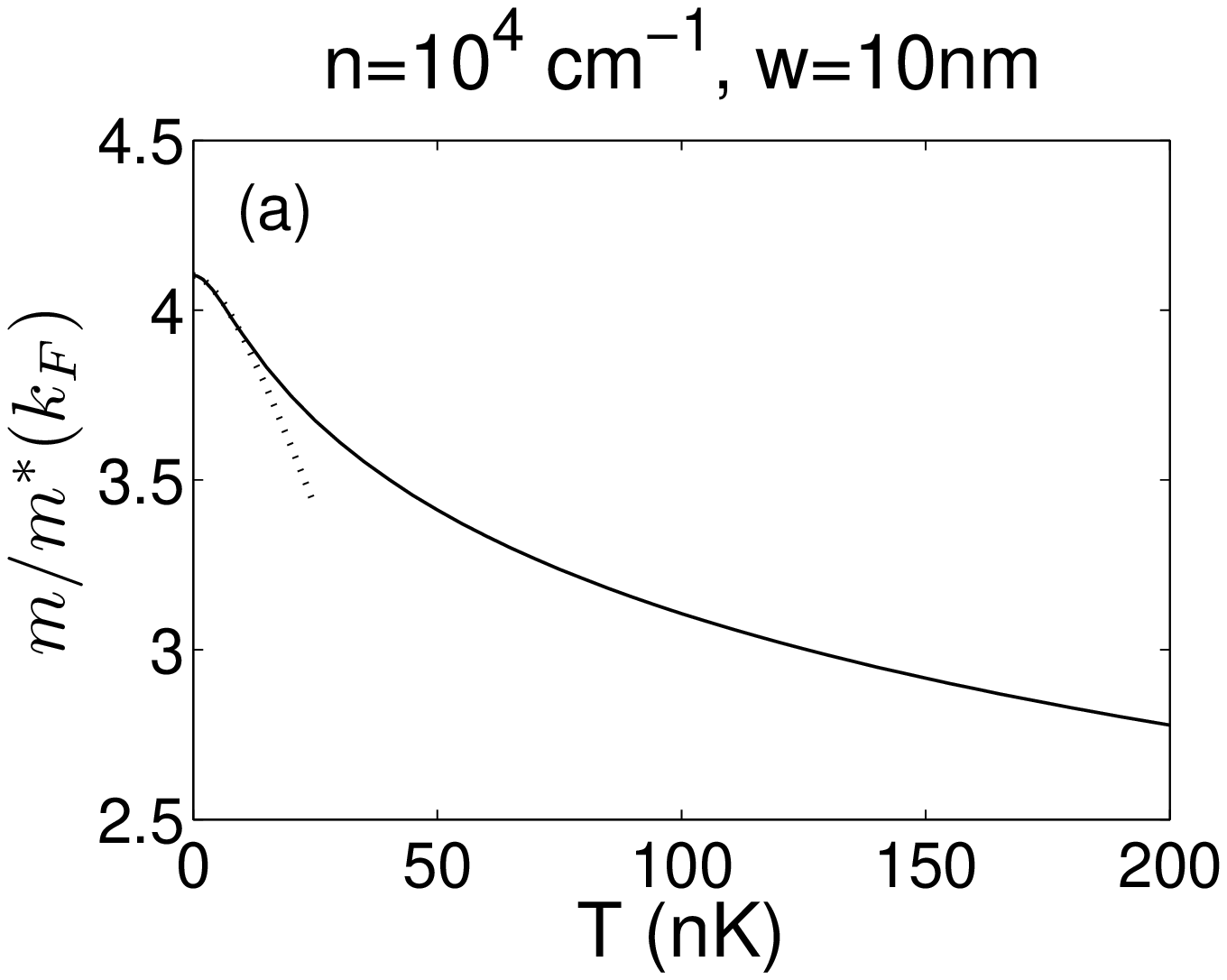}
  \\
  \includegraphics[width=\columnwidth]{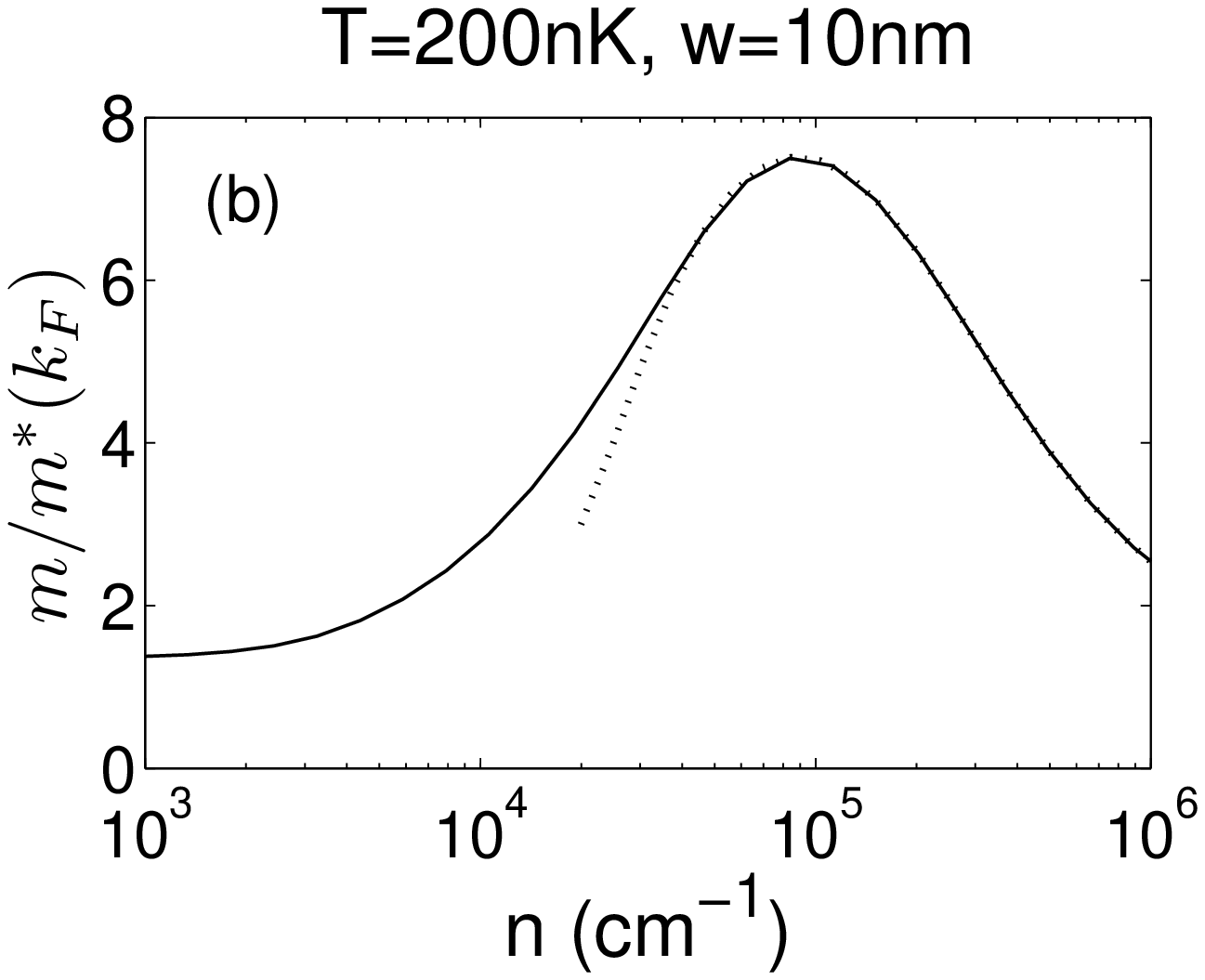}
\caption{Effective mass at the Fermi surface for quasi-1D KRb with $w=10$nm (a) vs temperature at fixed density and (b) vs density at fixed temperature.  Dotted lines show the low $T/T_F$ approximations.}\label{fig:1Dmeff}
\end{figure}

The effective mass is plotted in Fig.~\ref{fig:1Dmeff} as a function of density and temperature.  The effective mass is again diminished by the interaction and slowly increases towards its bare value at high temperatures.  At low densities, as in 2D, the reduction of the effective mass by the interaction increases with density.  However, as the density increases further the effective mass eventually returns to its bare value since the effective 1D interaction saturates while the Fermi velocity continues to increase.  In order to clearly show the trend, we have included in Fig.~\ref{fig:1Dmeff}(b) extremely high densities.  The analytic low temperature results from Eq.~\eqref{eq:1Ddsigmadk} are shown as dotted lines.

\subsubsection{Zero sound mode}
\begin{figure}[tbp]
  \includegraphics[width=\columnwidth]{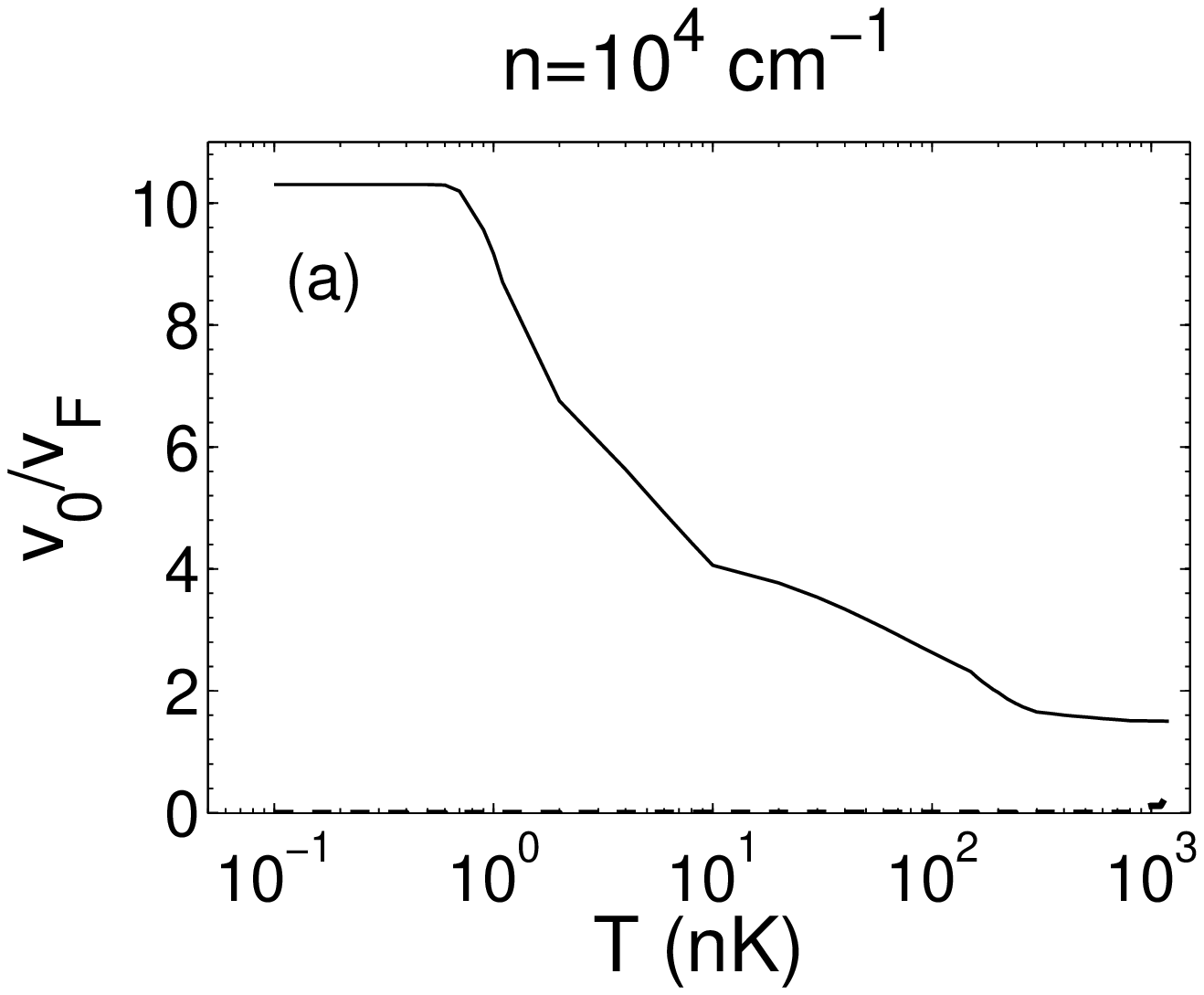}
  \\
  \includegraphics[width=\columnwidth]{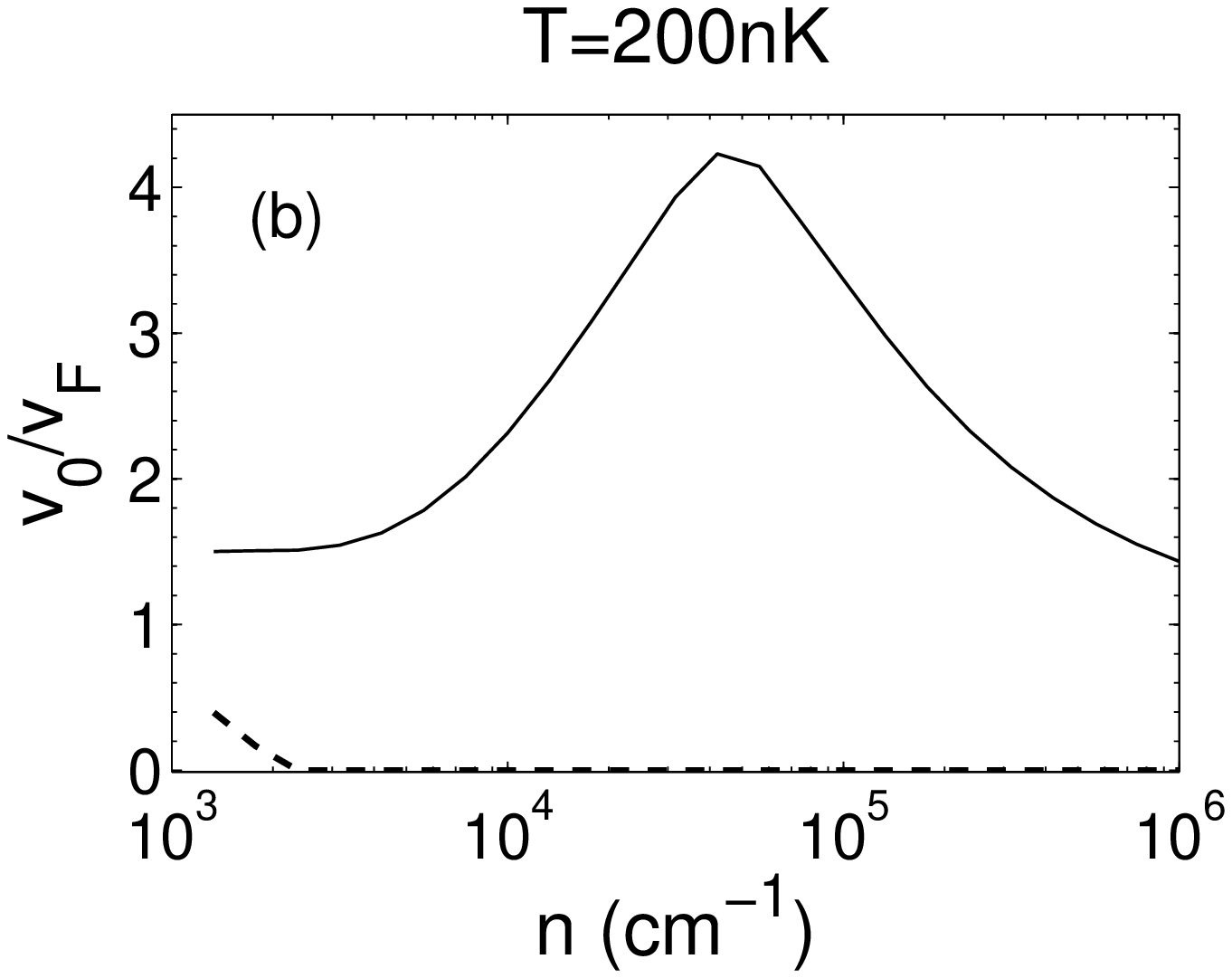}
\caption{Real (solid) and imaginary (dashed) parts of the zero sound speed for quasi-1D KRb with transverse size 10nm (a) vs temperature (note the log scale) at fixed density and (b) vs density at fixed temperature.}\label{fig:1Dzerosound}
\end{figure}
As in the 2D case, we consider the zero sound dispersion given by $1-V_{1D}\left(0\right) \chi \left(q\rightarrow 0, \omega\rightarrow v_0 q\right) = 0$, where
\begin{multline}
\chi \left(q\rightarrow 0, c_0 q \right) = \frac{k_F}{E_F} \int \frac{d x}{2\pi} \frac{- e^{\left(x^2 - \tilde{\mu_0} \right)/t}}{t \left[e^{\left(x^2 - \tilde{\mu_0} \right)/t}+1\right]^2}
 \\
 \times \left[1 - \frac{\frac{v_0}{v_{F}}}{\frac{v_0}{v_{F}} - x + i0} \right] .
\end{multline}
Once again, the finite transverse size which appears in Eq.~\eqref{eq:V1Dq} is important in determining the zero sound speed.  The dispersion is plotted in Fig.~\ref{fig:1Dzerosound} as a function of temperature and density for $w = 10$nm and dipoles aligned perpendicular to the axis of the tube.  In the low-temperature limit, one can show that $v_0/v_{F} = \sqrt{d^2 m/6\pi w^2\hbar^2 k_{F}}$, and we numerically recover this limit to better than 1\%.  One cannot obtain this limit by going to high density instead of low temperature, since, unlike in 2D where $\chi$ approaches unity in the high density limit, here $\chi$ goes to zero like $1/n$ for $v_0 \neq v_F$.  Thus, at high density the zero sound speed decreases to unity.  The damping strength, $\gamma/q$, is shown as a dashed line and it is negligible except near where the mode vanishes.  The collective mode is even more robust than in the 2D case since the 1D Fourier transform of the interaction is even more divergent.  Again, these results will remain physical until $r_*$ becomes comparable to either of the other length scales, $k_F^{-1}$ or $w$.

\begin{figure*}[tbp]
  \includegraphics[width=.66\columnwidth]{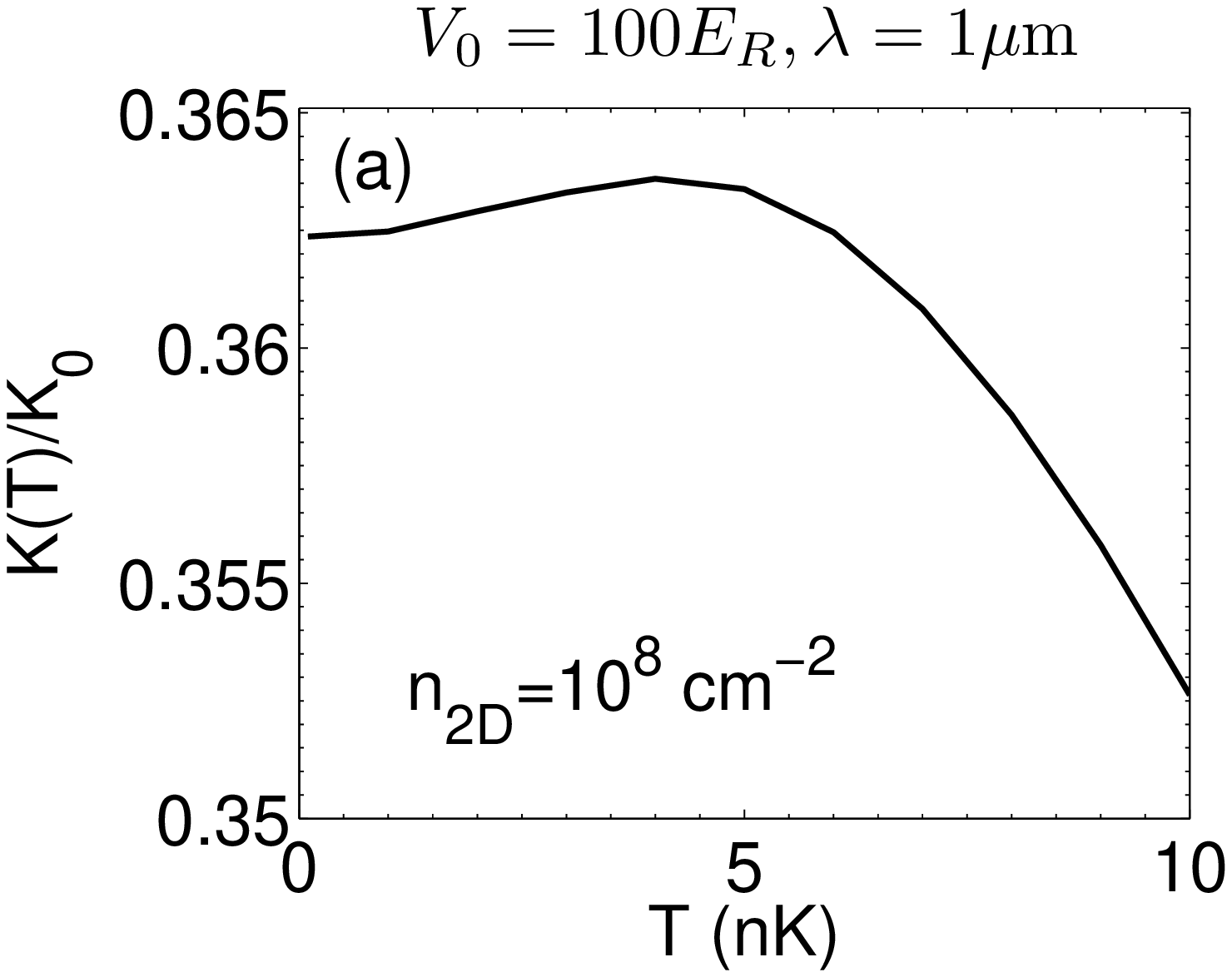}
  \includegraphics[width=.66\columnwidth]{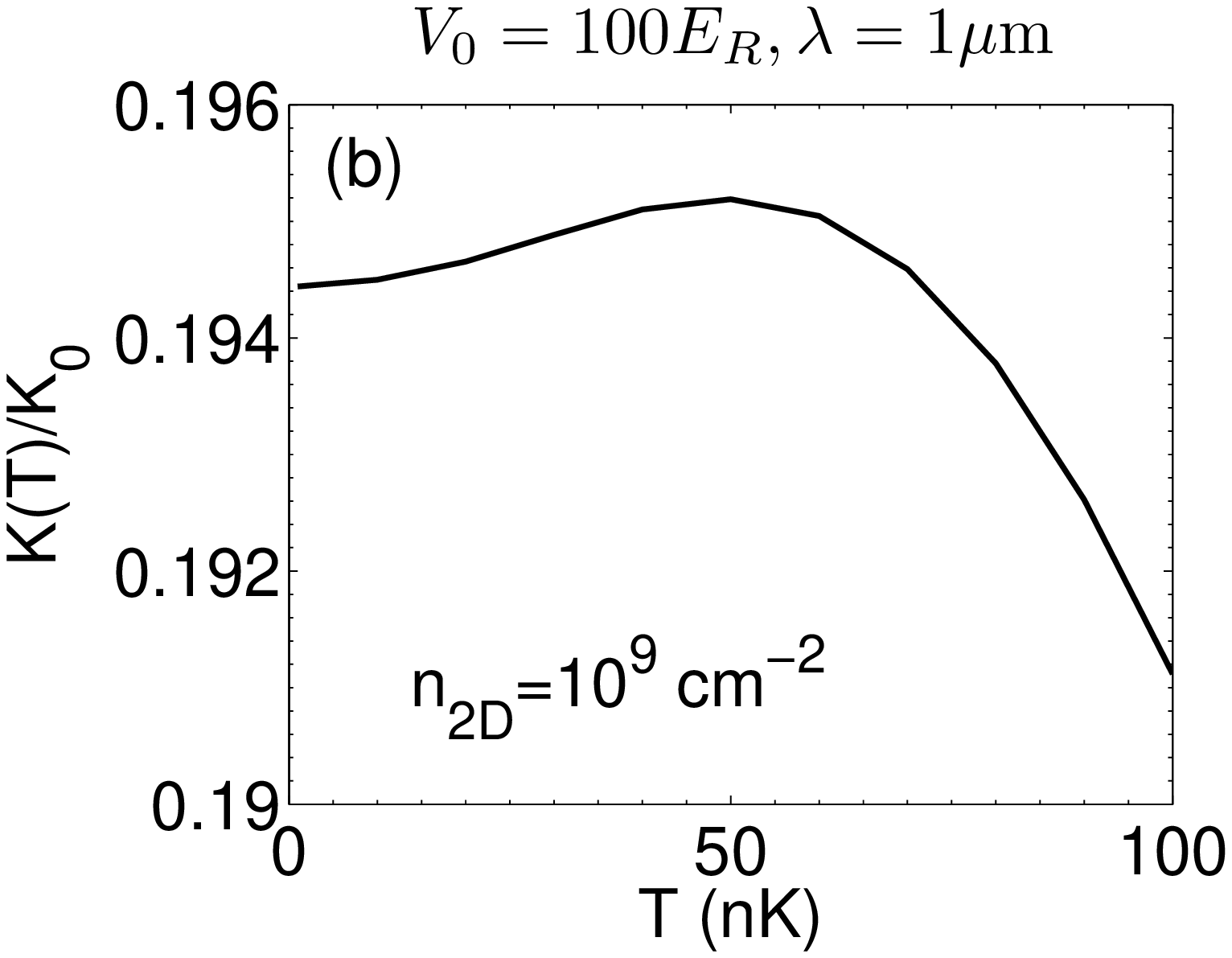}
  \includegraphics[width=.66\columnwidth]{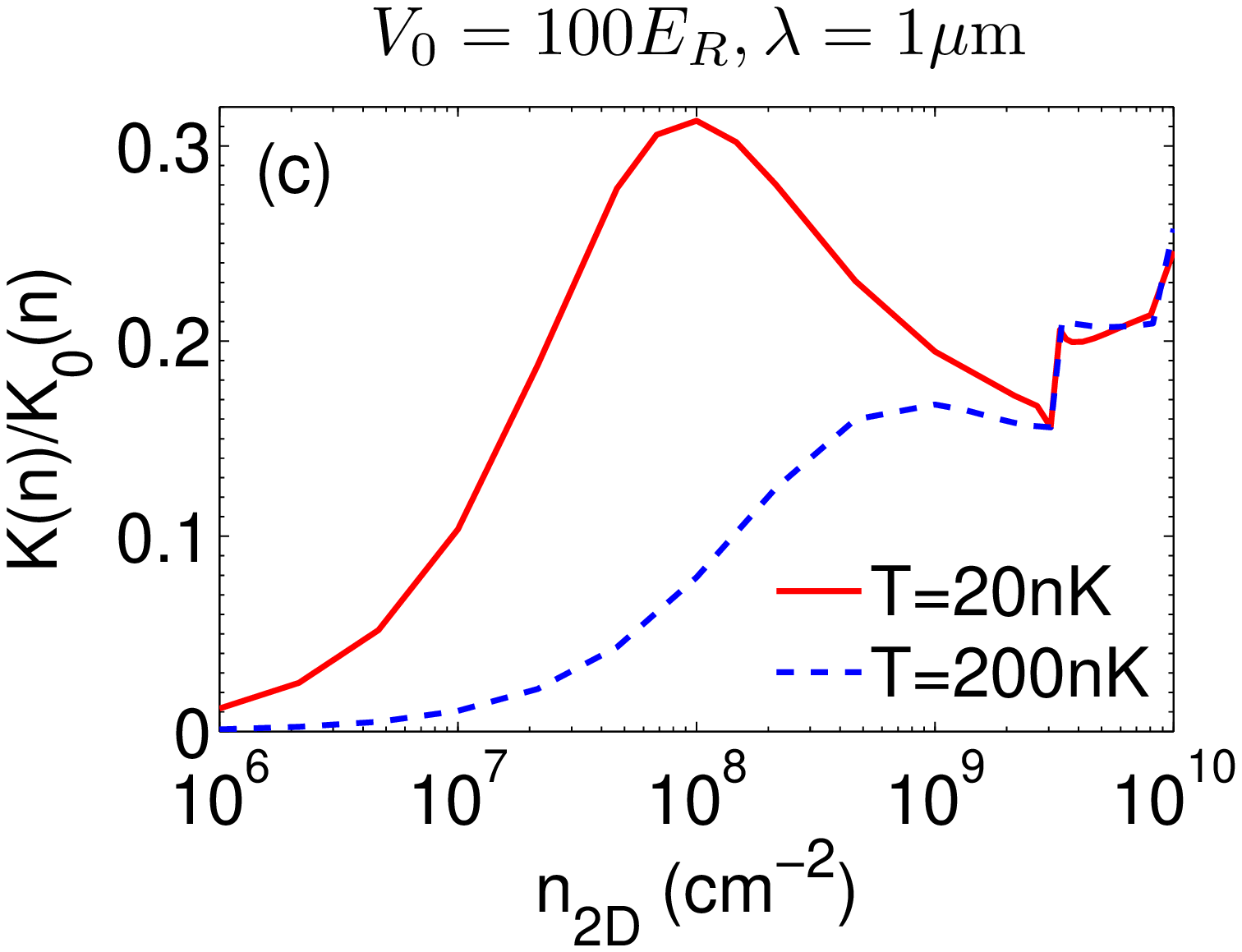}
\\
  \includegraphics[width=.66\columnwidth]{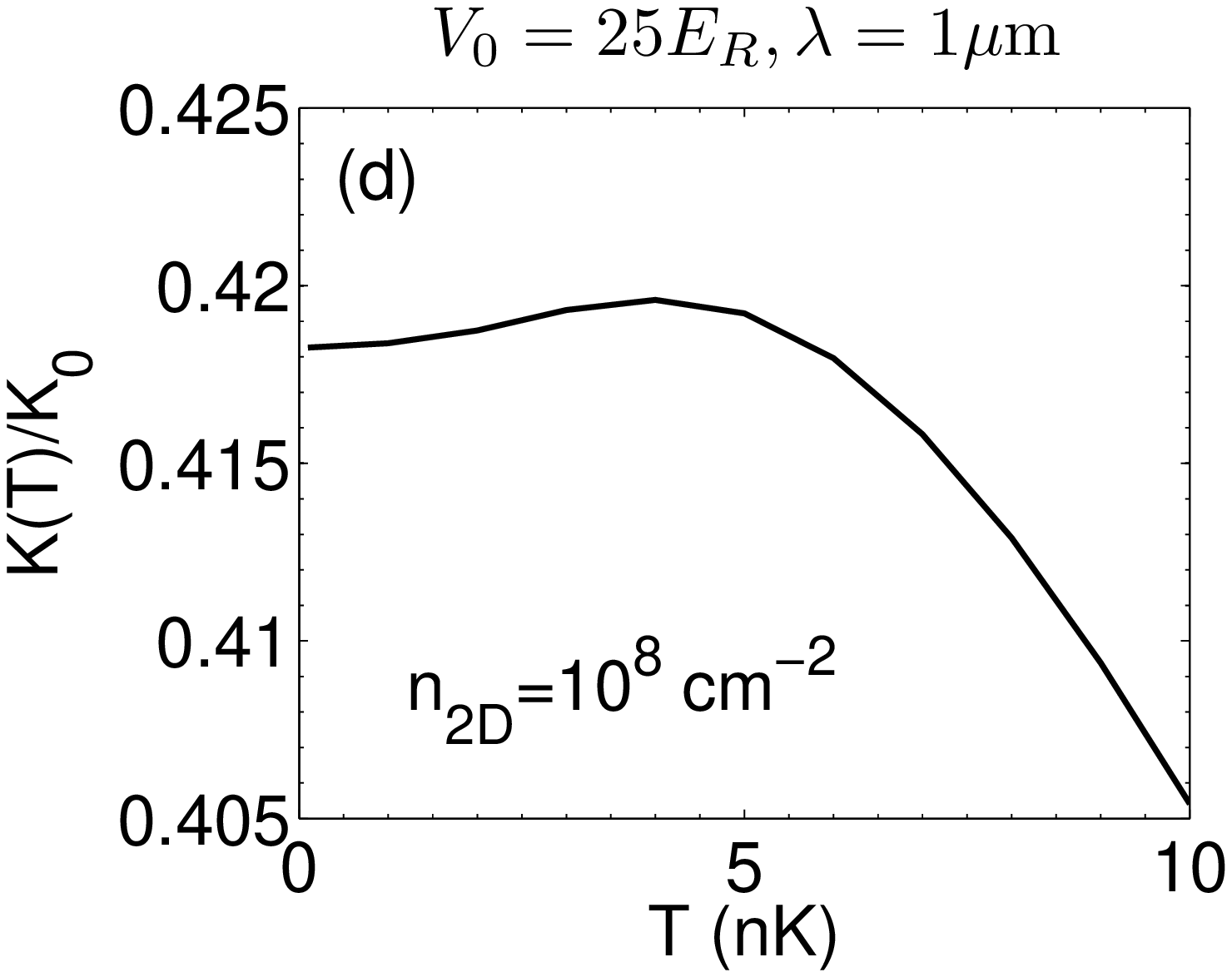}
  \includegraphics[width=.66\columnwidth]{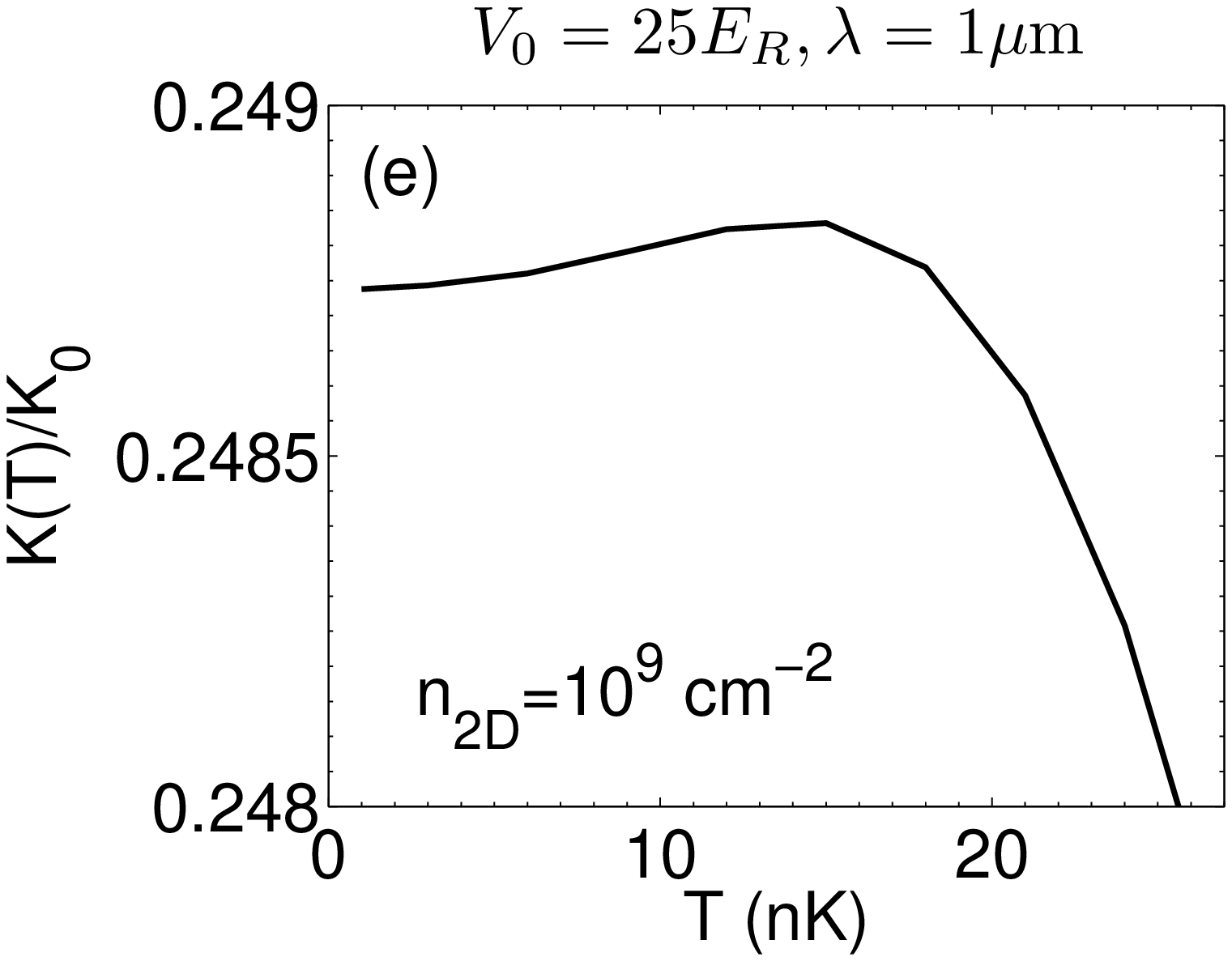}
  \includegraphics[width=.66\columnwidth]{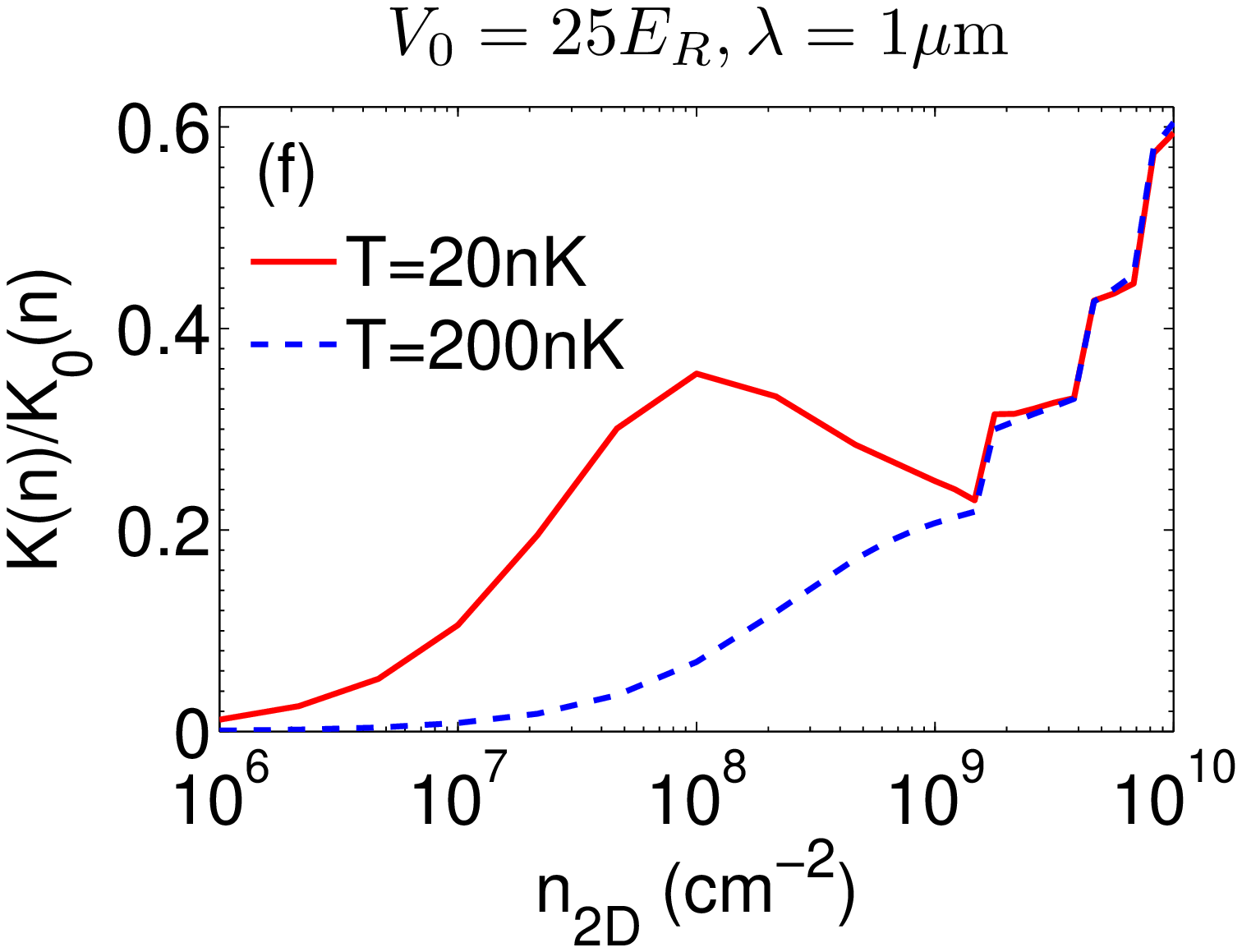}
\caption{(Color online.)  Compressibility for KRb in a periodic potential along $z$. (a-b,d-e) $\kappa/\kappa_0$ vs temperature at fixed density; (c,f) $\kappa/\kappa_0$ (\textit{i.e.}, $dn/d\mu$) vs density at fixed temperature.}\label{fig:multilayer}
\end{figure*}
\section{3D Multilayer self-energy and compressibility at finite temperature}\label{sec:multilayer}
\subsection{Compressibility}
Now we consider a 3D gas placed in a strong 1D optical lattice of depth $V_0$ and wavelength $\lambda$, resulting in a "stack of pancakes" configuration.  In the limit where $V_0$ and $\lambda$ become infinite, this is the same as the strictly 2D monolayer considered in Sec.~\ref{subsec:2D}.  Generally, however, we find that the nonmonotonic behavior discussed in Sec.~\ref{subsec:2D} persists in the presence of multilayer and finite thickness effects.  We will restrict our discussion to the case of dipoles aligned along the direction of the optical lattice.

Because of the external periodic potential along $z$, the effect of the interaction on the chemical potential to first order, $\mu = \mu_0 + \Delta \mu$, must now be written in real space:
\begin{equation}\label{eq:multilayermu}
\Delta \mu = \bigg\langle \int d^3 \mathbf{x} d^3 \mathbf{x'} \phi_{\mathbf{k_F}}^{\ast}\left(\mathbf{x}\right) \Sigma \left( \mathbf{x}, \mathbf{x'} \right) \phi_{\mathbf{k_F}}\left(\mathbf{x'}\right) \bigg\rangle_{\Omega_{\mathbf{k_F}}},
\end{equation}
where the angular averaging over the Fermi surface denoted by the brackets is necessary to separate the shift in chemical potential from the anisotropic deformation of the Fermi surface, and
\begin{multline}
\Sigma \left( \mathbf{x}, \mathbf{x'} \right) =  \int d^3 \mathbf{x''} \int \frac{d^3 \mathbf{k}}{\left(2\pi\right)^3} n_0\left(\mathbf{k}\right) V_{3D}\left(\mathbf{x}-\mathbf{x''}\right)
\\
\times \biggl[ \left|\phi_{\mathbf{k}}\left(\mathbf{x''}\right)\right|^2 \delta\left(\mathbf{x}-\mathbf{x'}\right)
- \phi^{\ast}_{\mathbf{k}}\left(\mathbf{x'}\right) \phi_{\mathbf{k}}\left(\mathbf{x}\right) \delta\left(\mathbf{x''}-\mathbf{x'}\right) \biggr] .
\end{multline}
The unperturbed single-particle wavefunction can be written in terms of the reciprocal lattice points, which are integer multiples of $2\pi/\lambda$, in the Bloch form
\begin{equation}\label{eq:multilayerphi}
\phi_{\mathbf{k}}\left(\mathbf{x}\right) = e^{i \mathbf{k}\cdot \mathbf{x}} \sum_{q=-N}^{N} u_{q} \left(k_z\right) e^{-i 2 \pi q x_z/\lambda}.
\end{equation}
The coefficients $u_q\left(k\right)$ are easily obtained numerically, as are the corresponding single-particle energies needed to evaluate $n_0\left(\mathbf{k}\right)$.  In the above we have introduced a cutoff, $2 N\pi/\lambda$, on the Bloch momentum in order to perform the sum numerically.  After some algebra, Eq.~\eqref{eq:multilayermu} can be recast as a sum over momenta,
\begin{widetext}
\begin{multline}\label{eq:multilayermu2}
\Delta \mu = \bigg\langle \int \frac{d^3 \mathbf{k}}{\left(2\pi\right)^3} n_0\left(\mathbf{k}\right) \sum_{q=-2N}^{2N} \biggl[ V_{3D}\left(0,\frac{2 \pi q}{\lambda} \right) \sum_{q'=q_<}^{q_>} u_{q'} \left(k_z\right) u_{q'+q}^{\ast} \left(k_z\right) \sum_{q''=q_<}^{q_>} u_{q''} \left(k_{F_z}\right) u_{q''+q}^{\ast} \left(k_{F_z}\right)
\\
- V_{3D}\left(k_{\perp}-k_{F_{\perp}}, k_z-k_{F_z}-\frac{2 \pi q}{\lambda} \right) \biggl|\sum_{q'=q_<}^{q_>} u_{q'} \left(k_z\right) u_{q'+q}^{\ast} \left(k_{F_z}\right) \biggr|^2 \biggr] \bigg\rangle_{\Omega_{\mathbf{k_F}}}
\end{multline}
\end{widetext}
where $V_{3D}\left(q_{\perp},q_z\right)$ is the 3D Fourier transformed interaction defined in Eq.~\eqref{eq:V3Dq} and $q_< \equiv \max\left(-N,-N-q\right)$, $q_> \equiv \min\left(N,N-q\right)$.  We typically take $N \sim 50$, which we find to be sufficient for the lattice depths treated.

To facilitate comparison with the strictly 2D case, we measure density in terms of the 2D density per layer, $n_{2D} = n_{3D}\lambda$, and compressibility in units of $\kappa_0 \equiv \frac{1}{n_{2D}^2} \frac{d n_{2D}}{d E_{2D}}$, where $E_{2D} = \hbar^2 4\pi n_{2D}/2m$.  By first analytically performing the azimuthal integration in Eq.~\eqref{eq:multilayermu2} and differentiating with respect to density, then numerically summing over reciprocal lattice points, averaging over the Fermi surface, and performing the remaining two-dimensional integral, the compressibility can be efficiently obtained.  In Fig.~\ref{fig:multilayer} we show the compressibility for KRb in a deep optical lattice (here the recoil energy $E_R = \hbar^2 \left(2\pi/\lambda\right)^2/2m$).  For strong lattices, the compressibility behaves similarly versus temperature to the 2D results of Fig.~\ref{fig:2D}, with a maximum at nearly the same temperature.  For weak lattices, the effect is less pronounced.

As density is varied, the ratio $\kappa/\kappa_0$ behaves as in the 2D case for low densities, with nonmonotonicity evident even for $T\sim T_F$, but at high density $E_{2D}$ becomes larger than the lattice bandgap and quasi-two-dimensionality breaks down.  Here the compressibility exhibits discontinuous jumps as the chemical potential enters higher bands, reminiscent of the effect seen in Ref.~\citep{Mazzarella09}.  At very large densities, the recoil energy is negligible compared to the kinetic energy of the gas, and $\kappa/\kappa_0$ should increase with density as in the 3D case rather than decreasing with density as in the 2D case \footnote{In the 3D case, $\kappa/\kappa_0$ goes to unity as $n$ goes to infinity.  However, since we have defined $\kappa_0$ in this section in terms of the 2D energy scale $E_{2D} \propto n$ rather than the 3D energy scale $E_{3D} \propto n^{2/3}$, the ratio plotted in Fig.~\ref{fig:multilayer}(c) should not go to unity, but should go like $n^{1/3}$ as $n$ goes to infinity.}.  So, in addition to the interaction-induced local maximum analogous to the 2D case, there can also be a local minimum as the system crosses over to three-dimensional behavior.  This is similar to the finite confinement effect seen in the 1D case above.

\subsection{Collective modes}
\begin{figure*}[tbp]
  \includegraphics[width=\columnwidth]{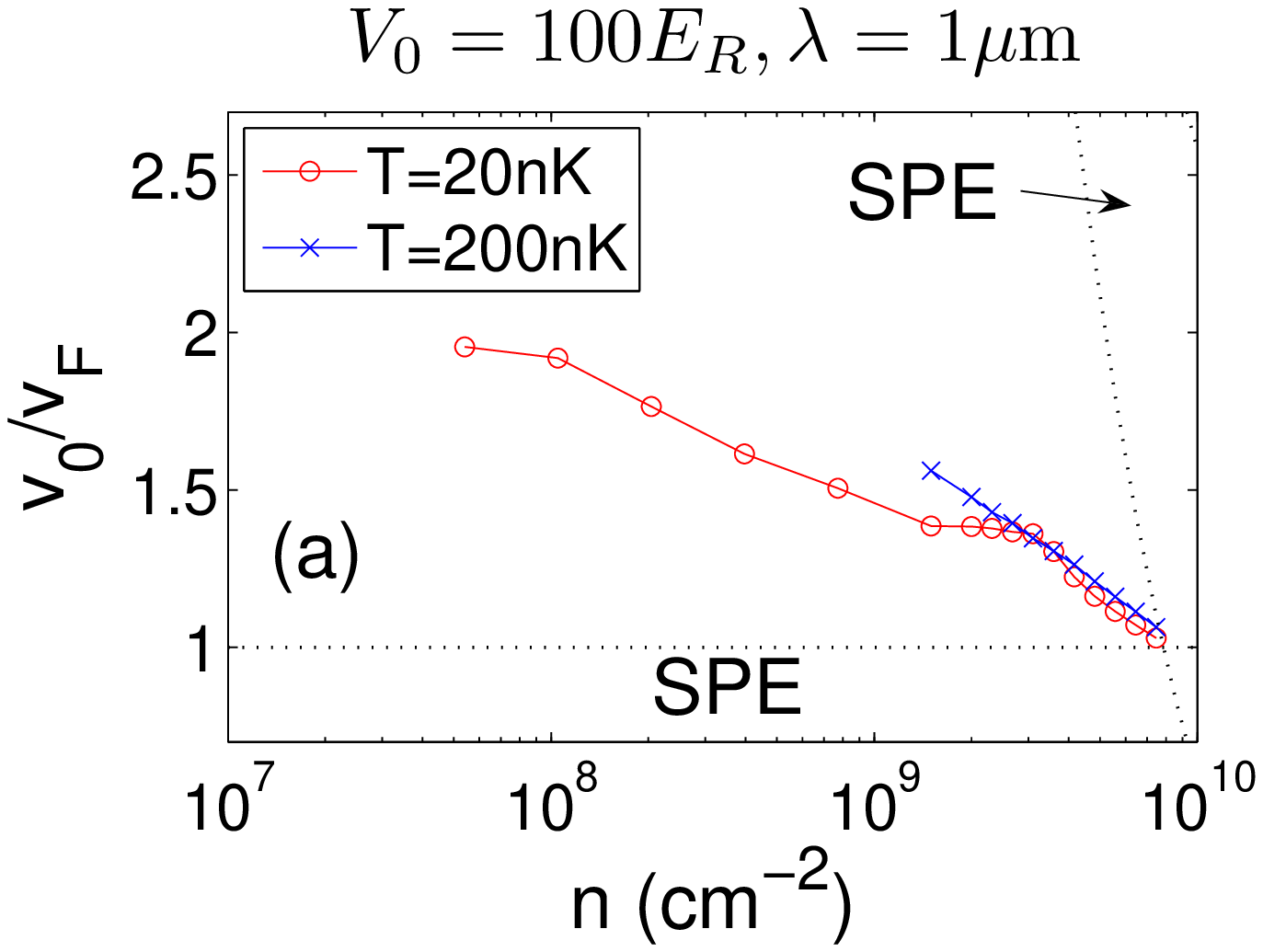}
  \includegraphics[width=\columnwidth]{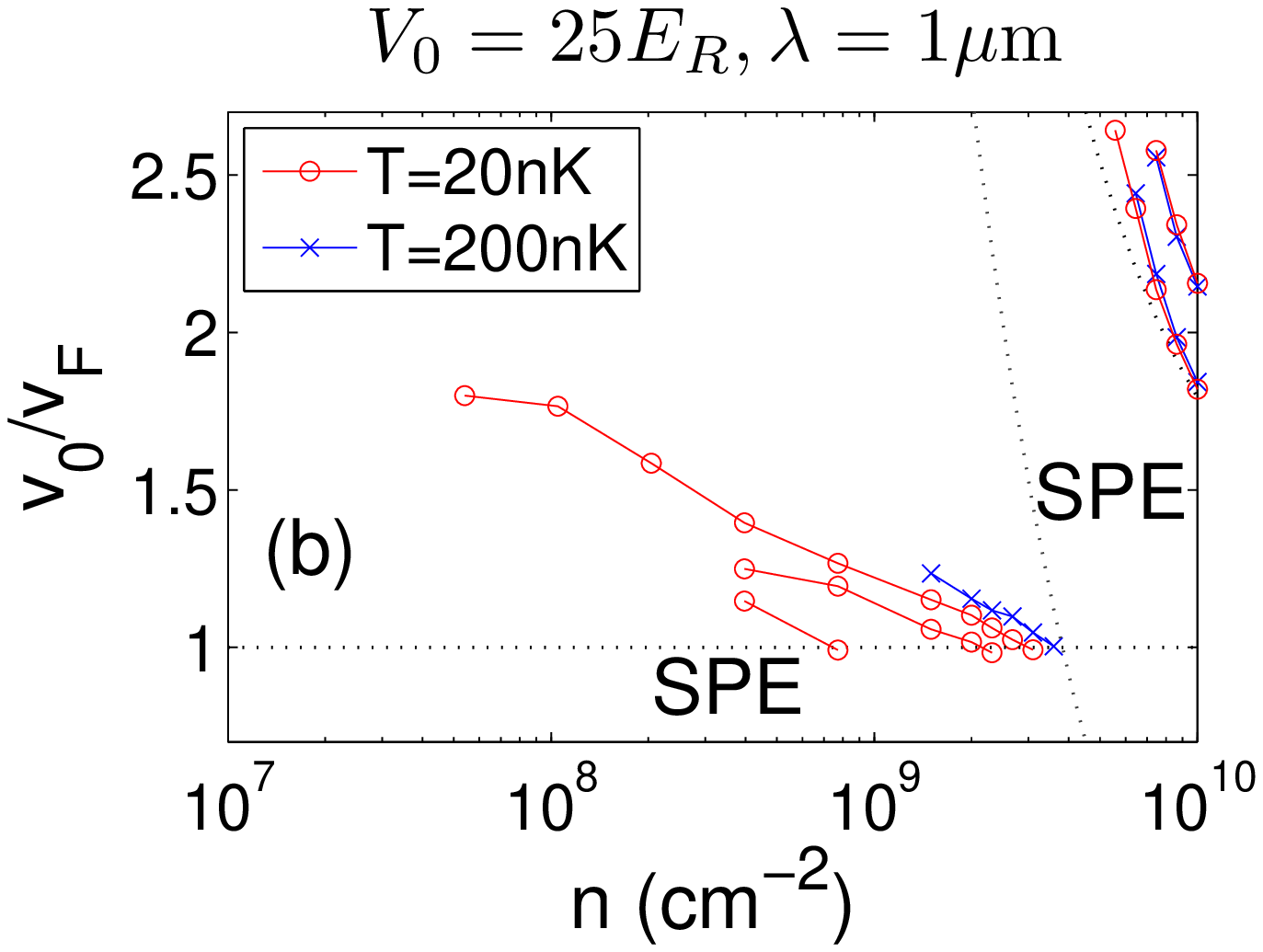}
\caption{(Color online.) Zero sound speed vs density for KRb in a periodic potential along the $z$-axis.}\label{fig:multilayerzerosound}
\end{figure*}
For the 3D multilayer system described above we now calculate the low-lying collective modes at finite temperature, neglecting coupling between partial waves.  In the limit of a large bandgap, such couplings should be suppressed, and outside this limit the description should still be qualitatively valid.  However, because of the spatial inhomogeneity, the 3D polarizability can no longer be written as a scalar function of the momentum transfer.  In terms of the Bloch wavefunctions defined in Eq.~\eqref{eq:multilayerphi}, the interaction and the zeroth order polarizability can be written in momentum space as
\begin{equation}
V_{\mathbf{k,k'}} = \int d^3 \mathbf{x} d^3 \mathbf{x'} \phi^{\ast}_{\mathbf{k}}\left(\mathbf{x}\right) V\left(\mathbf{x}-\mathbf{x'} \right) \phi_{\mathbf{k'}}\left(\mathbf{x'}\right) ,
\end{equation}
\begin{equation}
\chi_{\mathbf{k,k'}} \left(\omega\right) = \sum_{\mathbf{k_1,k_2}} a_{\mathbf{k}}^{\mathbf{k_1,k_2} \ast} a_{\mathbf{k'}}^{\mathbf{k_1,k_2}} \frac{n_0 \left(\mathbf{k_1}\right)-n_0 \left(\mathbf{k_2}\right)}{\omega - \epsilon_0 \left(\mathbf{k_1}\right) + \epsilon_0 \left(\mathbf{k_2}\right) + i0} ,
\end{equation}
where $\epsilon_0 \left(\mathbf{k}\right)$ is the bare single-particle energy associated with $\phi_{\mathbf{k}}$, and
\begin{equation}
a_{\mathbf{k}}^{\mathbf{k_1,k_2}} = \int d^3 \mathbf{x} d^3 \mathbf{x'} \phi^{\ast}_{\mathbf{k_1}}\left(\mathbf{x}\right) \phi_{\mathbf{k_2}}\left(\mathbf{x}\right) \phi_{\mathbf{k}}\left(\mathbf{x}\right) .
\end{equation}
The collective modes are given by the zeros of the determinant $\det \left|1-M \left(\omega\right) \right| = 0$, where $M_{\mathbf{k,k'}}  \left(\omega\right) = \left(V \chi \left(\omega\right) \right)_{\mathbf{k,k'}}$.  Conservation of momentum in the $x-y$ plane and of quasi-momentum along $z$ reduces $M$ to a matrix function of the momentum transfer, $M_{m,n} \left(\mathbf{q}, \omega \right)$, where $m$ is an integer denoting a reciprocal lattice vector $K_m = 2\pi m/\lambda$ and $q_z$ is in the first Brillouin zone.  Some algebra results in a computationally feasible expression for the matrix elements,
\begin{widetext}
\begin{multline}
M_{m,n} \left(\mathbf{q}, \omega \right) = \sum_{Q=-3N-m}^{3N-m} \int \frac{d^3 \mathbf{k}}{\left(2\pi\right)^3} \frac{n_0 \left(k_{\perp} - q_{\perp}, k_{z}-q_z +  2\pi Q/\lambda \right) - n_0 \left(k_{\perp},k_z \right)}{\omega - \epsilon \left(k_{\perp} - q_{\perp}, k_{z}-q_z +  2\pi Q/\lambda \right) + \epsilon \left(k_{\perp},k_z \right) + i0}
\\
\times \left[\sum_{m'n'} u_{Q+m+m'-n'}^{\ast} \left(k_z-q_z+\frac{2\pi Q}{\lambda}\right) u_{m'} \left(k_z\right) u_{n'}^{\ast} \left(q_z+\frac{2\pi m}{\lambda}\right) \right]
\\
\times \left[\sum_{m'n'} u_{Q+n+m'-n'} \left(k_z-q_z+\frac{2\pi Q}{\lambda}\right) u_{m'}^{\ast} \left(k_z\right) u_{n'} \left(q_z+\frac{2\pi n}{\lambda}\right) V_{3D} \left(q_{\perp}, q_z+\frac{2 \pi \left(n-n'\right)}{\lambda} \right) \right].
\end{multline}
\end{widetext}
The matrix elements can be evaluated numerically for a given frequency and momentum.  To proceed, one must obviously truncate the matrix at some finite reciprocal lattice vector, and we have used a $17\times 17$ matrix, as this seems to include the dominant terms for large potential depth, $V_0$.  As $V_0$ is decreased, the off-diagonal terms of the matrix vanish while the number of nonnegligible diagonal elements increases.

In Fig.~\ref{fig:multilayerzerosound} we plot the lowest-lying underdamped collective modes for long-wavelength, in-plane momentum transfer, $q_z=0, q_{\perp} \rightarrow 0$.  The zero sound speed is shown in units of the 2D Fermi speed, $v_{F} = \hbar^2 \sqrt{4\pi n_{2D}}/m$.  In addition to the Landau damping due to single particle excitations in the lowest band for $v_0<v_{F}$, there are also regions of damping due to single-particle excitations in excited bands which occur for $v_0>v_{F}$.  The boundaries of these regions are shown in dotted lines.  The modes plotted are weakly thermally damped.

It is instructive to first consider the 20nK curve in Fig.~\ref{fig:multilayerzerosound}(a), where one can identify three distinct regions.  At densities below $\sim 10^9 cm^{-2}$, the thermal energy is comparable to the 2D energy scale, $E_{2D} = \hbar^2 4\pi n_{2D}/2m$.  For higher densities, one might expect to see a density independent speed, as in Sec.~\ref{subsubsec:2Dzerosound} and the zero temperature calculation of Ref.~\citep{Li10}.  This appears to be the case for an intermediate region, with the value $v_0>v_{F} \sim 1.4$ in qualitative agreement with the value $\sim 1.6$ obtained using the multilayer result of Ref.~\citep{Li10} and inserting an estimate for the transverse size in the effective 2D interaction \eqref{eq:V2Dq} based on the harmonic approximation to the local wells of the optical lattice.  However, for even higher densities the speed abruptly drops towards the continuum.  This begins happening where $E_{2D}$ becomes comparable to the bandgap, signalling a transition to 3D behavior.  As we have seen in Sec.~\ref{subsubsec:3Dzerosound}, the mode with momentum transfer perpendicular to the electric field must vanish in 3D.  In this region, the behavior is roughly independent of $T$, since the reduced temperature $T/T_F$ is effectively zero.

For $T=200$nK, the mode does not propagate except at high densities, and we cannot reach the low temperature limit without increasing the density into the 3D transition.  For the weaker lattice in Fig.~\ref{fig:multilayerzerosound}(b), even at $T=20$nK the intermediate plateau has vanished because higher bands are already becoming active as we increase the density out of the large $T/T_F$ regime.  In Fig.~\ref{fig:multilayerzerosound}(b) we see multiple modes entering the continuum, although only one is stable at low density.  Also, at high density, faster zero sound modes appear above the first excited SPE band.

\begin{figure}[tbp]
  \includegraphics[width=\columnwidth]{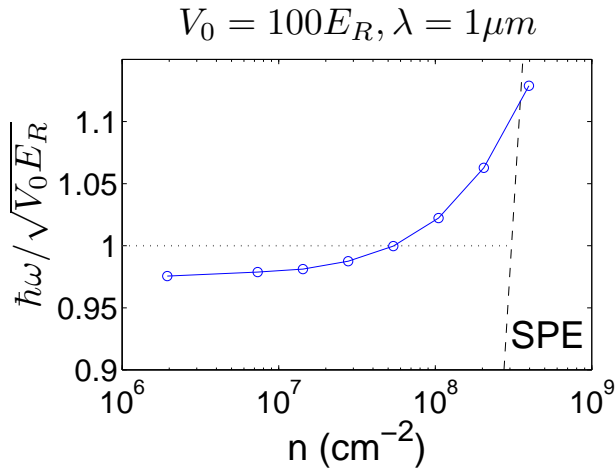}
\caption{Sloshing frequency vs density for KRb in a periodic potential along the $z$-axis.}\label{fig:wzV100}
\end{figure}

In Fig.~\ref{fig:wzV100} we show the lowest underdamped collective excitation for momentum transfer along the $z$-axis in units of the frequency of the local harmonic well approximation to the optical lattice.  In a harmonic trap this is the Kohn (or sloshing) mode \citep{Kohn61}, and it is independent of the interaction and the momentum.  Here the density dependence is due to anharmonicity of the potential.  The curve shown is independent of temperature and momentum.  For $V_0=25E_R$, we do not find this mode.

So, although one can observe a Kohn-like collective mode along the direction of the optical lattice at currently attainable temperatures and densities, the observation of zero sound in a reasonably strong optical lattice with standard site separation on the order of a micron requires an order of magnitude improvement in the attainable temperature or density.  Based on the 2D calculations of Sec. \ref{subsubsec:2Dzerosound}, it is reasonable to expect this requirement could be eased by using a more tightly spaced lattice.

\section{Experimental detection of compressibility}\label{sec:exp}
\begin{figure*}
  \includegraphics[width=.95\columnwidth]{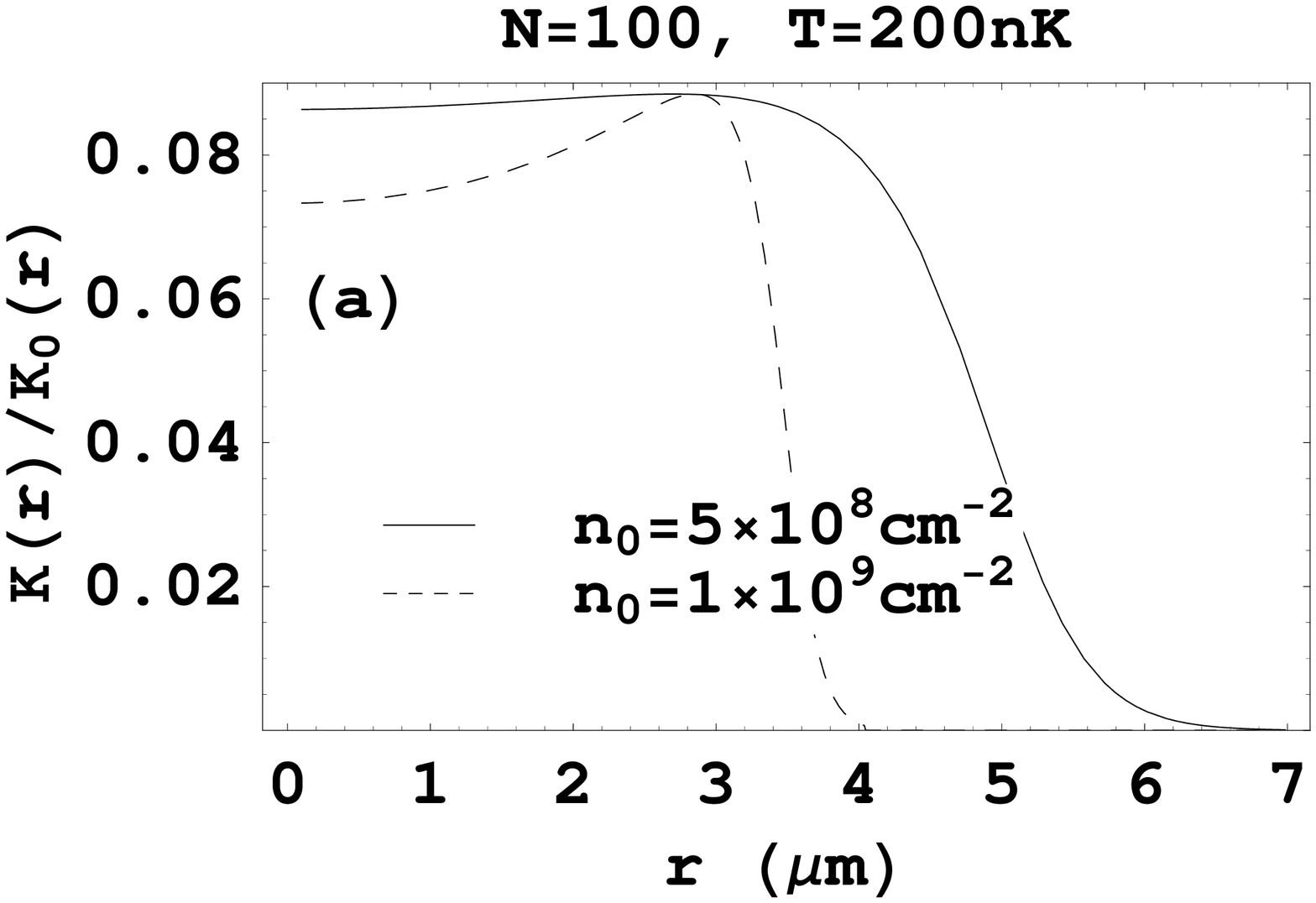}
  \includegraphics[width=.95\columnwidth]{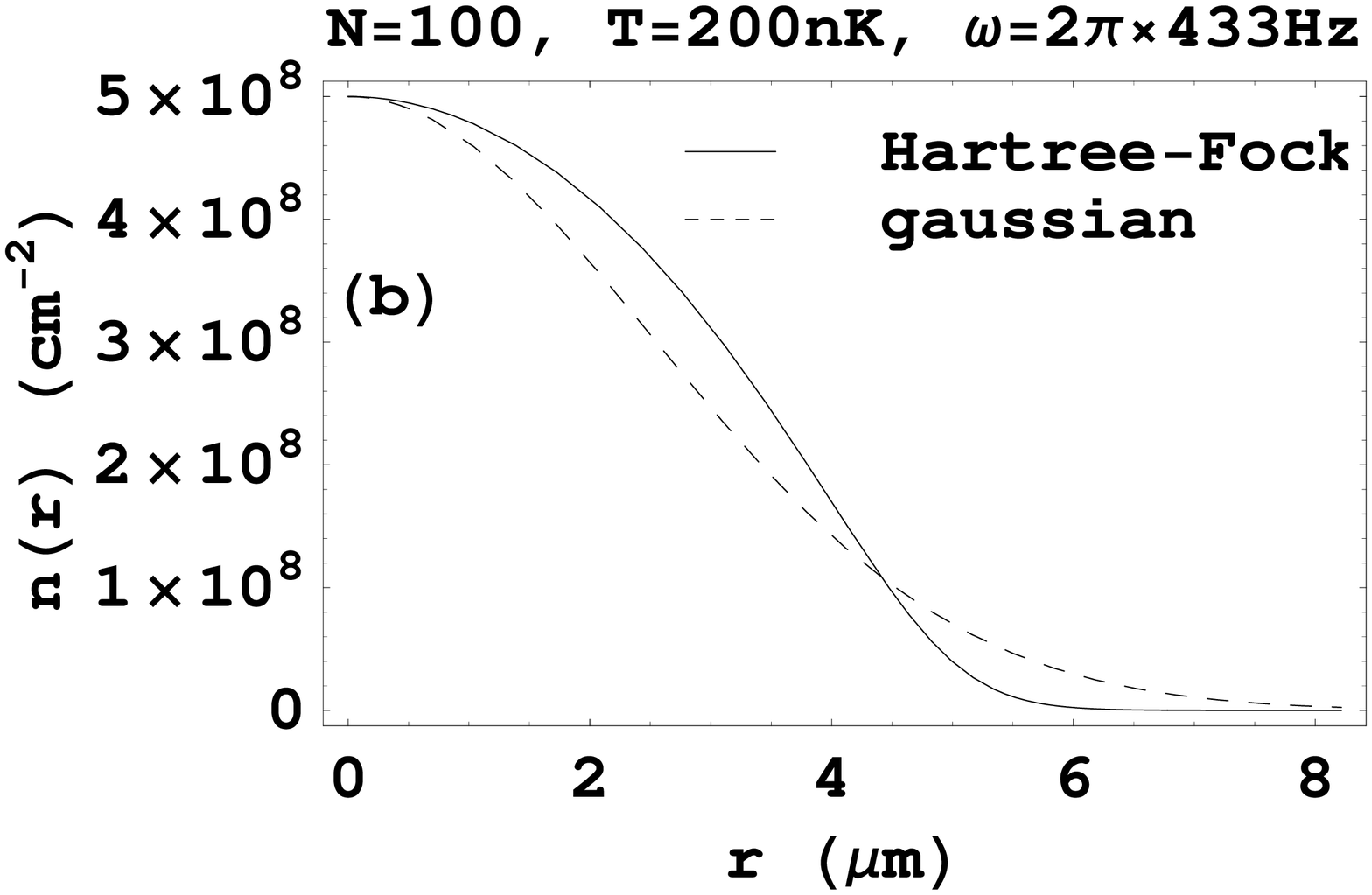}
  \\
  \includegraphics[width=.95\columnwidth]{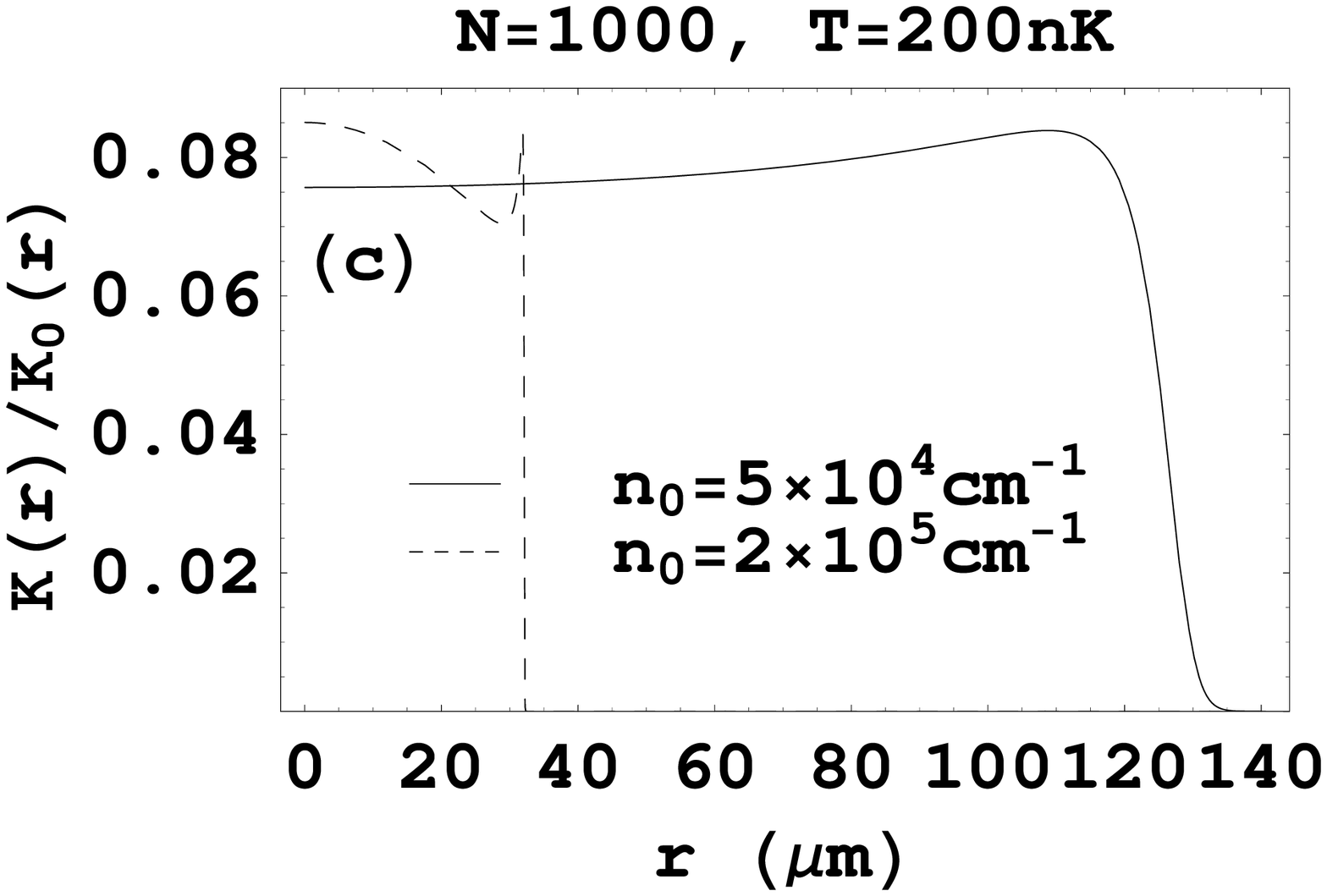}
  \includegraphics[width=.95\columnwidth]{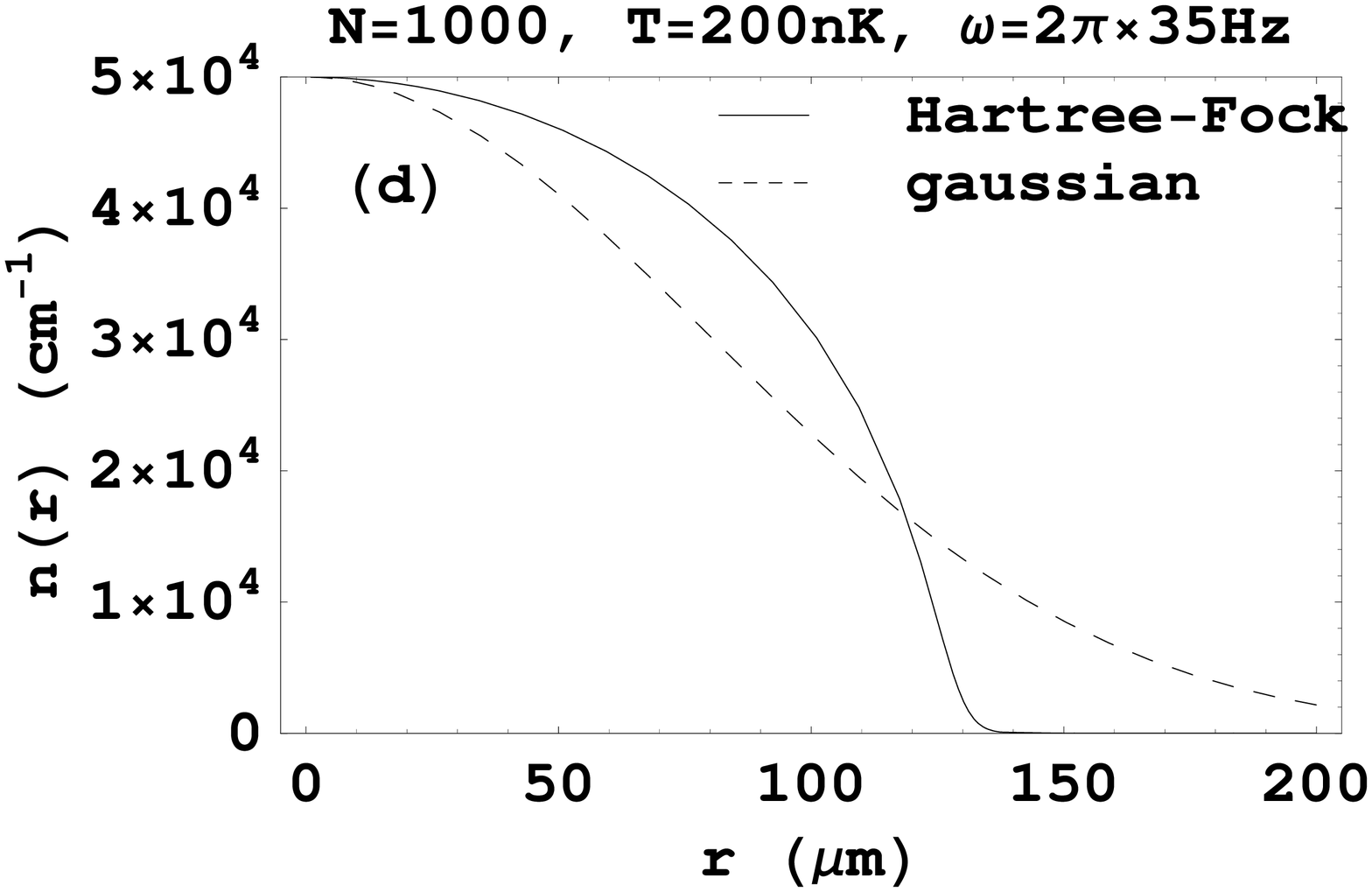}
\caption{2D results shown in panels (a,b) and quasi-1D results with $w=10$ nm in panels (c,d) for KRb.  (a,c) Local $\kappa/\kappa_0$ vs distance from trap center, using LDA.  (b,d) Density profile, using LDA, with a gaussian profile shown for reference.}\label{fig:trapped}
\end{figure*}
We have shown that a quasi-low-dimensional experiment may be able to observe an interaction-induced maximum in the ratio of the finite-temperature interacting compressibility and the zero-temperature noninteracting compressibility, particularly if the measurement is performed with variable density.  Now, the results we have shown are for homogeneous systems, while there is always a global confining potential in experiments.  However, if we incorporate a weak harmonic trap through a simple local density approximation (LDA), $\mu \left(r\right) \approx \mu\left(0\right) - m \omega^2 r^2/2$, where one can think of a local compressibility, $\kappa\left(r\right) = \frac{1}{n^2 \left( r \right)} \frac{\partial n\left( r \right)}{\partial \mu\left(r\right)} = - \frac{1}{n^2 \left( r \right) m \omega^2 r} \frac{\partial n\left(r\right)}{\partial r}$, then the trap can be beneficial in that precise \textit{in-situ} measurement of the density profile automatically allows the simultaneous measurement of compressibility across a wide range of densities.  This technique has recently been used to measure local compressibility in an optical lattice~\citep{Gemelke09}.  The ratio $\kappa\left(r\right)/\kappa_0\left(r\right) \sim dk_F^2\left(r\right)/d\mu\left(r\right)$ is similarly accessible.  In Fig.~\ref{fig:trapped} we have shown the ratio as a function of radial distance in the LDA, and the associated density profile, for KRb at reasonable temperature and peak density.  A gaussian profile with the same peak density and number of molecules as the calculated profile is shown in Figs.~\ref{fig:trapped}(b) and \ref{fig:trapped}(d) for reference.  The 2D results shown are generated from the strictly 2D calculations of Sec.~\ref{subsec:2D}, rather than the more complicated multilayer case.

If the strictly 2D case is realized by applying intense counter-propagating beams of wavelength $\sim 1 \mu$m along the axis of a cigar-shaped cloud that contains $\sim 10^5$ molecules and is $\sim 600 \mu$m in length, with radius $\sim 10 \mu$m~\citep{Zirbel08, Ni09}, then a typical 2D plane will contain $\sim 100$ molecules.  As shown in Fig.~\ref{fig:trapped}(a), to reach suitably large densities with such a small number requires the in-plane trapping frequency to be strong enough that the spatial extent of the cloud is rather small.  This may preclude \textit{in-situ} measurement of the profile, so for that specific method, it is preferable to apply the lattice perpendicular to the axis of the dipole trap in order to deal with larger clouds.  Rather than a cigar-shaped dipole trap, one might instead apply a pancake-shaped trap as in Ref.~\citep{Ni10}, but it would be rather difficult to reach the 2D regime without using an optical lattice.

If the 1D case is realized by applying two orthogonal pairs of counter-propagating beams perpendicular to the axis of the initial cigar-shaped trap, then a typical quasi-1D tube may contain $\sim 1000$ molecules.  This case is shown in Figs.~\ref{fig:trapped}(c) and \ref{fig:trapped}(d).  In addition to the interaction-induced maximum, there is a local minimum for sufficiently high densities, as discussed above.  For those higher trap frequencies, the density decreases more rapidly and the features are correspondingly narrowed.

However, there is the possible complication that in the presence of a trap, the Hartree term is no longer just a constant.  Due to the long-range interactions, one may need to explicitly take the nonuniformity into account in a self-consistent manner with a spatially dependent Hartree term, as was done, for example, in Refs.~\citep{Lin09,Zhang10}.  It is not clear how important this effect is, although one may estimate that it will be negligible if the interaction energy, $d^2/\ell_t^3$, at distances on the order of the characteristic length scale of the trap, $\ell_t$, is very small compared to the local Fermi energy.  Since we have already assumed the local Fermi energy is much larger than $h^2/2m\ell_t^2$ by taking the LDA, this implies $2md^2/h^2 \ll \ell_t$ is a sufficient condition for neglecting spatial variation of the Hartree term.  For KRb, $2md^2/h^2 \sim 30$nm, so it is reasonable to treat the system the way we have as a first step. Further investigation is required, though, to fully resolve the issue.

\section{Conclusions}\label{sec:conclusions}
In this work, our goal has been to propose well-defined experiments which can see quantum many-body effects in a stable, ultracold gas of dipolar molecules at finite temperature.  Our work is admittedly a leading-order approximate calculation, however it should be generally valid qualitatively and, for the coupling strengths accessible to experiment, essentially exact.  Our results indicate that the compressibility of a quasi-low-dimensional gas should exhibit a nonmonotonic temperature dependence for low temperatures.  (Note that a similar maximum in the Hartree-Fock compressibility occurs in an electron gas with $1/r$ interactions, although there it is due to a $t^2 \ln t$ term in the self-energy~\citep{Li09}; see Appendix A.)  Furthermore, the closely related quantity $dE_F/d\mu$, which is the ratio of the finite-temperature interacting compressibility and the zero-temperature noninteracting compressibility, should also exhibit a quantum many-body effect, namely a nonmonotonic \emph{density} dependence, which persists for $T \sim T_F$.

We have also considered the finite temperature behavior of the effective mass and zero sound dispersion.  In the 3D case, the zero sound shows a nonmonotonic temperature dependence, in addition to angular anisotropy.  However, as expected, we have seen that the damping increases rapidly with temperature, so that the mode does not propagate at currently feasible temperatures and densities.  However, in quasi-low-dimensional scenarios, the mode is far more robust, propagating at relatively high temperatures for tight transverse confinement, assuming the length scale at which the interaction deviates from $1/r^3$ behavior is small compared to the transverse size.

We have performed a numerical calculation for a 3D dipolar gas in a 1D periodic potential, and in addition to finding similar behavior as in the 2D case in the presence of multilayer effects, we observe effects of excited bands at high density and the breakdown of two-dimensionality.  In particular, at moderate lattice depths we see discontinuous jumps in the compressibility and interesting collective mode structure.

Finally, we have discussed a method for experimentally observing the characteristic nonmonotonicity of the compressibility in the presence of a global confining potential via high-resolution \textit{in-situ} imaging.

JPK acknowledges helpful discussions with D.-W. Wang.  This work is supported by JQI-AFOSR-MURI.

\appendix
\section{}
Below we briefly compare the results given above for a spinless 2D dipolar gas to results similarly derived for a 2D spinless Coulomb gas with $V\left(r\right) = e^2/r$.  In the latter case, the low-temperature behavior within Hartree-Fock is~\citep{Li09},
\begin{equation}
\frac{\kappa_0}{\kappa} \approx 1 + \frac{r_s}{\pi} \left(- 1 + 0.13 t^2 + \frac{\pi^2}{32} t^2 \ln t \right)
\end{equation}
where $r_s = 1/a_B \sqrt{\pi n}$, with $a_B$ the Bohr radius.  The $t^2 \ln t$ term is not an artifact of the dimensionality, as a similar result has also been obtained for the 3D Coulomb gas~\citep{Li09}.  Here again, the compressibility behaves nonmonotonically, due to the interaction-dependent term first decreasing, then increasing with temperature.  However, in contrast to the dipolar gas, this nonmonotonicity is evident already at leading order in $t$.


\begin{thebibliography}{99}

\bibitem{Lewenstein07} M. Lewenstein, A. Sanpera, V. Ahufinger, B. Damski, A. Sen, U. Sen, Adv. Phys. \textbf{56}, 243 (2007).

\bibitem{Bloch08} I. Bloch, J. Dalibard, W. Zwerger, Rev. Mod. Phys. \textbf{80}, 885 (2008).

\bibitem{Ketterle08} W. Ketterle and M. W. Zwierlein, arXiv:0801.2500v1.

\bibitem{Lang08}  F. Lang, K. Winkler, C. Strauss, R. Grimm, and J. Hecker Denschlag, Phys. Rev. Lett. \textbf{101}, 133005 (2008).

\bibitem{Deiglmayr08}  J. Deiglmayr, A. Grochola, M. Repp, K. Mortlbauer, C. Gluck, J. Lange, O. Dulieu, R. Wester, and M. Weidemuller, Phys. Rev. Lett. \textbf{101}, 133004 (2008).

\bibitem{Ospelkaus08}  S. Ospelkaus, A. Pe'er, K.-K. Ni, J. J. Zirbel, B. Neyenhuis, S. Kotochigova, P. S. Julienne, J. Ye, and D. S. Jin, Nat. Phys. \textbf{4}, 622 (2008).

\bibitem{Zirbel08} J. J. Zirbel, K.-K. Ni, S. Ospelkaus, T. L. Nicholson, M. L. Olsen, P. S. Julienne, C. E. Wieman, J. Ye, and D. S. Jin, Phys. Rev. A \textbf{78}, 013416 (2008).

\bibitem{Ni08}  K.-K. Ni, S. Ospelkaus, M. H. G. de Miranda,  A. Pe'er, B. Neyenhuis, J. J. Zirbel, S. Kotochigova,  P. S. Julienne, D. S. Jin, and J. Ye, Science \textbf{322}, 231 (2008).

\bibitem{Ni09} K.-K. Ni, S. Ospelkaus, D.J. Nesbitt, J. Ye and D. S. Jin, Phys. Chem. Chem. Phys \textbf{11}, 9626 (2009).

\bibitem{Ospelkaus09}  S. Ospelkaus, K.-K. Ni, G. Quemener, B. Neyenhuis, D. Wang, M. H. G. de Miranda, J. L. Bohn, J. Ye and D. S. Jin, arXiv:0908.3931.

\bibitem{Ni10} K.-K. Ni, S. Ospelkaus, D. Wang, G. Quemener, B. Neyenhuis, M. H. G. de Miranda, J. L. Bohn, J. Ye, and D. S. Jin, arXiv:1001.2809v1.

\bibitem{Micheli06} A. Micheli, G. K. Brennen, and P. Zoller, Nat. Phys. \textbf{2}, 341 (2006).

\bibitem{Ortner09}  M. Ortner, A. Micheli, G. Pupillo, P. Zoller, New J. Phys. \textbf{11}, 055045 (2009).

\bibitem{Santos03} L. Santos, G. V. Shlyapnikov, and M. Lewenstein, Phys. Rev. Lett. \textbf{90}, 250403 (2003).

\bibitem{Ronen07}  S. Ronen, D. C. E. Bortolotti, and J. L. Bohn, Phys. Rev. Lett. \textbf{98}, 030406 (2007).

\bibitem{Wang08}  D.-W. Wang and E. Demler, arXiv:0812.1838.

\bibitem{Sengupta05} P. Sengupta, L. P. Pryadko, F. Alet, M. Troyer, and G. Schmid, Phys. Rev. Lett. \textbf{94}, 207202 (2005).

\bibitem{Boninsegni05} M. Boninsegni and N. Prokof'ev, Phys. Rev. Lett. \textbf{95}, 237204 (2005).

\bibitem{Capogrosso09}  B. Capogrosso-Sansone, C. Trefzger, M. Lewenstein, P. Zoller, G. Pupillo, arXiv:0906.2009v1.

\bibitem{Pollet09} L. Pollet, J. D. Picon, H. P. Buechler, M. Troyer, arXiv:0906.2126v1.

\bibitem{Burnell09} F. J. Burnell, M. M. Parish, N. R. Cooper, S. L. Sondhi, arXiv:0901.4366v1.

\bibitem{Tewari06} S. Tewari, V. W. Scarola, T. Senthil, S. Das Sarma, Phys. Rev. Lett. \textbf{97}, 200401 (2006).

\bibitem{Wang07}  D.-W. Wang, Phys. Rev. Lett. \textbf{98}, 060403 (2007).

\bibitem{Wang06}  D.-W. Wang, M. D. Lukin, and E. Demler, Phys. Rev. Lett. \textbf{97}, 180413 (2006).

\bibitem{Lutchyn09} Roman M. Lutchyn, Enrico Rossi, S. Das Sarma, arXiv:0911.1378v1.

\bibitem{Lin09}  Chien-Hung Lin, Yi-Ting Hsu, Hao Li, Daw-Wei Wang, arXiv:0907.3125v1.

\bibitem{Sogo09} T. Sogo, L. He, T. Miyakawa, S. Yi, H. Lu, and H. Pu, New J. Phys. \textbf{11}, 055017 (2009).  The original work contains an error in the stability criterion, with the corrected critical value being cited in Ref.~\citep{Ronen10}.

\bibitem{Miyakawa08} T. Miyakawa, T. Sogo, and H. Pu, Phys. Rev. A \textbf{77}, 061603(R) (2008).

\bibitem{Chan10} C.-K. Chan, C.-J. Wu, W.-C. Lee, and S. Das Sarma, Phys. Rev. A \textbf{81}, 023602 (2010).

\bibitem{Baranov05} M. A. Baranov, K. Osterloh, and M. Lewenstein, Phys. Rev. Lett. \textbf{94}, 070404 (2005).

\bibitem{Baranov08a}  M. A. Baranov, H. Fehrmann, and M. Lewenstein, Phys. Rev. Lett. \textbf{100}, 200402 (2008).

\bibitem{Negele88}  W. Negele and H. Orland, \textit{Quantum many-particle systems}, (Perseus Books (Sd), 1988).

\bibitem{Fregoso09}  B. M. Fregoso, K. Sun, E. Fradkin and B. L. Lev, New J. Phys. \textbf{11}, 103003 (2009).

\bibitem{Cooper09} N. R. Cooper and G. V. Shlyapnikov, Phys. Rev. Lett. \textbf{103}, 155302 (2009).

\bibitem{Sun09}  K. Sun, E. Zhao, W. V. Liu, arXiv:0908.0190v2.

\bibitem{Baranov08b}  M. A. Baranov, Physics Reports \textbf{464}, 71 (2008).

\bibitem{Pupillo08}  G. Pupillo, A. Micheli, H. P. Buchler, P. Zoller, in \textit{Cold molecules: Creation and applications}, edited by R. V. Krems, B. Friedrich and W. C. Stwalley (Taylor \& Francis, 2008), preprint arXiv:0805.1896v1.

\bibitem{Carr09}  L. D. Carr, D. DeMille, R. V. Krems and J. Ye, New J. Phys. \textbf{11}, 055049 (2009).

\bibitem{Micheli10} A. Micheli, Z. Idziaszek, G. Pupillo, M. A. Baranov, P. Zoller, and P. S. Julienne, arXiv:1004.5420v1.

\bibitem{Eisenstein94}  J. P. Eisenstein, L. N. Pfeiffer, and K. W. West, Phys. Rev. B \textbf{50}, 1760 (1994).

\bibitem{Rahimi03}  M. Rahimi, M. R. Sakr, S. V. Kravchenko, S. C. Dultz and H. W. Jiang, Phys. Rev. B \textbf{67}, 081302(R) (2003).

\bibitem{Martin08} J. Martin, N. Akerman, G. Ulbricht, T. Lohmann, J. H. Smet, K. von Klitzing and A. Yacoby, Nat. Phys. \textbf{4}, 144 (2008).

\bibitem{Hwang07}  E. H. Hwang, Ben Yu-Kuang Hu, and S. Das Sarma, Phys. Rev. Lett. \textbf{99}, 226801 (2007).

\bibitem{DasSarma09}  S. Das Sarma, E. H. Hwang, and Qi Li, Phys. Rev. B \textbf{80}, 121303(R) (2009).

\bibitem{Ronen10}  S. Ronen and J.L. Bohn, Phys. Rev. A \textbf{81}, 033601 (2010).

\bibitem{Mahan} G. D. Mahan, \textit{Many-Particle Physics}, (Plenum Press, New York, 1981).

\bibitem{Pethick73} C. J. Pethick and G. M. Carneiro, Phys. Rev. A \textbf{7}, 304 (1973).

\bibitem{Gradshteyn} Defined as in I. S. Gradshteyn and I. M. Ryzhik, \textit{Tables of Integrals, Series, and Products, 4th ed.}, (Academic Press, New York, 1980).

\bibitem{Li10} Qiuzi Li, E. H. Hwang, and S. Das Sarma, to be published.

\bibitem{Mazzarella09} G. Mazzarella, L. Salasnich, and F. Toigo, Phys. Rev. A \textbf{79}, 023615 (2009).

\bibitem{Kohn61} W. Kohn, Phys. Rev. \textbf{123}, 1242 (1961).

\bibitem{Gemelke09} N. Gemelke, X. Zhang, C.-L. Hung and C. Chin, Nature \textbf{460}, 995 (2009).

\bibitem{Zhang10}  J.-N. Zhang and S. Yi, arXiv:1001.0426v1.

\bibitem{Li09} Qiuzi Li, E. H. Hwang, and S. Das Sarma, to be published.

\end{thebibliography}
\end{document}